\documentclass[11pt]{article}
\usepackage{amssymb,cite,graphicx}
\newcommand\pbl[1]{\lfloor #1\rfloor}
\newcommand{\fa}[1]{\centerline{\includegraphics[width=0.75\linewidth,%
height=0.52\linewidth]{#1}}}
\newcommand{\fb}[1]{\centerline{\includegraphics[width=0.7\linewidth,%
height=0.52\linewidth]{#1}}}
\newcommand{\fc}[1]{\centerline{\includegraphics[width=0.6\linewidth,%
height=0.52\linewidth]{#1}}}
\newcommand{\pl}[1]{^{\phantom{j_j}}_{#1}}

\newcommand{\pll}[1]{_{#1}}
\newcommand{\blnkt}[1]{}

\newcommand\blank[1]{}

\newcommand\toline[1]{--#1}
\newcommand{\fract}[2]{{\textstyle\frac{#1}{#2}}}

\newcommand{\CS}{{\cal S}}

\newcommand\Zth{{\mathbb Z}_3}

\newcommand{\CA}{{\cal A}}
\newcommand{\CM}{{\cal M}}

\newcommand\eq{\begin{equation}}
\newcommand\en{\end{equation}}
\newcommand\bea{\begin{eqnarray}}
\newcommand\eea{\end{eqnarray}}
\newcommand\nn{\nonumber}
\newcommand\ba{\(\begin{array}}
\newcommand\ea{\end{array}\)}

\newcommand{\resection}[1]{\setcounter{equation}{0}\section{#1}}

\newcommand{\lo}{\ensuremath{\mathit{l}}}
\newcommand{\lt}{\ensuremath{\mathit{l}'}}
\newcommand{\lth}{\ensuremath{\mathit{l}''}}

\newcommand{\fra}[2]{\ensuremath{\mbox{\footnotesize $\frac{#1}{#2}$}}}
\begin{document}
\begin{titlepage}
\vskip 0.5cm
\begin{flushright}
DCPT-02/51 \\
{\tt hep-th/0208111} \\
\end{flushright}
\vskip 1.5cm
\begin{center}
{\Large{\bf
Integrable aspects
of the scaling $q$-state Potts models I:\\[5pt]
bound states and bootstrap closure
}}
\end{center}
\vskip 0.8cm
\centerline{Patrick Dorey%
\footnote{e-mail: {\tt p.e.dorey@durham.ac.uk}},
Andrew Pocklington$^2$%
\phantom{\footnote{e-mail: {\tt andrew@ift.unesp.br}}}
and Roberto Tateo%
\footnote{e-mail: {\tt roberto.tateo@durham.ac.uk}}}
\vskip 0.9cm
\centerline{${}^{1,3}$\sl\small Dept.~of Mathematical Sciences,
University of Durham, Durham DH1 3LE, UK\,}
\vskip 0.2cm
\centerline{${}^2$\sl\small IFT/UNESP, Instituto de Fisica Teorica, 
01405-900, Sao Paulo - SP, Brasil\,}
\vskip 1.25cm
\begin{abstract}
\noindent
We discuss the $q$-state Potts models for $q\le 4$, in the
scaling regimes close to their critical or tricritical points.
Starting from the kink S-matrix elements
proposed by Chim and Zamolodchikov,
the bootstrap is closed for the scaling regions of all critical
points, and for the tricritical points when $4>q\ge 2$. 
We also note a curious appearance of the extended last line of
Freudenthal's magic square in connection with the Potts models.
\end{abstract}
\end{titlepage}
\setcounter{footnote}{0}
\def\thefootnote{\fnsymbol{footnote}}
%
\resection{Introduction}
\label{intr}
The $q$-state Potts models directly generalise
the most well-known of all two-dimensional
integrable models, the Ising model. 
They have been much studied, both in their own right as interesting
statistical-mechanical systems, and because of their relations with other
models -- the limit $q\to 1$, for example, describes bond percolation.
However, they are by no means completely understood.

In this paper and its sequel we shall discuss the treatment of these
models in the framework of continuum field theory.
Such techniques are 
expected to be applicable in scaling regimes near to
critical points, though for the $q$-state Potts models
some elements of the treatment will be rather formal, reflecting
the nonlocal manner in which the models are initially defined on the lattice.
In this paper we focus on the description of the models in
terms of the on-shell data provided by an exact S-matrix. A number of
years ago, 
Chim and Zamolodchikov proposed a set of amplitudes for the
scattering of elementary kink-like excitations in the low-temperature
phase of the model~\cite{CZ}. 
(A different treatment had previously been
suggested by Smirnov \cite{Sm},
based on quantum-group reductions of the Izergin-Korepin S-matrix. 
In this article we shall work from the Chim-Zamolodchikov formulation,
as it more closely reflects the continuous nature of the 
Fortuin-Kasteleyn~\cite{KF}
definition
of the theory on a lattice, but we note that
the relationship between the two approaches has recently been
clarified, in \cite{FR}.)
The fundamental S-matrix elements form only part of the on-shell
description of the model, and
in order to complete the picture it is necessary to
find out if any further asymptotic states are present.
In the exact S-matrix framework this is commonly achieved by
an analysis of the pole structure of all S-matrix elements,  
a process which is known as closure of the bootstrap.
The case of
the $q$-state Potts models turns out to involve
a number of subtleties, which we attempt to highlight and resolve 
in this paper.
Since we have not been entirely successful in this enterprise,
we also include some details of the problems that we encountered. 
They only arose when continuing the S-matrices far into the regime of
tricritical models, but they may nonetheless be important signals of
new behaviour in the bootstrap programme.

In \S2, we review Chim and Zamolodchikov's proposal and its
background, stressing some of the non-standard features that the
models exhibit. Then in \S3 we close the bootstrap for the scaling
regions of all critical models, finding that we need to invoke the
Coleman-Thun mechanism in a novel way in order
to explain the full
pole structure of all S-matrix elements.
A continuation of Chim and Zamolodchikov's S-matrix is
expected to describe the scaling regions of the tricritical models,
and this is discussed in \S4.
Finally, \S5 contains our conclusions and 
some more detailed tables are collected in four appendices. 

In a companion paper \cite{pottsII}, these models will be discussed from a
complementary point of view, using finite-size effects.

\resection{Review}
\label{revsec}
\subsection{The models on the lattice and their continuum limits}
The standard definition of the lattice $q$-state Potts model is
through the Hamiltonian
\eq
{\cal H} =-J\sum_{\langle x,y\rangle}\delta_{\sigma(x),\sigma(y)}
\label{oldh}
\en
where $J$ is a coupling strength and
the spin $\sigma(x)$ associated with the lattice site $x$ may
take
any of $q$ distinct values. The summation is over all 
nearest-neighbour pairs of sites $\langle x,y\rangle$. 
In terms of ${\cal H}$, the partition function at temperature $T$
is given by
\eq
{\cal Z}  =  \sum_{\{\sigma\}}e^{-\frac{1}{kT}{\cal H}}~.
\label{oldpf}
\en
Notice that ${\cal H}$ is 
invariant under the group $S_q$ of permutations of the $q$ possible 
values of $\sigma$.

This definition
only applies for integer values of $q$, but
a reformulation 
due to Fortuin and Kasteleyn~\cite{KF}
allows the constraint to be lifted.
Expanding (\ref{oldpf}) as
\eq
{\cal Z} = \sum_{\{\sigma\}}\prod_{\langle x,y\rangle}
\left(1+(e^{\frac{J}{kT}}-1)\delta_{\sigma(x),\sigma(y)}\right)
\en
they observed that it could be written in the form
\eq
{\cal Z}
  =  \sum_{{\cal G}}K^{\nu}q^C \label{eq:Zsum}
\en
where $K=e^{\frac{J}{kT}}-1$, and  the sum is over all graphs 
${\cal G}$ on the lattice, with $\nu$ the number of bonds in ${\cal G}$
and $C$ the
number of connected components (sets of sites joined by bonds). 
When the model is defined in this way, $q$ is no longer restricted to 
integer values, and can be taken as a continuous parameter.
However, at general values of $q$, ${\cal Z}$ cannot be written in
terms of a local Hamiltonian, and the precise meaning of its `$S_q$'
symmetry is not clear. Nevertheless, the partition function
is certainly well-defined,
and we shall see later that many other
features which might be thought special to locally-defined
theories also make sense.

The model undergoes a phase transition at $K = K_c = \sqrt{q}$\,, 
which is second-order for
$q\le 4$. At these values of $q$ 
a continuum limit can be taken, and
if the limit is taken exactly at the critical point,
the resulting field
theory is conformal. Parametrising $q$ as
\eq
 \sqrt{q}=2\sin\gamma = 2\cos\biggl(\frac{\pi}{\xi{+}1}\biggr)
\label{gamdef}
\en
with $0\le \gamma\le \pi/2$ and $\xi=(\pi{+}2\gamma)/(\pi{-}2\gamma)$, 
its central charge
is \cite{Dot}
\eq
c(q)=1-\frac{6}{\xi(\xi{+}1)}~.
\label{cformu}
\en

A generalisation of the basic Potts model allows for the existence of
vacancies \cite{Nienh}. In addition to the critical points just discussed,
these models have tricritical points, with the critical and tricritical 
points coinciding at $q=4$. 
The exponents for the tricritical points can be 
obtained from those of the critical points by continuing $\gamma$ into
the range $\pi/2\le \gamma\le \pi$ \cite{Nienh} (see also \cite{Bu,Na}).
This shifts $\xi$ into the range $-\infty<\xi<-3$.

If $\xi$ is rational, the value of $c(q)$ coincides with the central
charge $c=1-6(p'{-}p)^2/pp'$ of a minimal model $\CM_{pp'}\,$, $p'>p$.
More precisely, the relationship is
\eq
\xi=\frac{p}{p'{-}p}
\label{xicrit}
\en
on the critical branch, and
\eq
\xi=\frac{p'}{p{-}p'}
\label{xitricrit}
\en
on the tricritical branch. This coincidence does {\em not\/} mean that
the conformal field theories of the Potts models at rational $\xi$ are
minimal models -- in particular, the relevant partition functions 
only agree when $q$ is an integer \cite{dFSZ} -- but it does mean that
connections with minimal models are to be expected at these points.

Other aspects of the critical and tricritical models, and of their
relationships with conformal field theories, are reviewed in
\cite{pottsII}.
However in this paper we are more concerned with the massive field
theories which arise if a continuum limit 
is taken with the temperature (and any other parameters)
tending to a critical value with
the correlation length held finite in physical units.
In many interesting cases (including the ones currently under
discussion \cite{Sm,CZ}) these field theories turn out to be 
integrable \cite{Zam1}, allowing them
to be understood in considerable detail. In turn, this
allows the scaling regions of the original lattice models near to 
their second-order transitions
to be explored; it is also relevant to the
computation of certain universal characteristics of the transitions
themselves~\cite{CD}. 

In practice, the field theories are usually found
directly in the continuum, either by a consideration of the symmetries
that they should inherit from the lattice, or
by starting from the continuum conformal field theories, and then adding
to their actions suitable continuum operators to describe the departure
from criticality -- the basic idea of perturbed conformal field theory,
as put forward by Zamolodchikov \cite{Zam1}.
For shifts in temperature, the operator to add is the local
energy density, which corresponds to $\phi_{21}$ for the critical Potts models
\cite{Dot}, and $\phi_{12}$ for the tricritical models \cite{DF}.
We shall refer to the massive theories as the scaling Potts models.
Strictly speaking, they are not
perturbations of minimal conformal field theories, 
except at integer values of $q$. However, in situations where the
infinite-volume ground states of the $q$-state Potts and minimal
models coincide, we expect the two to agree on issues such as mass
spectrum and (diagonal) S-matrix elements.
We shall
say a little more about this issue in later sections and in
\cite{pottsII}, but since it is
not directly relevant to our current discussion of the Potts S-matrices, 
we leave it to one side for now.

Irrespective of whether the unperturbed $c<1$ theories are minimal models,
Zamolodchikov's counting argument~\cite{Zam1} shows that generic
$\phi_{21}$ or $\phi_{12}$ perturbations 
preserve conserved charges with 
spins $s=1,~5,~7$ and $11$.
By an argument due to Parke~\cite{parke}, the existence of these 
charges
is enough to ensure that,
as quantum field theories defined in Minkowski space, the models
must have factorisable S-matrices, allowing the full multiparticle
S-matrix to be
given in terms of the set of two-particle scattering 
amplitudes\footnote{In all probability the models actually
have conserved charges at all integer spins not multiples of $2$ or $3$, but 
just two with spins $>1$ are enough for Parke's argument to go through.}.
The main result
of Chim and Zamolodchikov~\cite{CZ} was a conjecture for these
amplitudes for certain fundamental, kink-like excitations. To finish 
the story, amplitudes for
the scattering of all possible bound states of these kinks must be found, 
and this is the main goal of the present paper.
In the remainder
of this section, we shall review
the assumptions that went into Chim and Zamolodchikov's initial
proposal. 

\subsection{The fundamental S-matrix elements}
To begin, we discuss the scaling regions of the {\em critical\/} models.
The scattering theory is most easily treated by working in the
low-temperature, ordered phase, in which one would expect to find
$q$ degenerate vacua. It is then natural to postulate the existence of
a set of particles, or kinks, $K_{ab}(\theta)$
($a,b=1\ldots q\,;~a\neq b$), domain walls which interpolate
between different vacua. The fact that $q$ is not necessarily an integer
may appear worrysome, but following \cite{CZ} we can decide to treat $q$ 
formally, motivated at least in part by the fact that this can be given
a precise sense on the lattice through the Fortuin-Kasteleyn trick reviewed 
above.
The energy-momentum of $K_{ab}(\theta)$ is
parameterized by the rapidity $\theta$,
$p^{\mu}=(m\cosh\theta,m\sinh\theta)$, and asymptotic $n$-particle states
interpolating between vacua $a_0$ and $a_n$ correspond to the products
\eq
K_{a_0a_1}(\theta_1)K_{a_1a_2}(\theta_2)\ldots K_{a_{n-
1}a_n}(\theta_n)\ \ \ \ \ \ \ \ \ (a_i\neq a_{i+1})~.
\en
Scattering of these kinks is completely described by the two-particle
amplitudes ${\cal S}_{ac}^{bd}(\theta)$:
\eq
K_{ab}(\theta_1)K_{bc}(\theta_2) = 
\sum_{d\neq a,c}{\cal S}_{ac}^{bd}(\theta)K_{ad}(\theta_2)K_{dc}(\theta_1)~,
\qquad \theta=\theta_1-\theta_2~.
\en

\begin{figure}[t]
\[
\begin{array}{llll}
\includegraphics[width=0.15\linewidth]{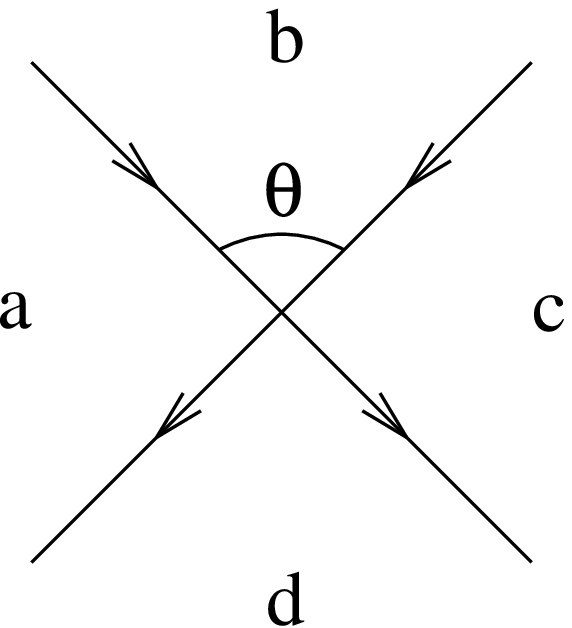}{~~~~~~} &
\includegraphics[width=0.15\linewidth]{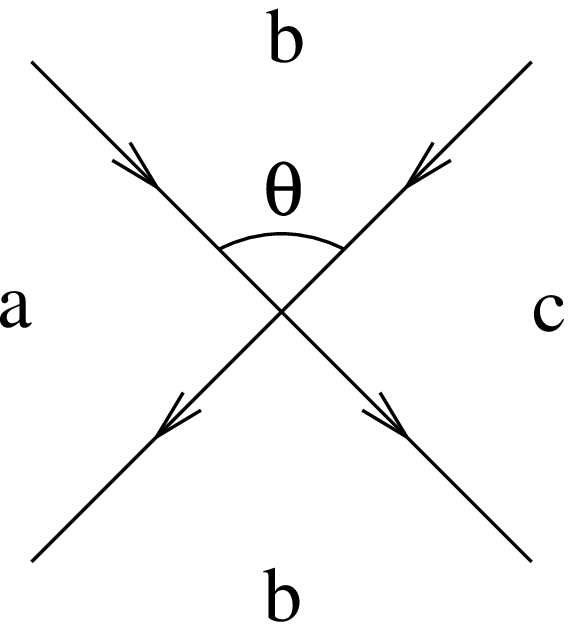}{~~~~~~} &
\includegraphics[width=0.15\linewidth]{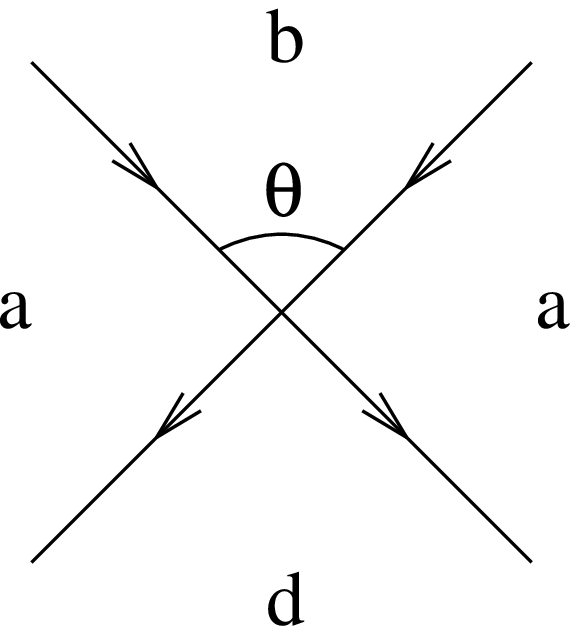}{~~~~~~} &
\includegraphics[width=0.15\linewidth]{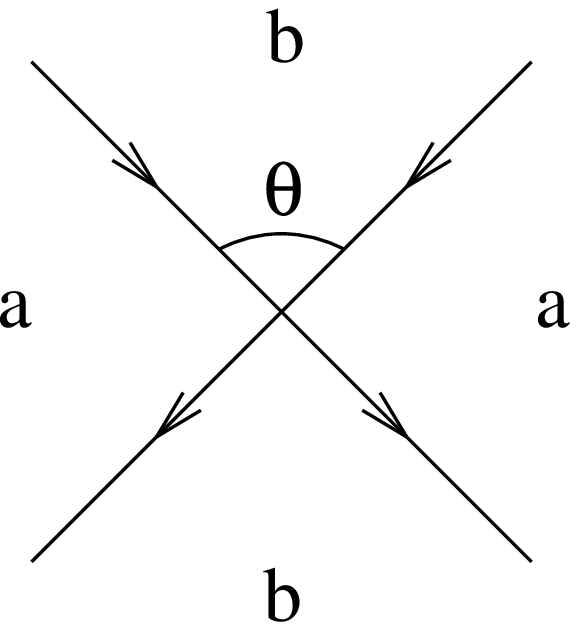} \\[4pt]
\parbox[t]{.17\linewidth}{\raggedright%
{~~~}$\CS^0(\theta)$}~&
\parbox[t]{.17\linewidth}{\raggedright%
{~~~}$\CS^1(\theta)$}~& 
\parbox[t]{.17\linewidth}{\raggedright%
{~~~}$\CS^2(\theta)$}~&
\parbox[t]{.17\linewidth}{\raggedright%
{~~~}$\CS^3(\theta)$
} 
\end{array} 
\]
\caption{ The four independent two-particle amplitudes}
\label{4elements}
\end{figure}

Due to the $S_q$ symmetry of the model, there are only four
independent two-particle amplitudes $\CS^n(\theta)$ ($n=0\ldots3$).
These are represented in figure~\ref{4elements}.

The  S-matrix 
must be unitary, crossing symmetric, and satisfy the Yang-Baxter
equations. Chim and Zamolodchikov found that the 
latter implies
\begin{eqnarray}
\CS^0(\theta) & = &
\sin(\gamma)\sin(i\lambda\theta)\sin(3\gamma+i\lambda\theta)R(\theta) 
\label{C0d}\\
\CS^1(\theta) & = &
\sin(2\gamma)\sin(\gamma+i\lambda\theta)
\sin(3\gamma+i\lambda\theta)R(\theta) 
\label{C1d}\\
\CS^2(\theta) & = &
\sin(2\gamma)\sin(i\lambda\theta)\sin(2\gamma+i\lambda\theta)R(\theta) 
\label{C2d}\\
\CS^3(\theta) & = &
\sin(3\gamma)\sin(\gamma+i\lambda\theta)\sin(2\gamma+i\lambda\theta)R(\theta)
\label{C3d}
\end{eqnarray}
where
$\gamma$ is defined by $2\sin\gamma= \sqrt{q}$ as in
(\ref{gamdef}) above, and $\lambda$ is an as-yet
undetermined parameter. Crossing symmetry requires that $\CS^0$ and
$\CS^3$ be unchanged under $\theta\to i\pi{-}\theta$ while
$\CS^1$ and $\CS^2$ swap over, which implies
$\pi\lambda = 3\gamma \ \bmod~\pi$ and
$R(i\pi-\theta)=R(\theta)$.
Unitarity then boils down to the condition
\eq
R(\theta)R(-\theta)  =  [\sin^2(\gamma)\sin(2\gamma +
i\lambda\theta)\sin(2\gamma - i\lambda\theta)\sin(3\gamma +
i\lambda\theta)\sin(3\gamma - i\lambda\theta)]^{-1}
\en
Together with crossing, this fixes the S-matrix as a function of
$\gamma$ and $\lambda$ up to a so-called
CDD factor, a $2\pi i$-periodic function $f$ satisfying
$f(\theta)f(-\theta)=1$, $f(i\pi{-}\theta)=f(\theta)$.
To resolve the remaining ambiguities, some further physical input
is required, and this comes from an initial
consideration of the pole structure.

\begin{figure}[h]
\[
\begin{array}{cc}
\includegraphics[width=0.129\linewidth]{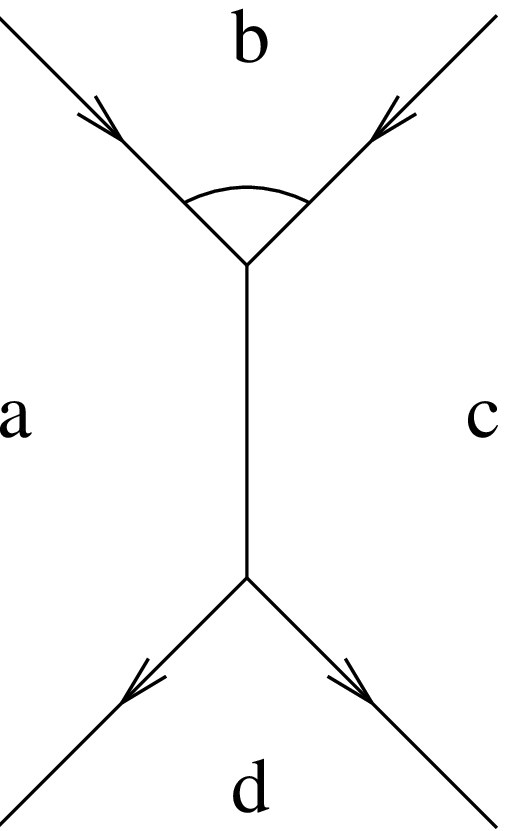} & \hspace{3cm}
\includegraphics[width=0.129\linewidth]{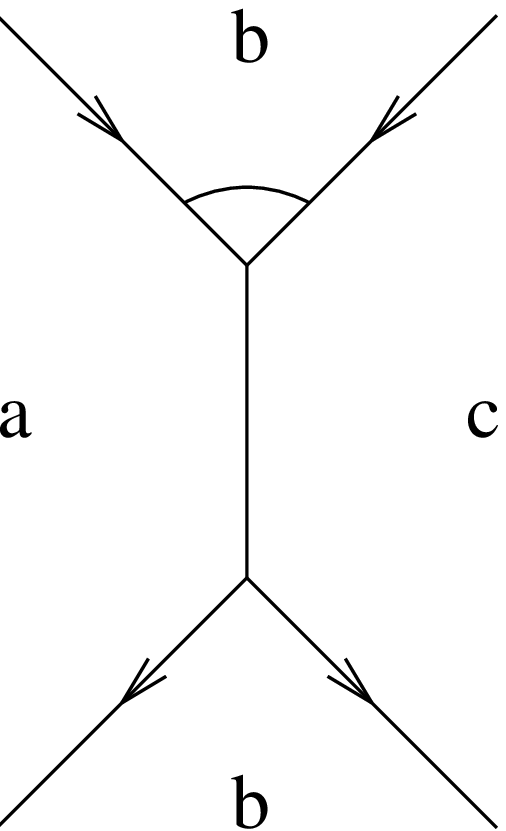} 
\\
\end{array}
\]
\caption{Direct channel $KK\rightarrow K$ bound states 
in $\CS^0$ and $\CS^1$}
\label{Direct}
\end{figure}

\begin{figure}
\[
\begin{array}{cc}
\includegraphics[width=0.22\linewidth]{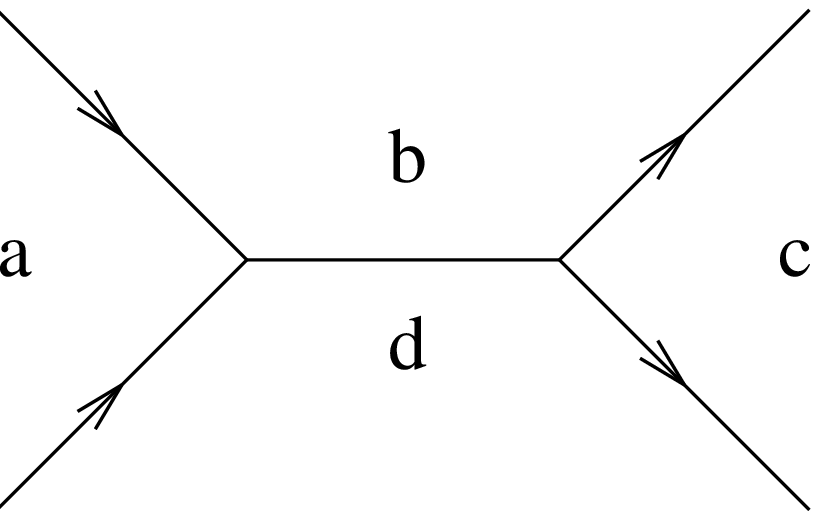} & \hspace{2cm}
\includegraphics[width=0.22\linewidth]{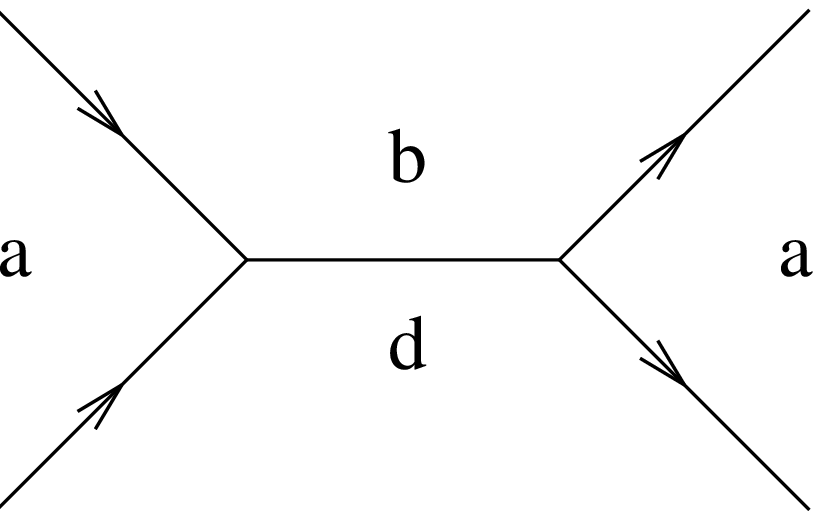}
\\
\end{array}
\]
\caption{Cross channel $KK\rightarrow K$ bound states
in $\CS^0$ and $\CS^2$}
\label{Cross}
\end{figure}

In general, it should be possible to explain all S-matrix
poles in the physical strip
\eq
0\le \Im m\,\theta\le \pi
\en
in terms of the bound-state structure of the model (though, as we
shall see, the mechanism can in some cases be quite involved).
Conversely, the fundamental
S-matrix elements should certainly exhibit
poles reflecting
the possibility to form an elementary kink as a bound state of two
other such kinks.
Taking
energy-momentum conservation and vacuum structure 
into account, direct-channel poles should appear in
$\CS^0(\theta)$ and
$\CS^1(\theta)$ at $\theta=\frac{2\pi i}{3}$, and cross-channel 
poles in $\CS^0(\theta)$ and $\CS^2(\theta)$ 
at $\theta=\frac{\pi i}{3}$. These are illustrated in 
figures~\ref{Direct} and \ref{Cross}.

A comparison
with (\ref{C0d}) -- (\ref{C2d}) shows that the common factor
$R(\theta)$ must therefore
have poles at $\frac{\pi i}{3}$ and $\frac{2\pi i}{3}$\,.
On the other hand, the pole at $\frac{2\pi i}{3}$ should be absent from
$\CS^2$ and $\CS^3$, and the pole at $\frac{\pi i}{3}$ absent from
$\CS^1$ and $\CS^3$. This can only be achieved by cancellations 
against
zeroes from the sine prefactors in (\ref{C1d}) -- (\ref{C3d}), 
which strengthens the previous condition on $\lambda$ to
$\pi\lambda=3\gamma \ \bmod~3\pi$.

\begin{figure}[hb]
\[
\begin{array}{c}
\includegraphics[width=0.68\linewidth]{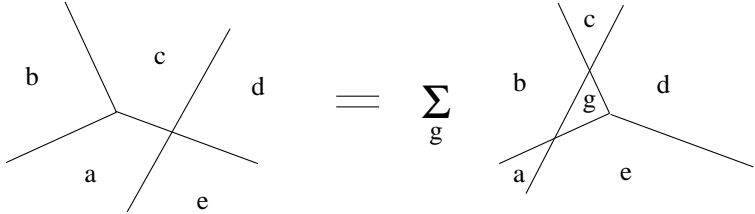}
\end{array}
\]
\caption{ $KK\rightarrow K$ bootstrap}
\label{kboot}
\end{figure}

The three-kink coupling leads to a set of
bootstrap equations, depicted in figure \ref{kboot}.
In contrast to the Yang-Baxter equations, these constraints
are felt by the CDD factor.
Combining these conditions with a prior knowledge of
the solutions for $q=2$ and $3$ led Chim and Zamolodchikov
to the simplest choice
$\pi\lambda=3\gamma$, and a minimal
solution for the S-matrix elements which, for the discussions to come, it
will be most convenient 
to rewrite in the following form:
\begin{eqnarray}
\CS^0(\theta) & = &
 \frac{\sinh(\lambda\theta)}%
{\sinh(\lambda(\theta{-}\frac{2\pi i}{3}))}
\,S(\theta)~, \label{eq:k1smtx0}\\[4pt]
\CS^1(\theta) & = & 
\frac{\sin(\frac{2\pi\lambda}{3})}%
{\sin(\frac{\pi\lambda}{3})}
\frac{\sinh(\lambda(\theta{-}\frac{\pi i}{3}))}%
{\sinh(\lambda(\theta{-}\frac{2\pi i}{3}))}
\,S(\theta)~, \label{eq:k1smtx1}\\[4pt]
\CS^2(\theta) & = & 
\frac{\sin(\frac{2\pi\lambda}{3})}%
{\sin(\frac{\pi\lambda}{3})}
\frac{\sinh(\lambda\theta)}%
{\sinh(\lambda(\theta{-}i\pi))}
\,S(\theta)~, \label{eq:k1smtx2}\\[4pt]
\CS^3(\theta) & = & 
\frac{\sin(\lambda\pi)}{\sin(\frac{\pi\lambda}{3})}
\frac{\sinh(\lambda(\theta{-}\frac{\pi i}{3}))}%
{\sinh(\lambda(\theta{-}i\pi))}
\,S(\theta)~.\label{eq:k1smtx3} 
\end{eqnarray}
The overall scalar factor $S(\theta)$ is
\eq
S(\theta)=
\frac{\sinh(\lambda(\theta{+}\frac{\pi
i}{3}))}%
{\sinh(\lambda(\theta{-}\frac{\pi i}{3}))}\,
e^{\CA(\theta)}
\en
with $e^{\CA(\theta)}$ defined in terms of the blocks
\eq
\mu(a)=
\frac{\Gamma(a-\frac{\lambda}{i\pi}\theta)}%
     {\Gamma(a+\frac{\lambda}{i\pi}\theta)}
\frac{\Gamma(a+\lambda+\frac{\lambda}{i\pi}\theta)}%
     {\Gamma(a+\lambda-\frac{\lambda}{i\pi}\theta)}
=
\exp\left[ 2\!\int_0^{\infty}\frac{dx}{x}%
\sinh(\fract{\lambda}{i\pi}\theta x)e^{-ax}
\frac{(1{-}e^{-\lambda x})}{(1{-}e^{-x})}\right]
\label{blockdef}
\en
as
\eq
e^{\CA(\theta)}=
\prod_{k=0}^{\infty}
\frac{\mu(1+2k\lambda)}{\mu((2k{+}1)\lambda)}
\frac{\mu(1+(2k{-}\frac{1}{3})\lambda)}{\mu((2k{+}\frac{4}{3})\lambda)}~.
\en
{}From the second equality in (\ref{blockdef}) and the 
formula
\eq
\frac{\sinh(\lambda(\theta{+}i\pi\alpha))}%
{\sinh(\lambda(\theta{-}i\pi\alpha))}\,
=\exp\left[-2\!\int_0^{\infty}\frac{dk}{k}%
\sinh(ik\theta)\frac%
{\sinh(\frac{\pi k}{2}(\frac{1}{\lambda}{-}2\alpha))}
{\sinh(\frac{\pi k}{2\lambda})}\right]
\en
it is 
a simple matter to recover integral representations
equivalent to those given in \cite{CD}\,:
\bea
\CA(\theta)&=&
-2\int_0^{\infty}\frac{dk}{k} \sinh(ik\theta)
\frac{\cosh(\frac{\pi k}{6})
\sinh(\frac{\pi k}{2}(\frac{4}{3}{-}\frac{1}{\lambda}))}%
{\cosh(\frac{\pi k}{2})
\sinh(\frac{\pi k}{2\lambda})}~;\\[3pt]
\log S(\theta)&=&
-2\int_0^{\infty}\frac{dk}{k} \sinh(ik\theta)
\frac{\cosh(\frac{\pi k}{2}(\frac{1}{3}{-}\frac{1}{\lambda}))
\sinh(\frac{\pi k}{3})}%
{\cosh(\frac{\pi k}{2})
\sinh(\frac{\pi k}{2\lambda})}~.
\eea
In addition, $e^{\CA(\theta)}$ and $S(\theta)$ satisfy
\bea
e^{\CA(-\theta)}= e^{-\CA(\theta)}&,&
e^{\CA(i\pi -\theta)}=
\frac{\sinh(\lambda\theta)}%
{\sinh(\lambda(\theta{-}i\pi))}
\frac{\sinh(\lambda(\theta{+}\frac{\pi i}{3}))}%
{\sinh(\lambda(\theta{-}\frac{4\pi i}{3}))}
\, e^{\CA(\theta)}~;\\[3pt]
S(-\theta)= 1/S(\theta)&,&
S(i\pi {-}\theta)=
\frac{\sinh(\lambda\theta)}%
{\sinh(\lambda(\theta{-}i\pi))}
\frac{\sinh(\lambda(\theta{-}\frac{\pi i}{3}))}%
{\sinh(\lambda(\theta{-}\frac{2\pi i}{3}))}
\, S(\theta)~.
\label{ucross}
\eea
The poles and zeroes of these
two functions which can appear in the physical strip 
for $0<\lambda<3$ are summarised in tables 
\ref{scalpolesA} and
\ref{scalpolesS}.
Note that $e^{\CA(\theta)}$ has no physical strip poles at all for
$\lambda\le 1$.

\begin{table}
\begin{center}
\begin{tabular}{|l||c|c|c|c|c||c|} 
\hline
&\multicolumn{5}{c||}{}&
\multicolumn{1}{c|}{}\\[-9pt]
&\multicolumn{5}{c||}{Poles}&
\multicolumn{1}{c|}{\,Zeroes~}\\[2pt]
\hline
& & & & & & \\[-7pt]
$~~~\frac{\theta}{i\pi}~=$ & $\frac{1}{\lambda}-\frac{1}{3}$ &
$\frac{1}{\lambda}$ & $\frac{2}{\lambda}-\frac{1}{3}$ &
$\frac{2}{\lambda}$ & $\frac{3}{\lambda}-\frac{1}{3}$ & $1$ \\[4pt]
\hline
& & & & & & \\[-8pt]
\parbox{1.6cm}{Physical\\ strip?}
& $\lambda>\frac{3}{4}$& $\lambda>1$ & $\lambda>\frac{3}{2}$ &
$\lambda>2$ & $\lambda>\frac{9}{4}$ & $\forall\lambda$ \\[9pt]
\hline
\end{tabular} 
\end{center}
\vskip -11pt
\caption{Physical strip poles  and zeroes of
$e^{\CA(\theta)}$, for $0<\lambda<3$}
\vskip 5pt
\label{scalpolesA}
\end{table}

\begin{table}
\begin{center}
\begin{tabular}{|l||c|c|c|c||c|} \hline
&\multicolumn{4}{c||}{}&
\multicolumn{1}{c|}{}\\[-9pt]
&\multicolumn{4}{c||}{Poles}&
\multicolumn{1}{c|}{\,Zeroes~~}\\[2pt]
\hline
& & & &  & \\[-7pt]
$~~~\frac{\theta}{i\pi}~=$ &
$~~~\frac{1}{3}~~~$&$\frac{1}{\lambda}$ &
$\frac{1}{\lambda}+\frac{1}{3}$ &
$\frac{2}{\lambda}$ &
$~~~1~~~$ \\[4pt]
\hline
& & & & &\\[-8pt]
\parbox{1.6cm}{Physical\\ strip?}
& $\forall\lambda$ & $\lambda>1$
& $\lambda>\frac{3}{2}$ & $\lambda>2$ 
& $\forall\lambda$\\[9pt]
\hline
\end{tabular}
\end{center}
\vskip -11pt
\caption{Physical strip poles  and zeroes of
$S(\theta)$, for $0<\lambda<3$}
\label{scalpolesS}
\end{table}

For the critical models, $\gamma$ lies between $0$ and $\pi/2$ 
and so the S-matrix parameter
$\lambda=3\gamma/\pi$ should be between $0$ and $3/2$. As mentioned
above, the exponents of the
tricritical models can be obtained by a continuation of $\gamma$
into the interval $[\pi/2,\pi]$,
and it is natural to suppose that the same should hold at the level
of S-matrices. This led Chim and Zamolodchikov to
conjecture that the S-matrix elements
(\ref{eq:k1smtx0}) -- (\ref{eq:k1smtx3})
with $\lambda$ in the range $[3/2,3]$ should correspond to the scaling 
tricritical models. 
Thus the general relation between $q$ and $\lambda$ is
\eq
\sqrt{q}=2\sin\!\left(\frac{\pi\lambda}{3}\right),
\quad\mbox{where}~~
\lambda\in
\left\{
\begin{array}{ll}
[0,3/2]~~&\mbox{(perturbed critical models);}\\[3pt]
[3/2,3] &\mbox{(perturbed tricritical models).}
\end{array}
\right.
\label{eq:lambda}
\en
Recall that $\lambda\in [0,3/2]$
corresponds to $\phi_{21}$ perturbations, and $\lambda\in [3/2,3]$  to
$\phi_{12}$.  In all cases the central charge of the unperturbed
theory is
\begin{equation}
c = 1-\frac{(3{-}2\lambda)^2\!}{(3{+}2\lambda)}~.
\end{equation}
For future reference, we also record the rational values of $\lambda$, 
following from (\ref{xicrit}) and (\ref{xitricrit}) and the relation
$\lambda=\frac{3}{2}(\xi{-}1)/(\xi{+}1)$, at which the off-critical
Potts models are related to $\phi_{21}$ and $\phi_{12}$ perturbed
minimal models $\CM_{pp'}$\,, $p'>p$\,:
\bea
\lambda&=&\frac{3p}{p'}-\frac{3}{2}\qquad~\mbox{($\phi_{21}$
perturbations);}
\label{lambdamincrit}\\
\lambda&=&\frac{3p'}{p}-\frac{3}{2}\qquad\mbox{($\phi_{12}$
perturbations).}
\label{lambdamintricrit}
\eea

\resection{Completing the bootstrap for $0\le \lambda\le 3/2$ }
\label{newsec}
We now return to the discussion of the pole structures of the S-matrix 
elements, in order to see if there is any need to introduce further one-particle
asymptotic states into the model, beyond the elementary kinks.
It is convenient
to phrase the discussion in terms of the variable
\eq
t=\frac{\theta}{i\pi}
\en
already used implicitly in tables \ref{scalpolesA} and \ref{scalpolesS}, so
that the segment of the imaginary $\theta$-axis lying in the physical
strip corresponds to real values of $t$ between $0$ and $1$.
$t$ will sometimes be referred to as an `angle', though strictly speaking
the term should be reserved for $\pi t$ rather than $t$.
The poles at $t=2/3$ in $\CS^0$ and $\CS^1$, and at $t=1/3$ in $\CS^0$ and 
$\CS^2$, have already been treated, and are due to the fundamental
kink itself.
In~\cite{CZ}, Chim and Zamolodchikov noted that 
extra poles enter the physical strip once $\lambda$ passes $1$.
These they assigned to a new particle $B$, 
a breather-like excitation over a single
vacuum, appearing as a bound state in the scattering of two kinks. The
associated S-matrix elements $\CS_{BK}$ and $\CS_{BB}$
also contain physical strip poles. Some of these correspond to the
kink and breather already seen, but others were conjectured in
\cite{CZ} to signal the presence of yet further particles.
In this section, we 
re-examine this analysis and find
that these further particles do not in fact appear in the spectrum
while the critical models are considered.
The range $3/2<\lambda\leq3$, corresponding to perturbations of the
tricritical models,
turns out to be far more complicated and we postpone its discussion
until section 4.

\begin{table}
\begin{center}
\begin{tabular}{|c|cc|c|c|c|c|} \hline
& & & & & &\\
&\multicolumn{2}{c|}{Poles: $t=$}&
&Poles: $t=$ & &Poles: $t=$ \\
& & & & & &\\[-3pt]
\hline
\hline
 & & & & & & \\
$\CS^0(\theta)$: & $\frac{2}{3}$ & $\frac{1}{3}$ &
$\CS^1(\theta)$: & $\frac{2}{3}$ & $\CS^2(\theta)$: &
$\frac{1}{3}$ \\
& & & & & & \\
& $\frac{2}{3}-\frac{1}{\lambda}$ & $\frac{1}{3}+\frac{1}{\lambda}$
& & $\frac{2}{3}-\frac{1}{\lambda}$ & & $\frac{1}{3}+\frac{1}{\lambda}$ \\
& & & & & & \\ \cline{1-3}
 & & & & & & \\
$\CS^3(\theta)$: & $\frac{1}{\lambda}$ & $1-\frac{1}{\lambda}$ &
& $\frac{1}{\lambda}$ & & $1-\frac{1}{\lambda}$ \\
& & & & & & \\
& $\frac{2}{\lambda}$ & $1-\frac{2}{\lambda}$ &
& $\frac{2}{\lambda}$ & & $1-\frac{2}{\lambda}$ \\
& & & & & & \\ \hline
\end{tabular}
\end{center}
\caption{Physical strip poles of $\CS^0$,
$\CS^1$, $\CS^2$ and $\CS^3$ for $0\le\lambda\le 3$}
\label{t:psp}
\end{table}

\subsection{The pole structure of the fundamental kink amplitudes}
\begin{figure}[hb]
\[
\begin{array}{c}
\includegraphics[width=0.86\linewidth]{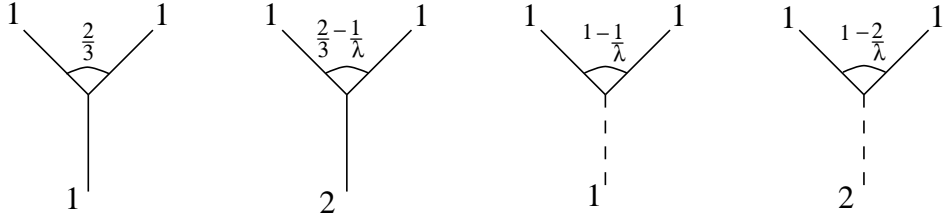}
\end{array}\]
\caption{$K_1K_1$ fusing vertices}
\label{d:k1k1v}
\end{figure}

Table \ref{t:psp} summarises the
physical strip pole structure of
the fundamental amplitudes (\ref{eq:k1smtx0})-(\ref{eq:k1smtx3}), for
$\lambda$  in the range $0\leq\lambda\leq3$.
Supposing each pole to correspond to a bound state particle in either
the direct or the crossed channel, a consideration of the 
patterns of vacua shown in figure~\ref{4elements}
allows them to
be classified as follows. All four vacua seen by
$\CS^0$ are different, so all of its poles
must correspond to kink type particles.
One out of each pair of
poles in $\CS^0$
appears in $\CS^1$\,, and the other in $\CS^2$. To be
consistent with the vacuum structure, the pole also appearing in $\CS^1$
must be direct channel, and that also in $\CS^2$ cross
channel. Similarly, all poles in $\CS^3$ must correspond to
excitations over a single vacuum (breathers). All poles which also
appear in $\CS^2$ must be direct channel, while those which
also appear in $\CS^1$
must be cross channel. The resulting particle
spectrum is summarised in table~\ref{t:4},
and the corresponding fusing vertices are depicted in
figure~\ref{d:k1k1v}.
\begin{table}
\begin{center}
\begin{tabular}{|l||c|c|c|c|} \hline
& & & & \\
Pole: $t=$ & $\frac{2}{3}$ &
$\frac{2}{3}-\frac{1}{\lambda}$ & $1-\frac{1}{\lambda}$ &
$1-\frac{2}{\lambda}$ \\
& & & & \\[-3pt]
\hline
\hline
& & & & \\
\parbox{1.6cm}{Physical\\ strip?}
& $\forall\ \lambda$& $\lambda>\frac{3}{2}$ & $\lambda>1$ &
$\lambda>2$ \\
& & & & \\ \hline
& & & & \\
Mass: & $m$ & $2m\cos\left(\frac{\pi}{3}-\frac{\pi}{2\lambda}\right)$ &
$2m\cos\left(\frac{\pi}{2}-\frac{\pi}{2\lambda}\right)$ &
$2m\cos\left(\frac{\pi}{2}-\frac{\pi}{\lambda}\right)$ \\
& & & & \\ \hline
& & & & \\
Particle: & $K_1\ (=K)$ & $K_2$ & $B_1\ (=B)$ & $B_2$ \\
& & & & \\ \hline
\end{tabular}
\end{center}
\caption{Direct channel pole assignments for 
$\CS^0$, $\CS^1$, $\CS^2$
and $\CS^3$}
\label{t:4}
\end{table}
Restricting to the range $0\le\lambda\le 3/2$, the only bound 
states
signalled in $K_1$\,$K_1$ scattering are the fundamental kink $K_1$
itself, and,
for $1<\lambda\le 3/2$, the breather $B_1$.

\subsection{Two particles: $1<\lambda\le 3/2$}
\begin{figure}[h]
\[
\begin{array}{cc}
\includegraphics[width=0.18\linewidth]{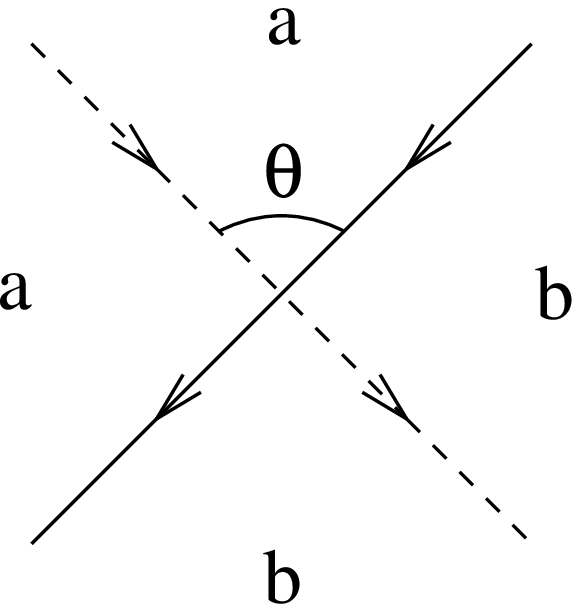} & \hspace{3cm} 
\includegraphics[width=0.18\linewidth]{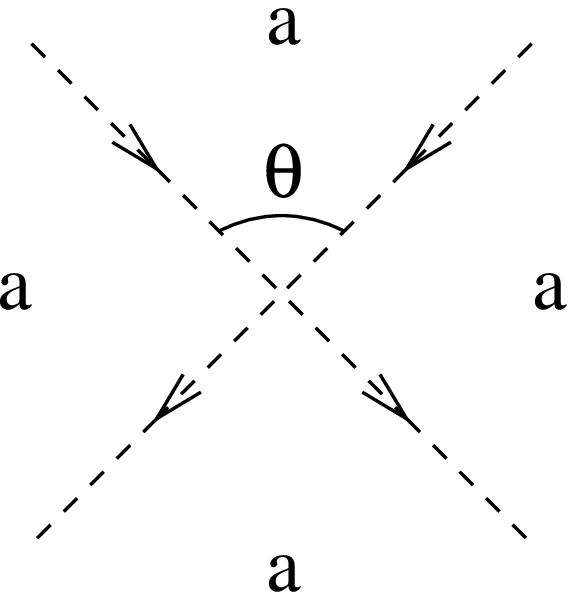} 
\end{array}
\]
\caption{$B_1K_1\rightarrow K_1B_1$ and $B_1B_1\rightarrow B_1B_1$
scattering}
\label{f:bscat}
\end{figure}

For $\lambda\le 1$, the bootstrap closes on the fundamental kink
alone and there is no more to be done, but once
$\lambda$ increases beyond $1$ the breather $B_1$ appears.
This extra particle brings with it two new scattering amplitudes,
${\cal S}_{B_1K_1}$ and ${\cal
S}_{B_1B_1}$\,:
\begin{eqnarray}
{\cal S}_{B_1K_1} & = & [\fra{1}{2}+\fra{1}{2\lambda}]
[\fra{1}{6}+\fra{1}{2\lambda}]~, \\[3pt]
{\cal S}_{B_1B_1} & = & [\fra{2}{3}][\fra{1}{\lambda}]
[\fra{1}{\lambda}-\fra{1}{3}]~.
\end{eqnarray}
Here we have rewritten the formulae from \cite{CZ} using the blocks
\eq
[a] =
(a)(1-a)~,\ \ \ \ \mathrm{where}\ \ \ \ 
(a)=\frac{\sinh\left(\fra{\theta}{2}+\fra{i\pi a}{2}\right)}
{\sinh\left(\fra{\theta}{2}-\fra{i\pi a}{2}\right)}~.
\label{bldef}
\en
The new amplitudes are illustrated in figure~\ref{f:bscat}.

The pole structures of these new S-matrix elements must be examined to
see if there are any further bound states. It is at this stage that our
analysis diverges from that of \cite{CZ}, since it turns out that
variants of the Coleman-Thun mechanism \cite{CT}
allow some poles to be explained,
for certain ranges of $\lambda$, without the need to
introduce new
particles. We shall treat the two new amplitudes in turn.

\subsection{The $B_1K_1$ scattering amplitude}
The relevant S-matrix element, which is also depicted in
appendix~\ref{polediags}, is
\eq
{{\cal S}_{B_1K_1} 
= [\fra{1}{2}{+}\fra{1}{2\lambda}][\fra{1}{6}{+}\fra{1}{2\lambda}]
= (\fra{1}{2}{-}\fra{1}{2\lambda})(\fra{1}{2}{+}\fra{1}{2\lambda})
(\fra{1}{6}{+}\fra{1}{2\lambda})(\fra{5}{6}{-}\fra{1}{2\lambda})
}
\en
The $K_1K_1B_1$ vertex 
allows the poles
from the factors 
$(\fra{1}{2}+\fra{1}{2\lambda})$ 
and $(\fra{1}{2}-\fra{1}{2\lambda})$ 
to be identified with the original
kink $K_1$ appearing as a bound state in the direct and crossed
channels respectively.
Assuming
that the poles in $[\fra{1}{6}+\fra{1}{2\lambda}]$ also correspond to a
bound state, we make an initial hypothesis that
the direct channel pole is at
$t=\fra{1}{6}+\fra{1}{2\lambda}$.
The associated particle must be an excited
kink of some sort, and calculating its mass we find
\eq
m=2m\cos(\fra{\pi}{3}-\fra{1}{2\lambda})~.
\en 
Comparing with the masses listed in table~\ref{t:4},
it is natural to identify this
with the $K_2$ state already seen in $K_1$\,$K_1$
scattering. This implies the existence of a vertex coupling
$K_1$, $K_2$ and $B_1$\,,
and the corresponding fusing angles are
shown in figure~\ref{d:k1b1k2v}. 
\[
\begin{array}{c}
\refstepcounter{figure}
\label{d:k1b1k2v}
\includegraphics[width=0.16\linewidth]{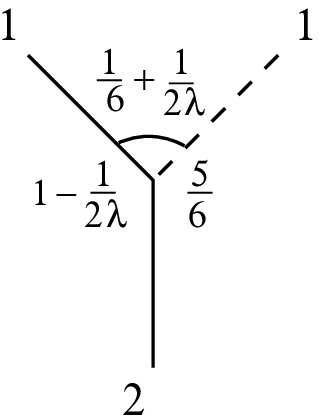} \\ 
\parbox[t]{.40\linewidth}{  
Figure \ref{d:k1b1k2v}:  the $K_1B_1\rightarrow K_2$ vertex}
\end{array}
\]
Unlike the $K_2$ bound state poles in $\CS^0$ and $\CS^1$,
which only enter the physical strip for $\lambda>\frac{3}{2}$, the
$(\fra{5}{6}-\fra{1}{2\lambda})(\fra{1}{6}+\fra{1}{2\lambda})$ poles in
${\cal S}_{B_1K_1}$ are already in the physical strip at
$\lambda=1$. 
This suggests
that $K_2$ should be present in the particle spectrum for all
${\lambda>1}$, and indeed this was proposed in \cite{CZ}.
We would like to advocate an alternative scenario, which begins with
the observation that, for $\lambda<3/2$,
one would expect the $B_1\,K_1$ scattering amplitude to exhibit an
anomalous threshold pole exactly at the location of the would-be
$K_2$ bound state pole, since
the on-shell diagram shown in figure~\ref{d:k1b1agb} is geometrically
possible. 
\[
\begin{array}{cc}
\refstepcounter{figure}
\label{d:k1b1agb}
\includegraphics[height=0.42\linewidth]{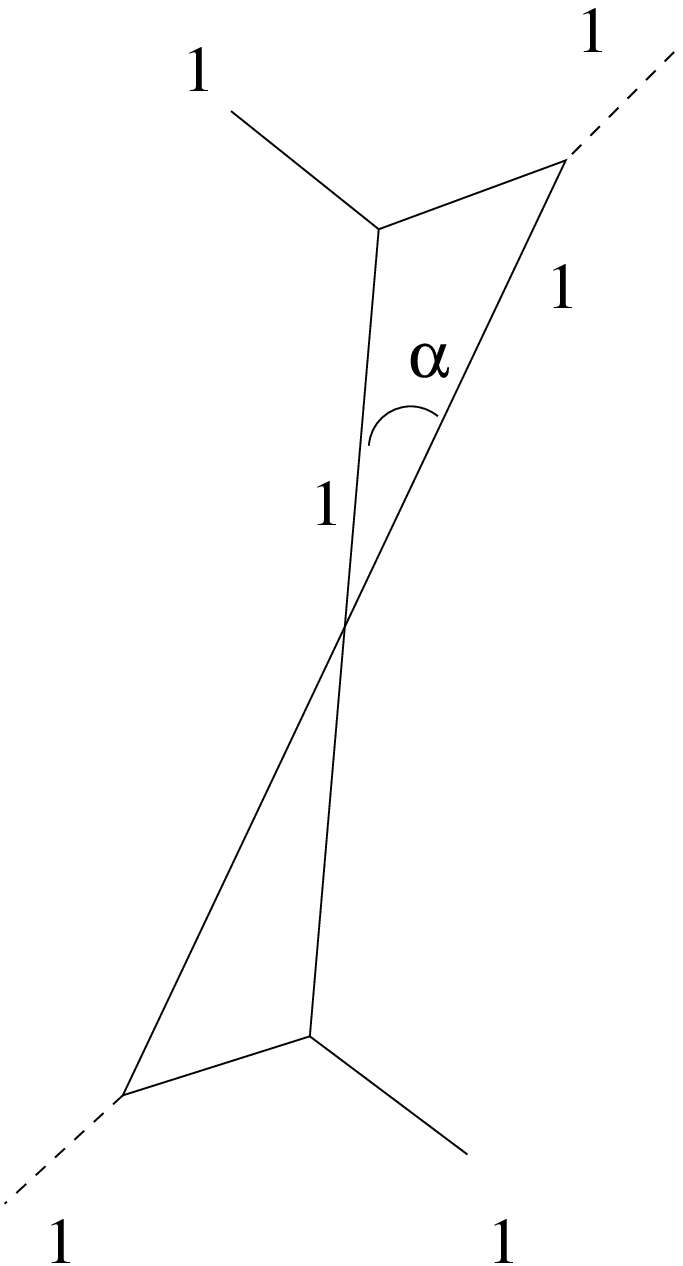} 
{}~~~~~~&~~~~~~
\refstepcounter{figure}
\label{d:b1b1bgb}
\includegraphics[height=0.42\linewidth]{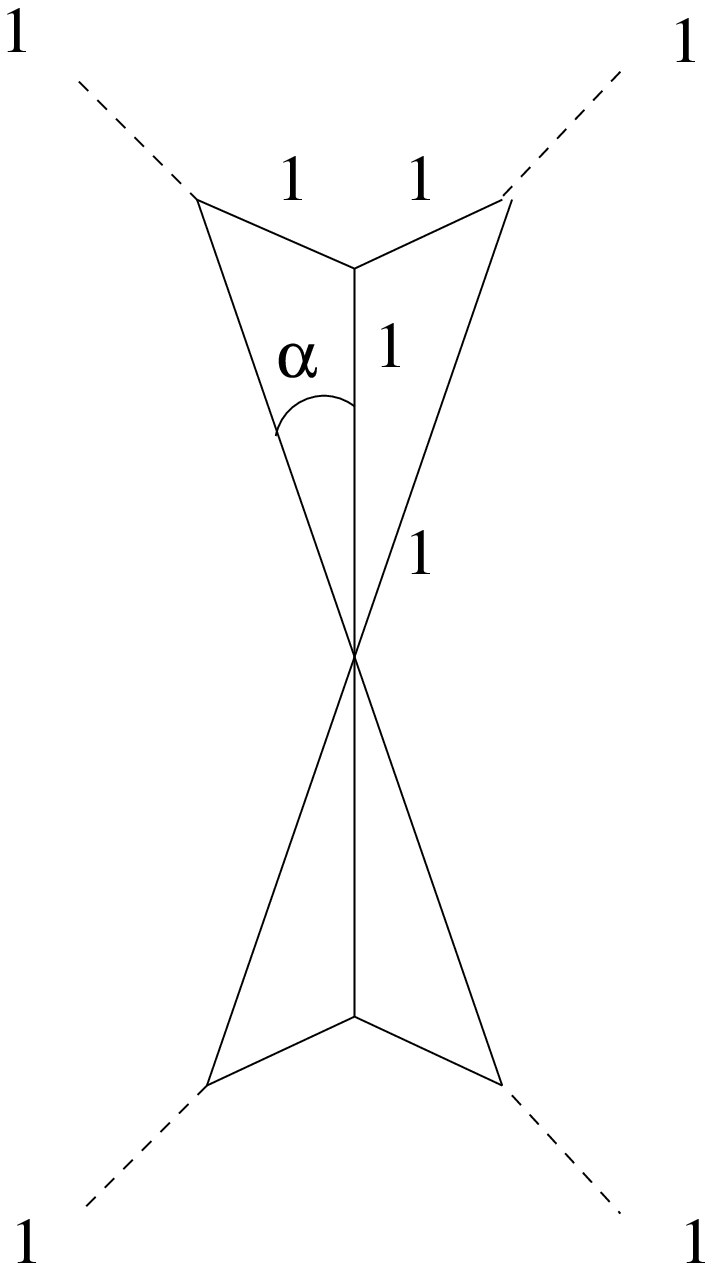} \\[5pt]
\parbox[t]{.37\linewidth}{  
Figure \ref{d:k1b1agb}:
$K_1B_1$ scattering: the\\
pole at $t=\fra{1}{6}+\fra{1}{2\lambda}$\,, for $\lambda<\frac{3}{2}$
}
{}~~~~~~&~~~~~~
\parbox[t]{.37\linewidth}{  
Figure \ref{d:b1b1bgb}: 
 $B_1B_1$ scattering: the\\ 
 pole at $t=\fra{1}{\lambda}-\fra{1}{3}$\,, for
$\lambda<\fra{3}{2}$
}
\end{array}
\]

The angle between the incoming particles $K_1$ and $B_1$ is
$t=\frac{1}{6}+\frac{1}{2\lambda}$, and the internal scattering angle
is
$\alpha=\frac{1}{\lambda}-\frac{2}{3}$. The diagram closes for
$0<\alpha<\frac{1}{3}$,
which translates as $1<\lambda<\frac{3}{2}$.
However, in two dimensions a diagram of this sort
would usually be expected to give rise to a
double pole, while ${\cal S}_{B_1K_1}$ only has a simple
pole at this value of $t$. 
The problem is resolved once the
contributions of the couplings and
S-matrix elements are taken into account. These are composed of
four three-particle couplings multiplying a sum over two-particle
amplitudes.
This sum, which we shall denote by ${\cal C}$, 
can be computed by considering the possibilities for the
internal vacua in the diagram.
The
`upper' internal vacuum must differ from both external vacua, and so
can take $(q{-}2)$ values. Once this vacuum has been fixed, there are
again $(q{-}2)$ possibilities for the lower internal vacuum, but they
are no longer equivalent. Either the lower vacuum is equal to the
upper one ($1$ possibility), in which case the relevant two-particle
amplitude needed to evaluate the diagram is $\CS^1(i\pi\alpha)$, or
else it is different ($(q{-}3)$ possibilities), in which case the
amplitude is rather $\CS^0(i\pi\alpha)$.
Adding everything up, we have
\bea
&&
\!\!\!\!\!\!
\!\!\!\!\!\!
{\cal C}=
(q-2)\left[\CS^1(i\pi\alpha) + (q-3)\CS^0(i\pi\alpha)
\right] \nn\\[4pt]
&&
\!\!\!\!\!\!
=  \frac{(q{-}2)S(i\pi\alpha)}%
{\sin((\fra{2}{3}{-}\alpha)\pi\lambda)
\sin(\fra{\pi}{3}\lambda)}
\left[\sin(\fra{2\pi}{3}\lambda)\sin((\fra{1}{3}{-}\alpha)\pi\lambda)
+\sin(\pi\lambda)\sin(\alpha\pi\lambda)\right]
\label{contra}
\eea
where use has been made of equation~(\ref{eq:lambda}). Substituting 
$\alpha=\frac{1}{\lambda}-\frac{2}{3}$,
and noting that
$S(i\pi\alpha)$ does not have any poles or zeroes at this
value of $\alpha$, reveals a remarkable cancellation:
\eq
{\cal C}=-\frac{(q{-}2)S(i\pi\alpha)}
{\sin(\fra{4\pi}{3}\lambda)\sin(\pi\lambda)}
[-\sin(\fra{2\pi}{3}\lambda)\sin(\pi\lambda)+
\sin(\pi\lambda)\sin(\fra{2\pi}{3}\lambda)]=0\,.
\en
The vanishing of ${\cal C}$ 
reduces the overall singularity
associated with the diagram
to a simple pole. Thus
the pole at
$t=\frac{1}{6}+\frac{1}{2\lambda}$ in ${\cal S}_{B_1K_1}$ can
be explained without 
the need to introduce the excited kink $K_2$, at least until
$\lambda=\frac{3}{2}$. Beyond this
point, the scattering process shown in figure~\ref{d:k1b1agb} is no
longer geometrically possible; and at the same
time, the presence of $K_2$ in the spectrum is also
signalled by
the appearance of its bound state pole in the kink-kink matrix elements
$\CS^0$ and $\CS^1$.

A very similar phenomenon was observed many years ago by Coleman and Thun
in the Sine-Gordon model~\cite{CT}, where the appearance of new
breathers is delayed until their poles are seen in the
soliton-soliton S-matrix (see 
section 4 of
\cite{pedrev} for a detailed
review of this story). It has also played a
r\^ole in the understanding of non self-dual affine Toda field
theories~\cite{CDS}, and crops up in the analysis of boundary
scattering~\cite{DTW}. A novel feature here is the formal treatment of the
vacua. When working out the combinatorics, $q$ is treated as an integer,
and the resulting formula is then taken to hold for general $q$. The
same spirit guided Chim and Zamolodchikov's original formulation of the
Yang-Baxter equations for the model; it can
be interpreted as a way to account for the statistics of
the kink states in a way that depends smoothly on $q$.
In principle it should be possible to phrase the discussion entirely
within the more standardly-defined 
formulation of Smirnov~\cite{Sm}, but since in this 
approach the vacuum structure depends in a 
discontinuous way on $q$, the arguments are likely to be considerably
more involved.

\subsection{The $B_1B_1$ scattering amplitude}
To complete the analysis we must consider
\eq
\CS_{B_1B_1} 
=[\fra{2}{3}][\fra{1}{\lambda}][\fra{1}{\lambda}{-}\fra{1}{3}]
=(\fra{1}{3})(\fra{2}{3})
 (\fra{1}{\lambda})(1{-}\fra{1}{\lambda})
 (\fra{1}{\lambda}{-}\fra{1}{3})(\fra{2}{3}{-}\fra{1}{\lambda})~.
\en
The poles from the blocks
$(\fra{2}{3})$ and $(\fra{1}{3})$ can be identified with
$B_1$ bound states in the direct and cross channels respectively. 
The poles from 
$(\fra{1}{\lambda})$ and
$(1-\fra{1}{\lambda})$ 
match
direct and cross channel bound states of $B_2$, while that in
$(\fra{1}{\lambda}-\fra{1}{3})$ is potentially
the direct channel pole for a
new particle, $B_3$, with mass
\eq
m_{B_3}=2m_{B_1}\cos(\fra{\pi}{2\lambda}-\fra{\pi}{6})=
4m\cos(\fra{\pi}{2}-\fra{\pi}{2\lambda})
\cos(\fra{\pi}{2\lambda}-\fra{\pi}{6})~.
\en
(This particle was denoted $B'$ 
in~\cite{CZ}.) 
The poles at $\frac{1}{\lambda}$ and $\frac{1}{\lambda}- \frac{1}{3}$
are both located in the physical strip 
for $\lambda>1$, when $B_1$ first appears.

However, the appearance of both $B_2$ and $B_3$  can
be delayed by invoking the Coleman-Thun mechanism. Most complicated
to treat is the would-be $B_3$ bound state pole.  The a-priori third order
diagram shown in figure~\ref{d:b1b1bgb} has internal scattering angle
$\alpha=\frac{1}{\lambda}-\frac{2}{3}$, and thus closes for
$\lambda<\frac{3}{2}$. Keeping track of all possible combinations 
we find that the third-order pole must be multiplied by a factor
\eq
{\cal C}=(q-1)(q-2)\left[\CS^1(i\pi\alpha)+
(q-3)\CS^0(i\pi\alpha)
\right]^2\left[\CS^1(i2\pi\alpha)+
(q-3)\CS^0(i2\pi\alpha)\right].
\label{contrb}
\en
We have already seen that 
$\CS^1(i\pi\alpha)+(q-3)\CS^0(i\pi\alpha)=0$
for $\alpha=\frac{1}{\lambda}-\frac{2}{3}$ in the treatment of
${\cal S}_{B_1K_1}$, and a simple calculation shows that the rest of the
function is finite and non-zero at this point. The diagram thus
contributes a simple pole, and $B_3$ need not show up as a
bound state until $\lambda=\frac{3}{2}$.

Turning to the pole 
at $t=\fra{1}{\lambda}$, 
provisionally assigned to $B_2$,
the relevant 
diagram is shown in figure~\ref{d:b1b1agb}. 
The internal scattering angle
$\alpha=\fra{2}{\lambda}-1$ is positive (making the diagram
geometrically possible) for
$\lambda<2$. Calculating the prefactor ${\cal C}$ we have
\bea
{\cal C}&=& (q-1)\left[\CS^3(i\pi\alpha)+(q-2){\cal
S}_1(i\pi\alpha)\right] 
\label{contrc}\\[4pt]
&=& (q-1)\frac{\sin((\alpha{-}\frac{1}{3})\pi\lambda)S(i\pi\alpha)}
{\sin(\frac{\pi}{3}\lambda)}
\left[\frac{\sin(\pi\lambda)}{\sin((\alpha{-}1)\pi\lambda)}+(q{-}
2)\frac{\sin(\frac{2\pi}{3}\lambda)}{\sin((\alpha{-}\frac{2}{3})\pi\lambda)}%
\right]\,.\nn
\eea
At $\alpha=\fra{2}{\lambda}-1$, the expression in square
brackets
vanishes and the overall singularity is reduced from a double to a
single  pole.
This means that there is no need to introduce the $B_2$
breather
until $\lambda$ passes $2$, at which stage
it also appears in the $K_1\,K_1$ amplitudes $\CS^2$ and $\CS^3$.
\[
\begin{array}{c}
\refstepcounter{figure}
\label{d:b1b1agb}
\includegraphics[width=0.28\linewidth]{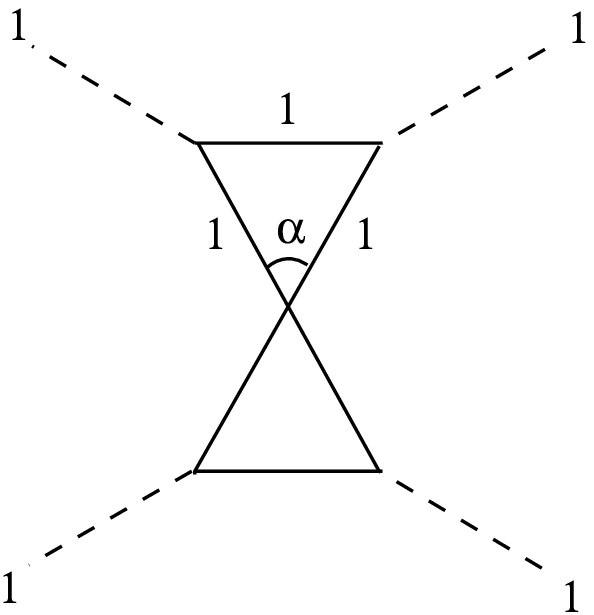}\\[4pt]
\parbox[t]{.60\linewidth}{  
Figure \ref{d:b1b1agb}: $B_1B_1$ scattering: $t=\fra{1}{\lambda}$, 
$\lambda<2$
}
\end{array}
\]

\subsection{Checks on the spectrum for $0<\lambda\le 3/2$}
To summarise, by invoking the Coleman-Thun mechanism we have seen
that the only particles forced to appear for $0<\lambda<\frac{3}{2}$ 
are the fundamental kink
$K_1$\,, present for all $\lambda$, and the breather
$B_1$\,, which appears for
$\lambda>1$. The two bound states $K_2$ and $B_3$ proposed
by Chim and
Zamolodchikov need not appear until $\lambda>\frac{3}{2}$. We 
also saw evidence 
for a further new particle
$B_2$\,, but its appearance was postponed until
$\lambda>2$.

While we have shown that our proposed spectrum is consistent, strictly
speaking 
the larger spectrum
initially suggested by Chim and Zamolodchikov
has not been completely ruled out -- genuine bound state poles could
be hiding behind the contributions provided by the Coleman-Thun 
diagrams.
Some
reassurance comes from some work by Delfino and Cardy~\cite{CD}. 
They calculated a number of universal quantities
using the form-factor
approach, which is sensitive to the massive particle
spectrum.
Some of these quantities (for example the central charge) can be
compared with known results. While the need to add in the first
breather state 
at $\lambda=1$
is clearly signalled, the data
up to $\lambda=3/2$, as illustrated by figure~5 of
\cite{CD},
shows no signs of any further
missing particles. 

Another 
check comes from the limiting point $\lambda=\frac{3}{2}$, $q=4$,
where the S-matrix for the minimal
$D_4$ related field theory discussed in~\cite{D4} should
be reproduced. The spectrum of this model consists of three light 
particles $\lo$, $\lt$, $\lth$ with the same mass $m_{\lo}$, and one
particle $h$ of mass $m_h=\sqrt{3}m_{\lo}$. The mass ratio $m_h:m_{\lo}$
equals that of $m_{B_1}:m_{K_1}$, and $h$ is naturally identified with
$B_1$. There are three distinct ways of pairing up the four vacua and
each of these may be associated with one of the light particles. This is
shown in figure~\ref{d:d4}, where for example $\lo$ is identified with
domain walls $1\leftrightarrow2$ and $3\leftrightarrow4$ (vacua labeled
from $1$ to $4$). This identification is unique up to permutations of the
light particles amongst themselves --- also a property of the $D_4$
S-matrix. The two-particle amplitudes $\mathcal{S}_{\lo\lo}(\theta)$,
$\mathcal{S}_{\lo\lt}(\theta)$, $\mathcal{S}_{\lo h}(\theta)$ and
$\mathcal{S}_{hh}(\theta)$ then correspond to $\CS^3(\theta)$, 
$\CS^0(\theta)$, ${\cal S}_{B_1K_1}(\theta)$ and ${\cal S}_{B_1B_1}(\theta)$
respectively. Substituting for $\lambda$, a direct comparison can be made
with the results of~\cite{D4}. The S-matrices match, since
$\CS^1(\theta)$ and  $\CS^2(\theta)$ vanish for
$\lambda=\frac{3}{2}$. Notice that the set of spins of conserved charges 
for the $D_4$-related model 
starts $1$, $3$, $3$, $5\dots$ and is thus larger than the generic
$\phi_{21}$ spectrum
$s=1,~5,~7,~11\dots$
reported above.
Moving from the kink to the particle
picture, the $\phi^3$ property of the S-matrix is lost,
and it is this which allows the enlarged set of conserved charges
to be represented locally
on the multiparticle states. In the kink basis, the extra charges are also 
present for $\lambda=3/2$, but they do not act diagonally.

\[
\begin{array}{c}
\refstepcounter{figure}
\label{d:d4}
\includegraphics[width=0.35\linewidth]{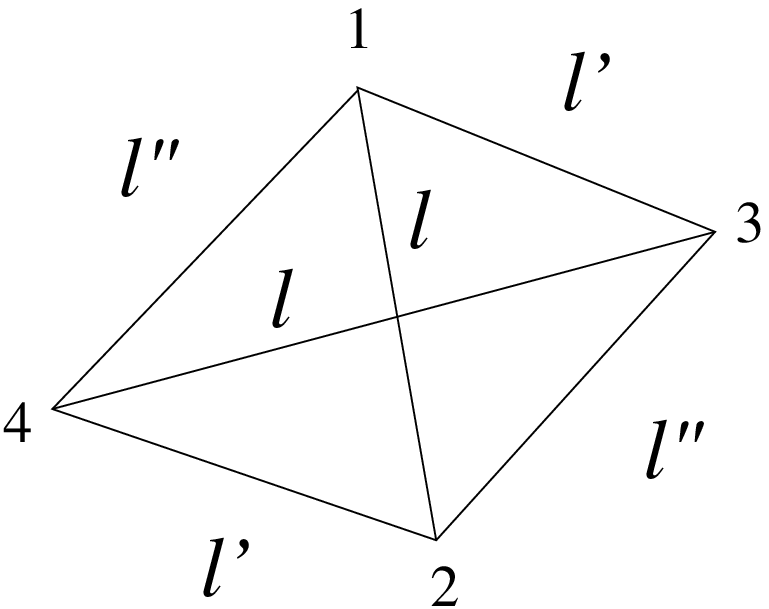} \\
\parbox[t]{.60\linewidth}{Figure \ref{d:d4}: 
Domain walls $\leftrightarrow$ particles at $q=4$ } 
\end{array}
\]
\resection{Perturbed tricritical models: $3/2<\lambda<3$}
We now move to the region
$\frac{3}{2}\leq\lambda\leq3$, 
suggested in \cite{CZ} to describe the 
scaling
tricritical $q$-state Potts models, and related to
$\phi_{12}$ perturbations
of $c<1$ conformal field theories.
The full S-matrix becomes extremely complicated as $\lambda$
increases,
and we have not completed our analysis for the whole tricritical
range of $\lambda$.
For this reason we will be a little more sketchy in our
descriptions in this section. 
Some further details are in the appendices, and in \cite{AP}.

\subsection{Four particles: $3/2<\lambda\le 2$}
\label{fourpart}
Two new particles, $K_2$ and $B_3$, 
must appear once  $\lambda$ passes $\frac{3}{2}$.
We found that no further
particles were
needed to explain the pole structure up to $\lambda = 2$. The
new S-matrix elements are as follows:
\bea
{\cal S}_{K_1K_2}^a
 & = &
t(\fra{1}{3}+\fra{1}{2\lambda})t(\fra{1}{3}-\fra{1}{2\lambda})
t(\fra{1}{2\lambda})t(\fra{2}{3}-\fra{1}{2\lambda})
\,\CS^a(\theta+\fra{i\pi}{2\lambda})\nn\\
{\cal S}_{K_2K_2}^a
& = &
[\fra{2}{3}]^2[\fra{1}{3}+\fra{1}{\lambda}][\fra{1}{\lambda}]
\,\CS^a 
\nn\\
{\cal S}_{B_1K_2}
& = &
[\fra{1}{2}][\fra{5}{6}][\fra{1}{6}+\fra{1}{\lambda}]
[\fra{1}{\lambda}-\fra{1}{6}]\nn\\
{\cal S}_{B_3K_1}
& = & [\fra{1}{3}]^2[\fra{1}{\lambda}]
[\fra{1}{\lambda}+\fra{1}{3}]\nn \\
{\cal S}_{B_3K_2}
& = & [1-\fra{1}{2\lambda}]^2
[\fra{1}{3}+\fra{1}{2\lambda}]^3[\fra{2}{3}+\fra{1}{2\lambda}]
[\fra{3}{2\lambda}-\fra{1}{3}][\fra{3}{2\lambda}]\nn\\
{\cal S}_{B_3B_3}
& = & [\fra{2}{3}]^3[\fra{1}{\lambda}]^3
[\fra{4}{3}-\fra{1}{\lambda}]^2[\fra{1}{3}+\fra{1}{\lambda}]
[\fra{2}{\lambda}-\fra{2}{3}][\fra{2}{\lambda}-\fra{1}{3}]\nn\\
{\cal S}_{B_3B_1}
& = & [\fra{1}{6}+\fra{1}{2\lambda}]^2
[\fra{1}{2}+\fra{1}{2\lambda}][\fra{7}{6}-\fra{1}{2\lambda}]
[\fra{3}{2\lambda}-\fra{1}{2}][\fra{3}{2\lambda}-\fra{1}{6}]
\eea
where $t(a)(t)=\tan(\fra{\pi}{2}(t+a))$, the blocks $[a]$ were
defined in equation (\ref{bldef}) above, and 
$\CS^a$ are the amplitudes for the scattering of the
fundamental kink $K_1$. (Note the shift in the argument 
in the first formula.)
Once again there
are four possible amplitudes corresponding to the four different vacuum
structures for kink-kink scattering. 
In tables~\ref{t:pspk1k2}
and~\ref{t:pspk2k2} we summarise the physical-strip poles of the new
kink-kink amplitudes for $3/2<\lambda\leq 3$.

\begin{table}[t]
\begin{center}
\begin{tabular}{|c|c|c|c|c|c|c|} \hline
&\multicolumn{6}{c|}{}\\[-2pt]
$t=\theta/i\pi$ &\multicolumn{6}{c|}{Poles: ~~$t=$}\\
&\multicolumn{6}{c|}{}\\[-5pt]
\hline
\hline
 & & & & & & \\
${\cal S}_{K_2K_1}^0(\theta):$ & $\frac{2}{3}+\frac{1}{2\lambda}$ &
$(\frac{1}{3}+\frac{1}{2\lambda})^2$ & $\frac{2}{3}-\frac{3}{2\lambda}$
& & & \\[7pt]
& $\frac{1}{3}-\frac{1}{2\lambda}$ & $(\frac{2}{3}-\frac{1}{2\lambda})^2$ &
$\frac{1}{3}+\frac{3}{2\lambda}$ & & & \\
& & & & & & \\ \hline
& & & & & & \\
${\cal S}_{K_2K_1}^1(\theta):$ & $\frac{2}{3}+\frac{1}{2\lambda}$ &
$\frac{1}{3}+\frac{1}{2\lambda}$ & $\frac{2}{3}-\frac{3}{2\lambda}$
& & & \\[7pt]
& & $(\frac{2}{3}-\frac{1}{2\lambda})^2$ & &
$\frac{1}{2\lambda}$ & $\frac{3}{2\lambda}$ & $\frac{5}{2\lambda}$ \\ 
& & & & & & \\ \hline
& & & & & & \\
${\cal S}_{K_2K_1}^2(\theta):$ & & $(\frac{1}{3}+\frac{1}{2\lambda})^2$ &
& $1-\frac{1}{2\lambda}$ & $1-\frac{3}{2\lambda}$ &
$1-\frac{5}{2\lambda}$ \\[7pt]
& $\frac{1}{3}-\frac{1}{2\lambda}$ & $\frac{2}{3}-\frac{1}{2\lambda}$ &
$\frac{1}{3}+\frac{3}{2\lambda}$ & & & \\
& & & & & & \\ \hline
& & & & & & \\
${\cal S}_{K_2K_1}^3(\theta):$ & & $\frac{1}{3}+\frac{1}{2\lambda}$ & &
$1-\frac{1}{2\lambda}$ & $1-\frac{3}{2\lambda}$ & $1-\frac{5}{2\lambda}$
\\[7pt]
& & $\frac{2}{3}-\frac{1}{2\lambda}$ & & $\frac{1}{2\lambda}$ &
$\frac{3}{2\lambda}$ & $\frac{5}{2\lambda}$ \\
& & & & & & \\ \hline
\end{tabular}
\end{center}
\caption{Physical strip poles of $\CS_{K_1K_2}^0$,
$\CS_{K_1K_2}^1$, $\CS_{K_1K_2}^2$ 
and $\CS_{K_1K_2}^3$ for $3/2<\lambda\le 3$}
\label{t:pspk1k2}
\end{table}

\begin{table}[t]
\begin{center}
\begin{tabular}{|c|cc|c|cc|} \hline
&\multicolumn{5}{c|}{}\\[-2pt]
$t=\theta/i\pi$ &\multicolumn{5}{c|}{Poles: ~~$t=$}\\
&\multicolumn{5}{c|}{}\\[-5pt]
\hline
\hline
& & & & & \\
${\cal S}_{K_2K_2}^0(\theta):$ & $(\frac{2}{3})^3$ & $(\frac{1}{3})^3$ &
${\cal S}_{K_2K_2}^1(\theta):$ & $(\frac{2}{3})^3$ & $(\frac{1}{3})^2$ \\
& & & & & \\
& $(\frac{2}{3}-\frac{1}{\lambda})^2$ & $(\frac{1}{3}+\frac{1}{\lambda})^2$
& & $(\frac{2}{3}-\frac{1}{\lambda})^2$ & $\frac{1}{3}+\frac{1}{\lambda}$ \\
& & & & & \\
& $1-\frac{1}{\lambda}$ & $\frac{1}{\lambda}$ & &
$1-\frac{1}{\lambda}$ & $(\frac{1}{\lambda})^2$ \\
& & & & & \\
& & & & & $\frac{2}{\lambda}$ \\
& & & & & \\ \hline
& & & & & \\
${\cal S}_{K_2K_2}^2(\theta):$ & $(\frac{2}{3})^2$ & $(\frac{1}{3})^3$ &
${\cal S}_{K_2K_2}^3(\theta):$ & $(\frac{2}{3})^2$ & $(\frac{1}{3})^2$ \\
& & & & & \\
& $\frac{2}{3}-\frac{1}{\lambda}$ & $(\frac{1}{3}+\frac{1}{\lambda})^2$
& & $\frac{2}{3}-\frac{1}{\lambda}$ & $\frac{1}{3}+\frac{1}{\lambda}$ \\
& & & & & \\
& $(1-\frac{1}{\lambda})^2$ & $\frac{1}{\lambda}$ & &
$(1-\frac{1}{\lambda})^2$ & $(\frac{1}{\lambda})^2$ \\
& & & & & \\
& & $1-\frac{2}{\lambda}$ & & $1-\frac{2}{\lambda}$ & $\frac{2}{\lambda}$ \\
& & & & & \\ \hline
\end{tabular}
\end{center}
\caption{Physical strip poles of $\CS_{K_2K_2}^0$,
$\CS_{K_2K_2}^1$, $\CS_{K_2K_2}^2$ and $\CS_{K_2K_2}^3$ 
for $3/2<\lambda\le 3$}
\label{t:pspk2k2}
\end{table}

The introduction of new particles  leads to further bootstrap equations. 
Here we quickly sketch those 
for the new kink-kink amplitudes. $K_2$ appears as a bound state
in ${K_1\,K_1}$, ${K_1\,B_1}$, 
and ${K_1\,B_3}$ scattering. This allows 
${\cal S}_{K_1K_2}^a$ to be obtained via a number of
a-priori distinct bootstrap equations. Consider first the $K_1\,K_1\to
K_2$ fusing. The general kink bootstrap equation illustrated in
figure~\ref{kboot} implies
\eq {\cal S}_{K_1K_2}^a(\theta)  =  \sum{\cal
S}_{K_1K_1}(\theta-\fra{i\pi}{3}+\fra{i\pi}{2\lambda}){\cal 
S}_{K_1K_1}(\theta+\fra{i\pi}{3}-\fra{i\pi}{2\lambda}) \label{k12boota}
\en
where the terms to be summed on the right-hand side depend both on the
particular matrix element being evaluated, and on the choice of vacuum
$b$ in figure \ref{kboot}. For example:
\bea
{\cal S}_{K_1K_2}^0(\theta) 
& = & \CS^3\CS^0 + \CS^2\CS^2 +
(q-4)\CS^2\CS^0 \nn\\
 & = & 
\CS^1\CS^1 + \CS^0\CS^3 + 
(q-4)\CS^0\CS^1 \nn\\
 & = & 
\CS^1\CS^0
+ \CS^0\CS^2 + (q-5)\CS^0\CS^0.
\eea
The compatibility of these formulae provides 
constraints on the non-scalar parts of the $K_1\,K_1$
amplitudes, which turn out to be just the original 
bootstrap equations,
with $\theta$ shifted by $\frac{i\pi}{2\lambda}$. 

Alternatively, the $K_1\,K_2$ scattering amplitude could have been found using
either the $K_1\,B_1\to K_2$ fusing: 
\bea
{\cal S}_{K_1K_2}^a(\theta)  
 & = & {\cal S}_{K_1B_1}(\theta-\fra{i\pi}{6}){\cal S}_{K_1K_1}^a(\theta+
\fra{i\pi}{2\lambda})\nn\\
 & = & {\cal S}_{K_1B_1}(\theta+\fra{i\pi}{6}){\cal
S}_{K_1K_1}^a(\theta-\fra{i\pi}{2\lambda}) \label{k12bootb}
\eea
or the $K_1\,B_3\to K_2$ fusing:
\bea
{\cal S}_{K_1K_2}^a(\theta)  
 & = & {\cal S}_{K_1B_3}(\theta-\fra{i\pi}{3}+\fra{i\pi}{2\lambda}){\cal
S}_{K_1K_1}^a(\theta+\fra{3\pi i}{2\lambda})\nn\\
 & = & {\cal S}_{K_1B_3}(\theta+\fra{i\pi}{3}-\fra{i\pi}{2\lambda}){\cal
S}_{K_1K_1}^a(\theta-\fra{3\pi i}{2\lambda}). \label{k12bootc}
\eea
The equality of the two expressions in (\ref{k12bootb})
can be checked using
\begin{equation}
S(\theta+\fra{i\pi}{\lambda}) = -\frac{\sinh(\fra{\theta}{2}-\fra{i\pi}{2})
\sinh(\fra{\theta}{2}-\fra{i\pi}{2}+\fra{i\pi}{2\lambda})
\sinh(\fra{\theta}{2}-\fra{i\pi}{3})
\sinh(\fra{\theta}{2}+\fra{i\pi}{3}+\fra{i\pi}{2\lambda})}{
\sinh(\fra{\theta}{2})\sinh(\fra{\theta}{2}+\fra{i\pi}{2\lambda})
\sinh(\fra{\theta}{2}+\fra{i\pi}{6})
\sinh(\fra{\theta}{2}-\fra{i\pi}{6}+\fra{i\pi}{2\lambda})}\,S(\theta)\,.
\label{scalarid}\end{equation}
Then~(\ref{k12bootc}) can be rewritten
in a form that can more easily be compared
with~(\ref{k12bootb}) by making use of~(\ref{scalarid}) and the fact that
\eq
{\cal S}_{K_1K_1}^a(\theta \pm\fra{i\pi}{\lambda})
= \frac{S(\theta\pm \fra{i\pi}{\lambda})}{S(\theta)}{\cal
S}_{K_1K_1}^a(\theta)\,.
\label{blob}
\en
As a last step, we need to check the compatibility of~(\ref{k12bootb})
with~(\ref{k12boota}). 
Starting from~(\ref{k12boota}) we can use (\ref{blob}) and~(\ref{scalarid}) 
to rewrite each
$S_{K_1K_1}(\theta+\fra{i\pi}{3}-\fra{i\pi}{2\lambda})$ as
$S_{K_1K_1}(\theta+\fra{i\pi}{3}+\fra{i\pi}{2\lambda})$ multiplied by
a common factor. The non-scalar parts of the formula are then
the original bootstrap 
equations for ${\cal
S}_{K_1K_1}^a(\theta+\fra{i\pi}{2\lambda})$, leaving 
it to be checked that the extra
factor is equal to ${\cal S}_{K_1B_1}(\theta-\fra{i\pi}{6})$. 
The bootstrap equations
for ${\cal S}_{K_2K_2}^a(\theta)$ 
can be treated in a similar manner, and all are
found to be satisfied.

While we omit the full details
here, we have checked that, for $3/2<\lambda<2$, all poles in the
S-matrix elements have a potential
field-theoretical explanation, often
via quite elaborate incarnations of the Coleman-Thun mechanism. To
give just one example, 
the triple pole
in ${\cal S}_{B_3B_3}$
at $t=\fra{1}{\lambda}$
can be associated with the  diagram shown in figure~\ref{following},
which closes for $\frac{3}{2}<\lambda<\frac{9}{4}$. 
The fusing angles needed to verify that the diagram does indeed close
as claimed can be found in appendix~\ref{fusings}.

\[
\begin{array}{cc}
\parbox{.41\linewidth}{
\refstepcounter{figure}
\label{following}
{}~~\includegraphics[width=0.9\linewidth]{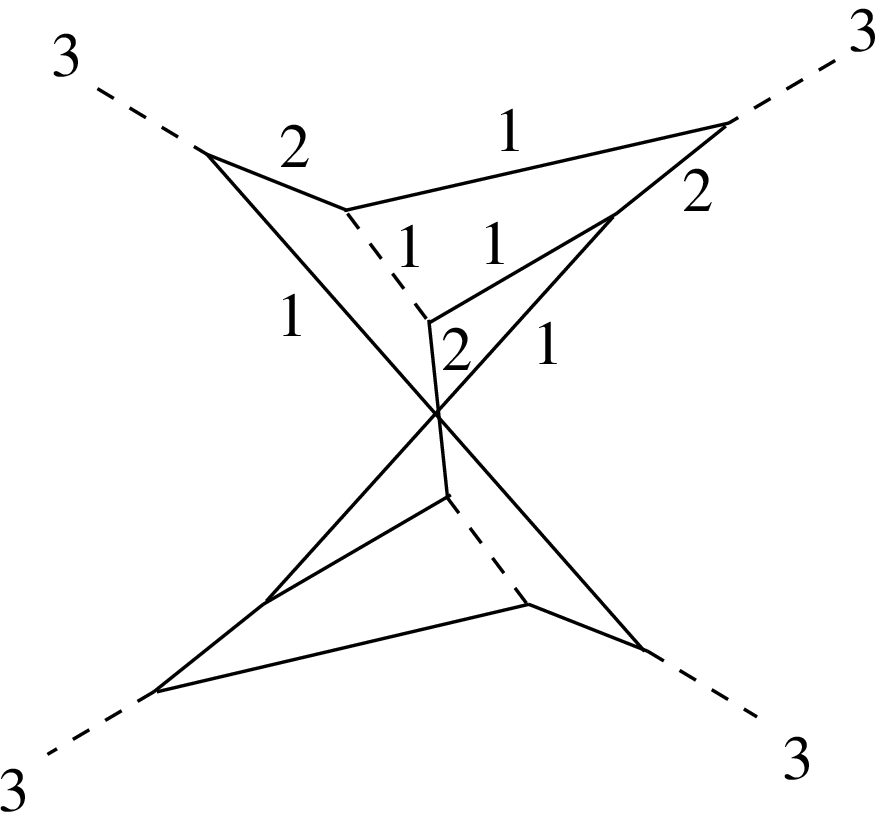}~~~~\\[15pt]
Figure~\ref{following}:  $B_3\,B_3$ scattering:
the triple  pole at $t=\fra{1}{\lambda}$\,, for
$\frac{3}{2}<\lambda<\frac{9}{4}$
}~~~~~
&~~~~~
\parbox{.41\linewidth}{
\refstepcounter{figure}
\vspace{-10pt}
\label{nfollowing}
{}~~~~~~~~~~~~~~\includegraphics[width=0.33\linewidth]{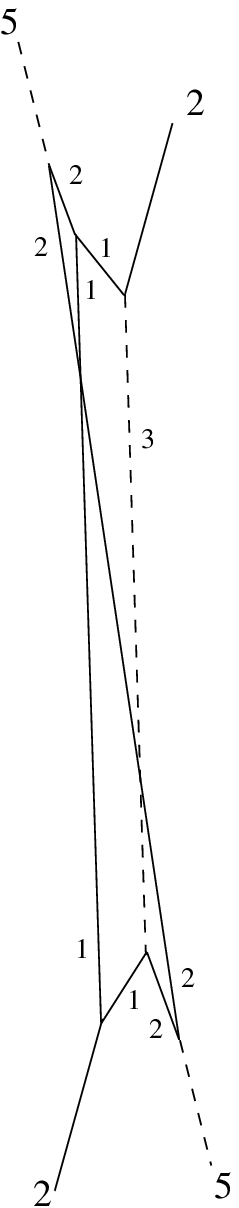}
 \\[8pt]
Figure~\ref{nfollowing}:  $B_5\,K_2$ scattering:
the double  pole at $t=\fra{1}{6}$\,,
for $2<\lambda<\frac{9}{4}$
}
\end{array}
\]

\subsection{Six particles: $2<\lambda\le 9/4$}
For $2<\lambda<\frac{9}{4}$\,, 
the already-advertised
$B_2$ together with
a further breather $B_5$ enter the spectrum. The
scattering amplitudes for the new particles
become increasingly
complicated, and greater reliance must 
be placed on the Coleman-Thun mechanism to explain the
pole structure. The new S-matrix elements in this region are:
\begin{eqnarray}
{\cal S}_{B_2B_1} & = & [1-\fra{1}{2\lambda}]
[\fra{2}{3}-\fra{1}{2\lambda}][\fra{3}{2\lambda}]
[\fra{3}{2\lambda}-\fra{1}{3}] \nn\\
{\cal S}_{B_2B_2} & = & [\fra{2}{3}][\fra{2}{3}-\fra{1}{\lambda}]
[\fra{1}{\lambda}-\fra{1}{3}][\fra{2}{\lambda}]
[\fra{2}{\lambda}-\fra{1}{3}][1-\fra{1}{\lambda}]^2 \nn\\
{\cal S}_{B_2B_3} & = & [\fra{1}{2}][\fra{5}{6}]
[\fra{2}{\lambda}-\fra{1}{6}][\fra{7}{6}-\fra{1}{\lambda}]^2
[\fra{2}{\lambda}-\fra{1}{2}][\fra{5}{6}-\fra{1}{\lambda}]^2 \nn\\
{\cal S}_{B_2K_1} & = & [\fra{1}{2}][\fra{1}{6}]
[\fra{1}{2}+\fra{1}{\lambda}][\fra{1}{6}+\fra{1}{\lambda}] \nn\\
{\cal S}_{B_2K_2} & = & [\fra{1}{2}-\fra{1}{2\lambda}]^2
[\fra{1}{6}+\fra{1}{2\lambda}]^2[\fra{1}{6}+\fra{3}{2\lambda}]
[\fra{3}{2\lambda}-\fra{1}{6}] \nn\\
{\cal S}_{B_5B_1} & = &
[\fra{1}{3}]^2[\fra{4}{3}-\fra{1}{\lambda}]
[\fra{1}{3}+\fra{1}{\lambda}][\fra{2}{\lambda}-\fra{2}{3}]
[\fra{2}{\lambda}-\fra{1}{3}][1-\fra{1}{\lambda}]^2 \nn\\
{\cal S}_{B_5B_2} & = & [\fra{4}{3}-\fra{3}{2\lambda}]^2
[1-\fra{3}{2\lambda}]^2[\fra{2}{3}+\fra{1}{2\lambda}]
[\fra{1}{3}+\fra{1}{2\lambda}]^3[1-\fra{1}{2\lambda}]^2
[\fra{5}{2\lambda}-\fra{1}{3}][\fra{5}{2\lambda}-\fra{2}{3}] \nn\\
{\cal S}_{B_5B_3} & = & [\fra{7}{6}-\fra{1}{2\lambda}]
[\fra{1}{6}+\fra{3}{2\lambda}][\fra{3}{2}-\fra{3}{2\lambda}]^2
[\fra{5}{2\lambda}-\fra{5}{6}][\fra{1}{2}+\fra{1}{2\lambda}]^3
[\fra{3}{2\lambda}-\fra{1}{6}]^3[\fra{5}{2\lambda}-\fra{1}{2}]
[\fra{5}{6}-\fra{1}{2\lambda}]^4 \nn\\
{\cal S}_{B_5B_5} & = &
[\fra{5}{3}-\fra{2}{\lambda}]^2[\fra{2}{3}]^5
[\fra{3}{\lambda}-1][\fra{3}{\lambda}-\fra{2}{3}][\fra{1}{\lambda}]^5
[\fra{1}{3}+\fra{1}{\lambda}]^3[\fra{4}{3}-\fra{2}{\lambda}]^3
[\fra{2}{\lambda}][\fra{4}{3}-\fra{1}{\lambda}]^2 \nn\\
{\cal S}_{B_5K_1} & = & [\fra{1}{2}-\fra{1}{2\lambda}]^2
[\fra{5}{6}-\fra{1}{2\lambda}]^2[\fra{1}{6}+\fra{3}{2\lambda}]
[\fra{3}{2\lambda}-\fra{1}{6}] \nn\\
{\cal S}_{B_5K_2} & = & [\fra{5}{6}]^2[\fra{1}{2}]^2
[\fra{7}{6}-\fra{1}{\lambda}]^2[\fra{5}{6}-\fra{1}{\lambda}]^3
[\fra{1}{2}+\fra{1}{\lambda}][\fra{2}{\lambda}-\fra{1}{6}]
[\fra{2}{\lambda}-\fra{1}{2}]
\end{eqnarray}
The complete mass spectrum is given in appendix~\ref{massspec}, and
the full set of fusings for generic values of $q$ is summarised in
appendix~\ref{fusings}.
Appendix~\ref{polediags} 
contains a detailed description of the pole structures of all
S-matrix amplitudes appearing for $\lambda\leq\frac{9}{4}$. Closed
scattering diagrams, such as the ones shown earlier, have been constructed
for all poles not associated with bound states in this region.
In figure \ref{nfollowing} we show one particularly-elusive example.

\subsection{Problems for $\lambda>9/4$}
\label{probsec}
For $\lambda>9/4$, we have not been able to close the bootstrap. The
only exception is the point $\lambda=5/2$, for which $q=1$, the kink
states decouple, and the breather sector reproduces the minimal $E_8$
S-matrix. Away from this point, our difficulties may simply be due to
a failure to spot the necessary Coleman-Thun diagrams; on the other
hand, they may hint at a genuine breakdown of the bootstrap programme.
In the following we shall mention some of the problems that we
encountered, in the hope of contributing to further work on these
issues.

Many of the poles in the already-described S-matrix elements can be 
explained all the way up to $\lambda=3$. However, the Coleman-Thun 
diagrams for some poles do not close for $\lambda>9/4$, a
sign that new particles may need to be introduced.
Consider first
the poles in $S_{B_3B_3}$ at $\frac{1}{\lambda}$, in $S_{B_2B_2}$ at
$\frac{1}{\lambda}{-}\frac{1}{3}$, and in $S_{B_2B_5}$ at
$\frac{1}{3}{+}\frac{1}{2\lambda}$. Once $\lambda$ passes $9/4$,
the residues of these poles do not change sign, and if all are
provisionally associated with forward-channel bound states, 
the masses of these states as calculated from the formula
\begin{equation}
m_c^2=m_a^2+m_b^2-2m_am_b\cos(\pi u^c_{ab})
\label{fusform}
\end{equation}
(where $u^c_{ab}$ is the fusing angle $t$ at which the pole occurs)
all coincide. It is therefore natural to assign these poles to a single new
breather state $B_6$. From (\ref{fusform}), its  mass is
$4m\cos(\frac{\pi}{2}{-}\frac{\pi}{\lambda})%
\cos(\frac{\pi}{6}{-}\frac{\pi}{2\lambda})$\,. The
other fusing angles 
for this putative particle follow from (\ref{fusform}) on permuting
the labels $a$, $b$ and $c$, and all 
turn out to have simple (constant plus linear) dependencies on
$\frac{1}{\lambda}$\,; they are listed explicitly
in appendix~\ref{efusings}.

A similar story can be told for the poles in $S_{B_1B_5}$ and
$S_{B_3B_3}$ at
$\frac{2}{\lambda}{-}\frac{2}{3}$, leading us to introduce a further
breather $B_7$ with mass 
$8m\cos(\frac{\pi}{2}{-}\frac{\pi}{2\lambda})%
\cos(\frac{\pi}{6}{-}\frac{\pi}{2\lambda})%
\cos(\frac{\pi}{3}{-}\frac{\pi}{\lambda})$\,. However,
the (fifth-order) pole in $S_{B_5B_5}$ at $\frac{1}{\lambda}$\,, which
corresponds to $B_8$ at $\lambda=5/2$\,, is more enigmatic.
First, we note that it overlaps with an
odd-order pole at $\lambda=12/5$, causing 
its residue to change sign. This
might suggest that the identification of 
the direct and cross channel poles should 
be swapped at this point,
though since the pole is of higher order it is not possible
to say definitively that this must happen (cf.~the discussions in 
\cite{BCDSb}).
As mentioned in appendix~\ref{polediags}, bound state poles 
in some kink scattering amplitudes also 
have crossovers, but for these the S-matrix contrives to 
preserve the signs of the residues. 
It is also worth noticing that some of the S-matrix
elements involving $B_8$ have physical-strip zeroes for
$\lambda<12/5$. There is no a-priori reason why this should not occur,
but it breaks the pattern seen for all other S-matrix elements up to
this value of $\lambda$\,. These two problems together make our
identification of the $B_8$ particle somewhat tentative for
$\lambda\le 12/5$\,. 

\begin{figure}
\[
\begin{array}{ccc}
\includegraphics[height=0.35\linewidth]{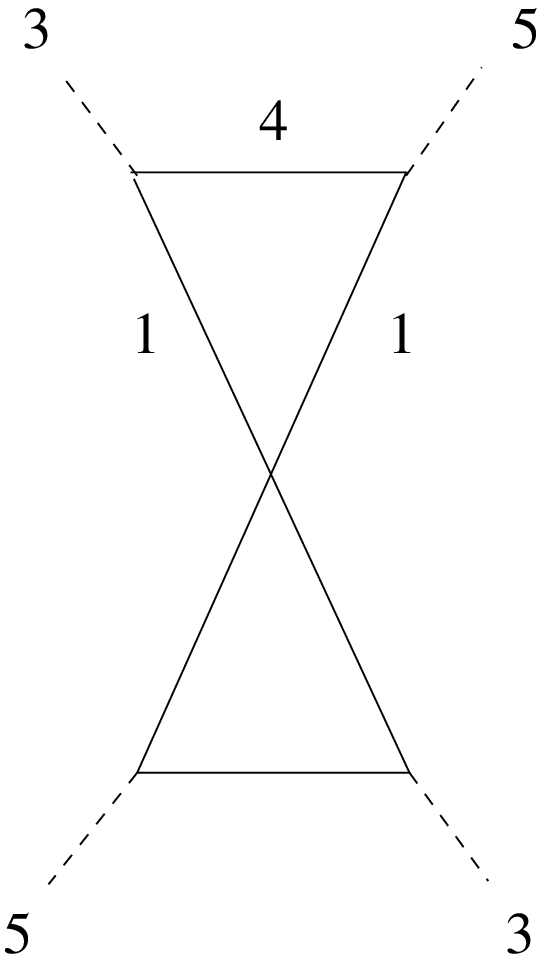} & \hspace{1cm}
\includegraphics[height=0.35\linewidth]{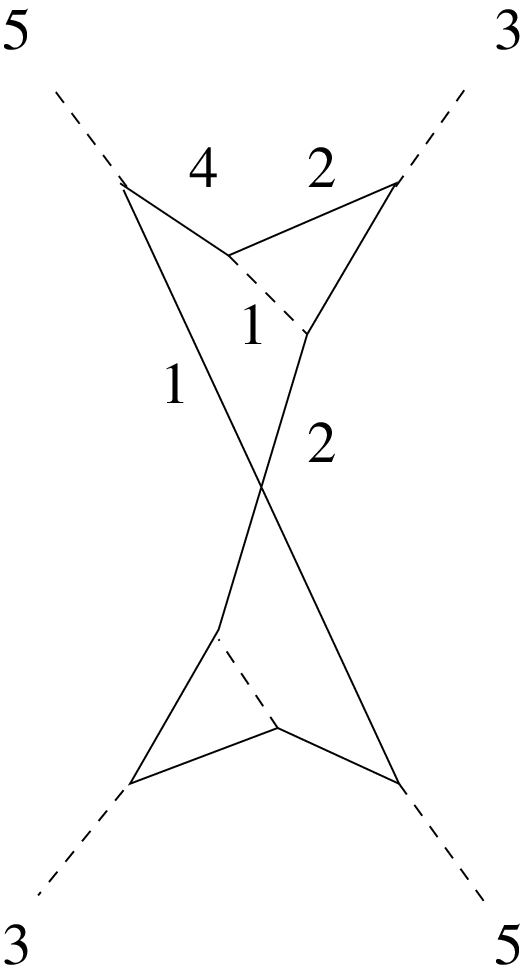} & \hspace{1cm}
\includegraphics[height=0.35\linewidth]{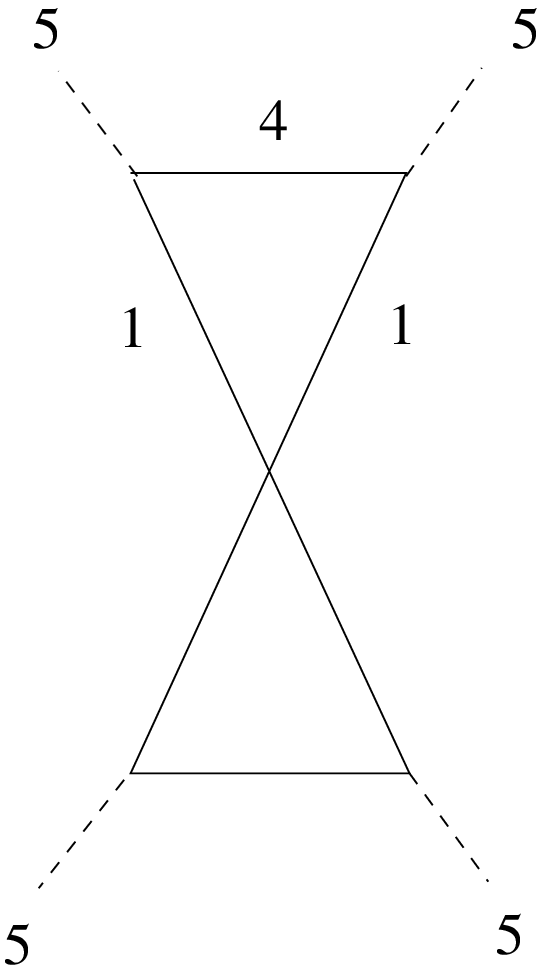} 
\end{array}
\]
\caption{Closed scattering diagrams 
without  corresponding poles in the S-matrix}
\label{k3mess}
\end{figure}

The worries about the $B_8$ particle would probably be resolved if
other, more serious, difficulties could be overcome. A number of poles
remain unexplained for $\lambda>9/4$ even after the introduction of
$B_6$, $B_7$ and $B_8$. Many have residues which change sign,
suggesting that for at least some range of $\lambda$ they will not
correspond to bound states. Of those which don't,
the poles at
$\frac{2}{3}-\frac{3}{2\lambda}$ in $S_{K_1K_2}$, $\frac{1}{6}$ in
$S_{B_2K_1}$, 
$\frac{1}{\lambda}$ in $S_{B_3K_1}$, 
$\frac{1}{6}+\frac{3}{2\lambda}$ in $S_{B_5K_1}$,
$\frac{1}{\lambda} - 
\frac{1}{6}$ in $S_{B_1K_2}$, and $\frac{1}{3} + \frac{1}{2\lambda}$ in
$S_{B_3K_2}$ give rise to kinks of 
the same mass. Calling this particle $K_4$, we could 
similarly identify another 
kink $K_4'$ with the poles at $\frac{3}{2\lambda}-\frac{1}{6}$ in
$S_{B_5K_1}$ and  in
$\frac{3}{2\lambda}-\frac{1}{3}$ in $S_{B_3K_2}$. 
However, the introduction of these new particles is problematic for a
number of reasons. Most fundamentally, and in contrast to the
situation for $B_6$, $B_7$ and $B_8$, some fusing angles
involving $K_4$ and $K_4'$, calculated using (\ref{fusform}), 
are {\em not\/} simple functions of $\frac{1}{\lambda}$. As a
result, the inclusion of $K_4$ or $K_4'$ in closed scattering diagrams 
for already-introduced particles
predicts poles which do not appear in their
S-matrix elements. For example, 
the diagrams shown in figure \ref{k3mess} can be drawn if $K_4$ is
included in the spectrum. 
In addition, the irrationality of the fusing angles leads to a
breakdown of the conserved charge bootstrap, forcing
higher-spin charges to be zero. This makes it highly unlikely that
$K_4$ and $K_4'$ should be added to the spectrum of the model.

On the other hand, 
we have not been able to account for the would-be $K_4$ and $K'_4$
poles in the manner of
Coleman and Thun, using the set of particles and fusings that we
have already identified.
Take the $\frac{2}{3}-\frac{3}{2\lambda}$ pole in $S_{K_1K_2}$ 
as an example. This pole is simple, so if the Coleman-Thun mechanism
is to be invoked, the naive order of poles from the relevant on-shell
diagrams must be reduced in some way.
One tactic is to search for closed scattering diagrams 
involving only kinks as internal states, in
order to get cancellations in sums over S-matrix elements of the sort
seen in earlier sections. Such diagrams must have the form shown in
figure \ref{12k3}.
\[
\begin{array}{c}
\refstepcounter{figure}
\label{12k3}
\includegraphics[width=0.25\linewidth]{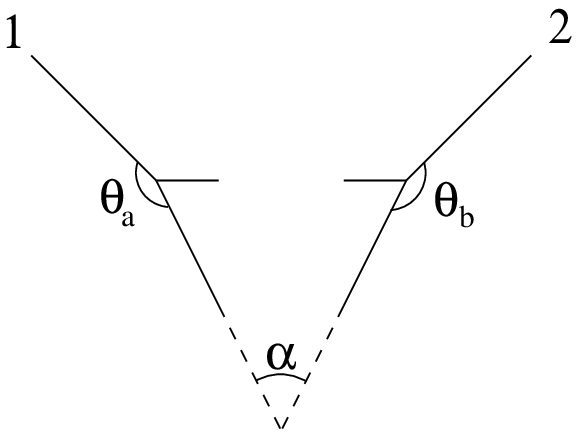} \\[4pt]
\parbox[t]{.8\linewidth}{
Figure~\ref{12k3}: Difficulties in finding Coleman-Thun 
diagrams for $S_{K_1K_2}$
}
\end{array}
\]
A quick check of the options for $\theta_a$ and $\theta_b$ shows that
the scattering angle between the outermost internal lines $\alpha =
-\fra{4}{3}-\fra{3}{2\lambda}+\theta_a+\theta_b$ is always less than
zero, and so no closed diagram can be constructed. 
Similar arguments hold for other poles, in particular those in 
the $S_{B_3K_1}$ and 
$S_{B_5K_1}$ amplitudes that were mentioned above.

In the absence of a satisfactory explanation for these poles, the 
closure of the bootstrap is still an open question. 
There remains one further possibility: new particles might
enter the spectrum which
are not simple bound states of any of the already-existing particles,
so that their masses would not be immediately visible in the existing
S-matrix elements. This would mimic the situation which would arise
in the sine-Gordon
model if one only knew of the breather particles, and wanted to
deduce the presence of the solitons by looking at breather scattering
alone.
It is very hard to rule this out, but further work will be needed 
before we can tell whether it offers a way to 
escape from the problems discussed in this section.

\subsection{Other work on $\phi_{12}$ perturbations}
\label{otherwork}
Since our results on $\phi_{12}$ perturbations are
incomplete, it is particularly important to compare them with other
work. We begin by mentioning the
points $\lambda = 2$, $\frac{9}{4}$ and $\frac{5}{2}$, which
correspond to the minimal $E_6, E_7$ and $E_8$ S-matrices.
These are rather special, as for these values of $\lambda$\,,
$q$ is an integer. 
Not only do various poles overlap, but 
also some instances of the Coleman-Thun mechanism break down.
This is because the total contribution associated with a given
on-shell diagram can
involve factors of $q-(\mbox{integer})$ (equations
(\ref{contra}), (\ref{contrb}) and (\ref{contrc}) provide some examples
of this phenomenon). This means that,
exceptionally, these poles do {\em not\/} have Coleman-Thun explanations
when $q$ hits
integer values, and this requires the introduction of `exceptional' bound
states in order to complete the bootstrap at these points. 
The $K_3$ kink state at the $E_7$ point, and the $B_4$
breather at the $E_8$ point, appear to have this evanescent status. 
In addition, the overlapping of the poles allows a number of extra
couplings to appear, and this should be borne in mind before worrying
that the tables of couplings in appendices~\ref{fusings} and
\ref{efusings} seems small in comparison
with tables for Toda models given in \cite{BCDSa}.
These subtleties aside, the S-matrices at the exceptional points are
perfectly self-consistent.
Their physical-strip pole
structures match those of affine Toda field theories,
and the extensive discussions contained in 
\cite{BCDSa,BCDSb} can be borrowed to verify that all higher poles
can be described via the Coleman-Thun mechanism. 

Away from integer values of $q$,
some general aspects of the spectrum of $\phi_{12}$-perturbed minimal
models were discussed by
Smirnov in \cite{Sm},
while the particular case of $\CM_{56}+\phi_{12}$ was treated 
by Martins in~\cite{Martins:1991jj} and
by Koubek in~\cite{Ko}.
In none of these papers was a complete analysis of pole structures
attempted, but other aspects permit comparisons to be made.
In the papers of Smirnov and Koubek,
$\lambda$ corresponds to $\frac{\pi}{\xi}$, and
$\theta\rightarrow\beta$. 

In \cite{Sm}, Smirnov gave an initial description of
the spectrum implied by his 
S-matrix, concentrating mainly on points related to perturbations
of unitary minimal models. 
The chief difference between Smirnov's  S-matrix and that of Chim and
Zamolodchikov is a matter of kink structure, so one would
expect the two to agree on 
the spectrum of bound states and the diagonal scattering amplitudes.
Indeed, while explicit
formulae for the remaining S-matrix elements 
were not given in
\cite{Sm}, Smirnov comments that the spectrum for $\phi_{12}$
perturbations settles down to four particles
for (his) $\xi\ge\pi/2$. In our notation
this is $\lambda\le 2$\,, and so this regime of $\phi_{12}$
perturbations corresponds to the range
$3/2<\lambda\le 2$ discussed in \S\ref{fourpart}, for which we did
indeed find four particles. Furthermore, the masses are easily checked
to agree, and so all is consistent.

Referring to (\ref{lambdamintricrit}), 
the $\phi_{12}$ perturbation of $\CM_{56}$ discussed 
in \cite{Martins:1991jj} and~\cite{Ko}, should correspond to
$\lambda=\frac{21}{10}$, which lies in the region where we predicted
six particles.
In~\cite{Martins:1991jj}, Martins made a detailed numerical study of the
finite-volume spectrum of $\CM_{56}+\phi_{12}$ using the truncated
conformal space approach. 
(He also discussed some subtleties relating
to the choice of modular invariant for
the unperturbed theory; as mentioned above, we do not expect this to affect
the full spectrum of particle masses when the theories are considered
on the infinite line.)
High-mass states were hard to detect, but he
was able to predict the presence of five particles, which in our
notation are $K_1$, $B_1$, $K_2$, $B_2$ and $B_3$, with 
numerically-obtained values for the masses which are consistent with
predictions from the exact S-matrix. The remaining particle, $B_5$,
has a much higher mass and so its absence from the numerical data of
\cite{Martins:1991jj} is no surprise.

This case was further examined in \cite{Ko}. 
Translating from our notation to that of~\cite{Ko}, 
$[a]\leftrightarrow {{<}a{>}}$,
$B_2^{\mbox{\tiny (ref.\cite{Ko})}}=B_3^{\mbox{\tiny (us)}}$ and
$B_4^{\mbox{\tiny (ref.\cite{Ko})}}=B_5^{\mbox{\tiny (us)}}$. However, 
the mass of 
the breather identified in \cite{Ko} as
$B^{\mbox{\tiny (ref.\cite{Ko})}}_3$ is also equal to that of 
$B_3^{\mbox{\tiny (us)}}$, which is already in the spectrum at $\lambda<2$. 
As can be seen from appendix~\ref{massspec}, the other extra breather
to enter the spectrum for $\lambda>2$, $B_2^{\rm (us)}$, has a
mass less than that of $B_3^{\mbox{\tiny (us)}}$, which perhaps explains why it
was missed in \cite{Ko}.  
As a result of this problem, the spectrum and 
S-matrix given in \cite{Ko} are
unfortunately incomplete, but insofar as they go and modulo some
further typos, they are otherwise consistent with our results,
specialised to $\lambda=\frac{21}{10}$\,.

Finally, we should mention that S-matrices for $\phi_{12}$
perturbations of minimal models $\CM_{2,2n{+}1}$ were discussed in
\cite{Koubek:1991qq}. These theories have $\lambda=3n$ in our
notation, and hence are a long way from the region 
$\lambda\in [\frac{3}{2},3]$ relevant to tricritical scaling Potts 
models -- in particular, the perturbing operator always has a negative
scaling dimension apart from the somewhat-trivial case of $n{=}1$.
Perhaps more to the point, they also all correspond formally to $q=0$,
and the simplifications of the scattering theory at such points
\cite{Sm} make it hard to draw any general lessons.
Nevertheless, it is interesting that the bootstrap can be closed at
least at some locations beyond the region that we were able to treat.

\resection{Conclusions}
In this paper we have given what to the best of our knowledge is
the first complete treatment of the pole structure and bootstrap
for the critical and tricritical Potts models. 
For all of the critical models, and for the tricritical models with
$4>q>2$, we have found a spectrum of particles such that {\em all\/}
S-matrix poles have potential field-theoretical explanations. This has
been achieved by a novel variant of the Coleman-Thun mechanism, taking
into account the formal counting of intermediate vacua at general
values of $q$. The cancellations are at times extremely intricate, and
we take this success as offering retrospective
justification of our approach, though a rigorous framework
is still lacking.

Our calculations have proceeded on a case-by case (or rather,
pole-by-pole) basis, which becomes increasingly laborious 
as the number of particles increases. It is tempting to suppose that
there must be a better way to do all of this, if only the relevant
underlying structure could be identified. For the much-simpler
examples of the ADE-related diagonal scattering theories, 
a universal understanding of bootstrap closure has been
achieved 
using a construction of the S-matrix elements based on the
theory of root systems \cite{PED,PEDb}, and it would be very interesting 
to have a similar treatment for general $q$-state Potts models. At
present this seems to be a long way off, though the results of Oota
\cite{Oota}, extending the root systems approach to cover
the non simply-laced Toda theories, may be a sign that
things are not completely hopeless.
(Further discussions of the hidden geometry of affine Toda field
theory can be found in
\cite{Fring:1991gh,Dorey:1992gr,pedrev,Fring:1999jk}.)

The most important outstanding question left by our work concerns the
situation for $\lambda>\frac{9}{4}$, where we were unable to
complete the bootstrap. Problems with bootstrap closure
have been encountered a couple of times before. In \cite{SalKau}, the
intricacy of the mass spectrum for general complex $a_2^{(1)}$ Toda theory
is discussed. In Smirnov's approach, the Potts S-matrices are
associated with $a^{(2)}_2$; given the relations between $a_2^{(1)}$
and $a_2^{(2)}$ it is reasonable to hope that our
results even for $\lambda<\frac{9}{4}$ may be of some relevance to these
issues. However, since the situation for $a^{(1)}_2$ is likely to be
at least as complicated as that for $a^{(2)}_2$, this may not be the
best place to look for hints as to how to close the bootstrap for
$\lambda>\frac{9}{4}$. Rather, it seems more promising to investigate
further perturbations of minimal models, for which there are at least
techniques such as the TCSA to fall back on. In situations where the
Potts ground state (generated by the identity operator) and the
minimal model ground state (generally generated by some
negative-dimension operator) become degenerate in infinite volume, we
would expect scaling Potts and perturbed minimal model results to be
directly related. Difficulties with the
bootstrap for the models $\CM_{3,5}$ and $\CM_{3,7}$ perturbed by
$\phi_{12}$ have been remarked by Mussardo and Takacs \cite{mussgab},
though as these theories have $\lambda=\frac{7}{2}$ and $\frac{11}{2}$,
they are not directly relevant to our current concerns. The interval
$\frac{9}{4}<\lambda<3$ corresponds, via (\ref{lambdamintricrit}),
to $\frac{5}{4}<\frac{p'}{p}<\frac{3}{2}$\,, and any information on
$\phi_{12}$ perturbations of these minimal models, away from the $E_8$
point $\frac{p'}{p}=\frac{4}{3}$, would be extremely interesting, as
would any indications of extra pathologies in such cases.

We end with a curious piece of numerology.
Recall that as the parameter $\lambda$ runs
from $0$ to $3$, we pass first through the critical and then the 
tricritical scaling Potts models. At rational values of
$\lambda$, the theories are also associated
with $\phi_{21}$ or $\phi_{12}$ perturbations of minimal models.
But some
points are even more special, in that their minimal models can be
realised as `diagonal'
coset conformal field theories of the form $g^{(1)}\times
g^{(1)}/g^{(2)}$, with field labelled by $(1,1,{\rm ad})$ always being
the perturbing operator. (See, for example, chapter 18 of
\cite{bigyellowbook} and references therein for more on the coset
construction.) These points, together with the corresponding
values of $q$ and $c$, are listed in
appendix~\ref{massspec}, and they suggest that the scaling Potts
models provide a structure which unifies the following sequence of
Lie algebras:
\eq
\{\,
A_1\,,~
A_2\,,~
G_2\,,~
D_4\,,~
F_4\,,~
E_6\,,~
E_7\,,~
E_8~
\label{list}
\}
\en
Remarkably, the same sequence has appeared in the pure mathematics
literature, in the work of Vogel,
Deligne, Cohen and de Man,
Cvitanovic and others -- see 
\cite{Vogel,Del,DelII,Cohman,Cvit,Macf,landsberg}
for a selection of references. 
This set of algebras, sometimes referred to as `the' exceptional
series, is picked out by
a number of special properties -- for example,
the tensor products ${\rm ad}\otimes{\rm ad}$ decompose in a uniform
way (Deligne proposed that this should extend to higher powers
$\otimes^k{\rm ad}$, and conjectured that this could be explained by
the existence of some more general class 
of objects, depending on a
parameter $t$, which specialise
to the representations of the members of the exceptional series at
certain values of $t$).  The algebras also make up
the extended last line of the Freudenthal
magic square (in this context, it is interesting that an appearance
of the magic square has been noted, on a 
case-by-case basis, in studies of $R$- and $K$- matrices
and the Yang-Baxter equation (see for example \cite{niall})).
However, the deep sense in which the algebras (\ref{list})
form a family is still
rather mysterious, and so the fact that they are found in
connection with the {\em continuous} set of Potts models seems to be quite
suggestive.
Inspired by this coincidence, we can consider an alternative
labelling of the Potts models, by setting
\eq
h^{\vee}(\lambda)=\frac{6\lambda}{3-\lambda}~.
\label{hvee}
\en
This parameter coincides
with the dual Coxeter number of the relevant algebra at the 
special points, where it is also related to the parameter $a$ used in
\cite{Del,Cohman} by $h^{\vee}=1/a$.
Two more interesting properties can now be noted - first, the
`natural' range for $\lambda$, $[0,3]$, is mapped to the range
$[0,\infty]$ for $h^{\vee}$. At the Lie algebra related points,
$h^{\vee}$ is in some senses a measure of the complexity of the
corresponding scattering theory, so this suggests that $\lambda=3$ may indeed
be some sort of natural boundary also from the S-matrix point of view. 
Second, there is one more special point for the Potts models,
labelled `P' in appendix~\ref{massspec}
and corresponding to
$\lambda=\frac{1}{2}$, which is $q=1$ on the
critical branch -- the percolation point. It is natural to add
this point to our list, as it then gives the special points a symmetry
under $\lambda\to 3{-}\lambda$. One might worry that this would spoil
the nice match with the exceptional series of Lie algebras, but in \cite{Del}, 
Deligne remarks that it is natural to add
the trivial group $e$ to the list
(\ref{list}), so long as it is assigned the dual Coxeter
number $6/5$. This is precisely the value 
that the Potts models would suggest, if
the percolation value $\lambda=\frac{1}{2}$
is substituted into (\ref{hvee}).

In fact, the exceptional series has (at least) one further member: in
\cite{DelII}, Deligne and de Man remark that the superalgebra
$OSp(1|2)$, with dual Coxeter number $3/2$,
can be added between $e$ and $A_1$.  (This to some degree explains the
fact, already noted by Cohen and de Man
in \cite{Cohman}, that for $h^{\vee}=3/2$ various
dimension formulae take integer values.) Can we find a r\^ole for this
superalgebra in the Potts story? The answer turns out to be yes.
Inverting (\ref{hvee}), $h^{\vee}=3/2$ corresponds to $\lambda=3/5$,
$q=(5{-}\sqrt{5})/2$ and $c=8/35$. The construction of diagonal coset
models based on Lie superalgebras seems to be relatively unexplored
territory, but some facts about the relevant
affine superalgebras are known. In
particular, the level $k$ $OSp(1|2)$ central charge is 
\eq
c(\,OSp(1|2)^{(k)})=\frac{2k}{2k{+}3}
\en
(see, for example, \cite{Ennes:1997vt} and references therein).
Assuming that no subtleties interfere with the usual calculation, we
then expect
\eq
c\left(\frac{OSp(1|2)^{(1)}\times
OSp(1|2)^{(1)}}{OSp(1|2)^{(2)}}\right) = \frac{8}{35}
\en
which is exactly the required value. Note that this is the central
charge of a non-unitary minimal model, $\CM_{7,10}$.
This does not contradict its being realised as a coset, since
superalgebras are involved.
While not all fields of the nonunitary model
appear in the corresponding Potts model, the
identity and the energy operator are present, and at least in this sense
we can claim to have found the superalgebra $OSp(1|2)$ to be embedded into 
the continuous family of $q$-state Potts models.

Cohen and de Man \cite{Cohman}
found one further point at which integers
appear in dimension formulae: translating into our
notations it is $h^{\vee}=24$, $\lambda=12/5$.
As far as we are aware, no
group-theoretical interpretation of this point is known, but
from the Potts perspective it should be related to the 
tricritical model at $q=(5{-}\sqrt{5})/2$, or the $\phi_{12}$
perturbation of the minimal model $\CM_{10,13}$ (with central charge 
$38/65$). This is particularly tantalising: lying between the $E_7$ and
$E_8$ points, $\lambda=12/5$
is in the region where we have not been able to close
the bootstrap, and any new information 
might give us the necessary hint to resolve the
problems described in \S\ref{probsec} above. 
Furthermore, there are independent reasons
to think that $\lambda=12/5$ might be special: 
as mentioned in \S\ref{probsec}, for $\lambda>12/5$
the S-matrix elements involving $B_8$ have no
physical-strip zeroes, perhaps signalling that this is the point at
which the $B_8$ particle enters the spectrum.

While we do not have any explanation as yet for why the link between
the Potts models and the exceptional sequence of Lie (super-)algebras
should be so neat,
we feel that the coincidences are sufficiently striking to merit
further investigation. A better understanding might shed
light both on the structure of the
Potts model S-matrices, and on the deeper meaning
of the exceptional series and the Deligne conjecture.

\bigskip

\noindent
{\bf Acknowledgements -- }
This paper has been a long time in the writing, during the course of
which we have benefitted from discussions and correspondence with
John Cardy, 
Aldo Delfino, 
Clare Dunning, 
Davide Fioravanti, 
Philippe di Francesco, 
Paul Fendley,
Alan Macfarlane,
Barry McCoy,
Giuseppe Mussardo,
Bill Oxbury,
Marco Picco, 
Hubert Saleur, 
Tony Sudbery,
Gabor Takacs, 
Ole Warnaar,
Gerard Watts
and 
Jean-Bernard Zuber. 
In addition, PED thanks the YITP, Kyoto, for hospitality.
Visits of PED to
YITP were funded by a Daiwa-Adrian Prize and a Royal Society / JSPS
Anglo-Japanese Collaboration grant, title `Symmetries and integrability'.
RT thanks the EPSRC for an Advanced Fellowship,
and AJP thanks the JSPS and the FAPESP for postdoctoral fellowships.
%
\newcommand\sL[1]{\!\!\!#1}
\newcommand\sR[1]{\,\,\,#1\!}

\appendix
\newpage
\resection{The mass spectrum}
\label{massspec}

\newcommand\hf{\hspace*{21pt}}
\hspace*{35pt}
\includegraphics[width=0.84\linewidth,height=0.62\linewidth]%
{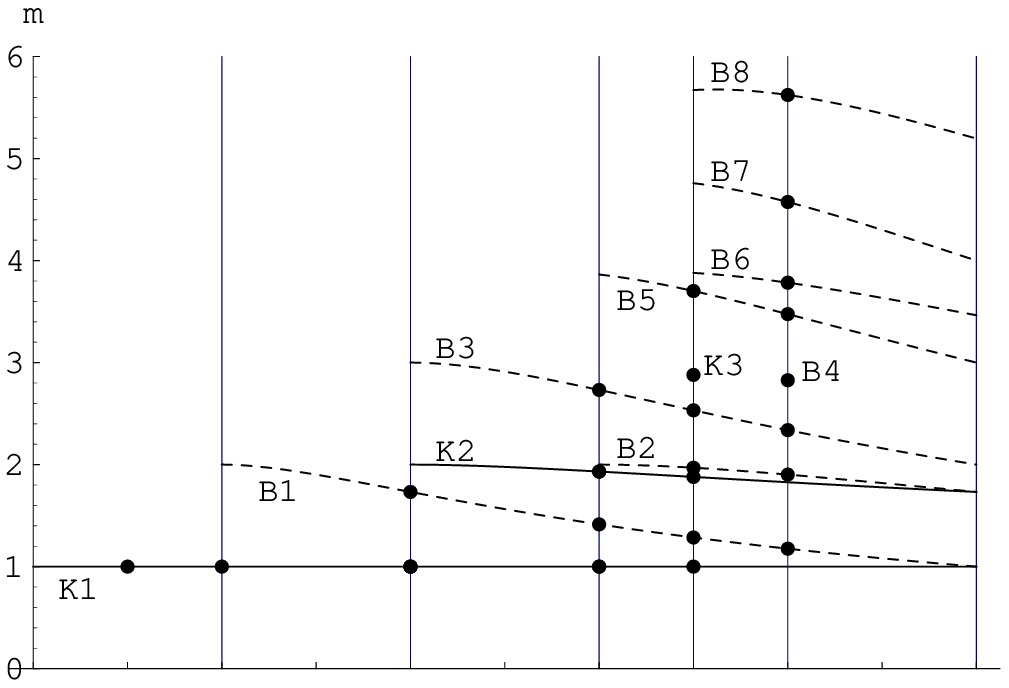}\\[4pt]
\hspace*{5pt}\parbox{0.95\linewidth}{
\begin{tabular}{lllllllllllllllll}
$~\lambda~:$
&$\frac{1}{2}$
&$\frac{3}{4}$
&$1$
&$\sL{\frac{6}{5}}$
&$\frac{3}{2}$
&$\sR{\frac{9}{5}}$
&$2$
&$\frac{9}{4}$
&$\frac{5}{2}$&&$3$\\[6pt]
$~q~:$ &$1$ &$2$ &$3$ & $\sL{\!\!\!\!\scriptstyle\frac{5{+}\sqrt{5}}{2}}$
 &$4$ & $\sR{\!\!\!\!\scriptstyle\frac{5{+}\sqrt{5}}{2}}$
&$3$ &$2$ &$1$&&$0$\\[6pt]
$~c~:$ &$0$ &$\fract{1}{2}$ &$\fract{4}{5}$ &$\sL{\fract{14}{15}}$
 &$1$ &$\sR{\fract{52}{55}}$ &$\fract{6}{7}$ 
&$\fract{7}{10}$ &$\fract{1}{2}$&&$0$\\[6pt]
$\!\!\!\!\!$Model:
&P
&$A_1$
&$A_2$
&$\sL{G_2}$
&$D_4$
&$\sR{F_4}$
&$E_6$
&$E_7$
&$E_8$
& \\
&\hf &\hf &\hf &\hf &\hf &\hf &\hf &\hf &\hf &\hf &\hf &\hf &\hf &\hf\hf
\end{tabular}\hfil
}

\vspace{-15pt}

\bea
m_{K_1} & = & m \nn\\[6pt]
m_{B_1} & = & 2m\cos(\fra{\pi}{2}-\fra{\pi}{2\lambda}) \nn\\[6pt]
m_{K_2} & = & 2m\cos(\fra{\pi}{3}-\fra{\pi}{2\lambda}) \nn\\
m_{B_3} & = & 4m\cos(\fra{\pi}{2}-\fra{\pi}{2\lambda})
\cos(\fra{\pi}{2\lambda}-\fra{\pi}{6}) \nn\\[6pt]
m_{B_2} & = & 2m\cos(\fra{\pi}{2}-\fra{\pi}{\lambda}) \nn\\
m_{B_5} & = & 4m\cos(\fra{\pi}{2}-\fra{\pi}{\lambda})
\cos(\fra{\pi}{3}-\fra{\pi}{2\lambda})\nn\\[6pt]
m_{B_6} & = & 4m\cos(\fra{\pi}{2}-\fra{\pi}{\lambda})
\cos(\fra{\pi}{6}-\fra{\pi}{2\lambda})\nn\\
m_{B_7} & = & 8m\cos(\fra{\pi}{2}-\fra{\pi}{2\lambda})
\cos(\fra{\pi}{6}-\fra{\pi}{2\lambda})\cos(\fra{\pi}{3}-\fra{\pi}{\lambda})
\nn\\
m_{B_8} & = & 8m\cos(\fra{\pi}{2}-\fra{\pi}{\lambda})
\cos(\fra{\pi}{3}-\fra{\pi}{2\lambda})\cos(\fra{\pi}{2\lambda})\nn
\,.\nn
\eea

\noindent
The breather states are labelled so as to be mass-ordered at the $E_8$
point $\lambda=\frac{5}{2}$; note also that the states $K_3$ and $B_4$ 
are only present in the model for $\lambda=\frac{9}{4}$ and
$\frac{5}{2}$ respectively, for reasons that were explained in
\S\ref{otherwork}. The solid dots indicate the mass spectra of the
diagonal scattering theories which occur at integer values of $q$.
For $\lambda>\frac{9}{4}$, the mass spectrum is almost certainly
incomplete, save for the $E_8$ point $\lambda=\frac{5}{2}$\,; 
in addition, as discussed in \S\ref{probsec}, it is
possible that the appearance of the $B_8$ particle should be postponed
to $\lambda>\frac{12}{5}$.
In the list of model identifications, `P' indicates that the theory at
$\lambda=\frac{1}{2}$ is related to the percolation problem.
The remaining entries are Lie algebras $g$, and
signal that the corresponding model is related
to a perturbation of the
$g^{(1)}\times g^{(1)}/g^{(2)}$ coset conformal field theory 
by its $(1,1,{\rm adj})$ operator.

\resection{Fusings and fusing angles for $\lambda<9/4$}
\label{fusings}
This table summarises the fusings and fusing angles between particles 
in the region where we have been able to close the bootstrap. Each
entry shows the particles which can be found as bound states of the
particles listed along the top and left of the table, together
with the angle $t$ at which each fusing occurs. 
Depending on the value of $\lambda$, some of the bound states may be
absent from the spectrum. In such cases, the corresponding pole is
either off the physical strip, or else has a Coleman-Thun explanation.

\newcommand\mc[1]{\multicolumn{1}{l||}{#1}}

\begin{center}
\begin{tabular}{|l||l|l|l|l|l|l||}
\hline
\hline
 &\mc{} &\mc{} & &\mc{} & & \\[-7pt]
 &\mc{$K_1$} &\mc{$B_1$} &$K_2$ &\mc{$B_3$} &$B_2$ &$B_5$ \\[5pt]
\hline
 &\mc{} &\mc{} & &\mc{} & & \\[-12pt]
\hline
\mc{}&\mc{}&\mc{}&&\mc{}&&\\[-10pt]
$K_1$ 
& \mc{${\scriptstyle K_1}~~~\fract{2}{3}$}
& \mc{${\scriptstyle K_1}~~~\fract{1}{2}{+}\fract{1}{2\lambda}$}
& ${\scriptstyle K_1}~~~\fract{2}{3}{+}\fract{1}{2\lambda}$
& \mc{${\scriptstyle K_2}~~~\fract{1}{\lambda}{+}\fract{1}{3}$}
& ${\scriptstyle K_1}~~~\fract{1}{2}{+}\fract{1}{\lambda}$
& \\[4pt]
& \mc{${\scriptstyle B_1}~~~1{-}\fract{1}{\lambda}$}
& \mc{${\scriptstyle K_2}~~~\fract{1}{6}{+}\fract{1}{2\lambda}$}
& ${\scriptstyle B_1}~~~1{-}\fract{1}{2\lambda}$
& \mc{}
&
& \\[4pt]
& \mc{${\scriptstyle K_2}~~~\fract{2}{3}{-}\fract{1}{\lambda}$}
& \mc{}
& ${\scriptstyle B_3}~~~1{-}\fract{3}{2\lambda}$
& \mc{}
&
& \\[4pt]
& \mc{${\scriptstyle B_2}~~~1{-}\fract{2}{\lambda}$}
& \mc{}
& 
& \mc{}
&
& \\[4pt]
\cline{1-2}
&&\mc{}&&\mc{}&&\\[-12pt]
\hline
&&&&&&\\[-10pt]
$B_1$ 
& 
& \mc{${\scriptstyle B_1}~~~\fract{2}{3}$}
& ${\scriptstyle K_1}~~~\fract{5}{6}$
& \mc{${\scriptstyle B_1}~~~\fract{7}{6}{-}\fract{1}{2\lambda}$}
& ${\scriptstyle B_1}~~~1{-}\fract{1}{2\lambda}$
& ${\scriptstyle B_3}~~~\fract{4}{3}{-}\fract{1}{\lambda}$\\[4pt]
&
& \mc{${\scriptstyle B_3}~~~\fract{1}{\lambda}{-}\fract{1}{3}$}
& 
& \mc{${\scriptstyle B_2}~~~\fract{1}{2}{+}\fract{1}{2\lambda}$}
& ${\scriptstyle B_3}~~~\fract{2}{3}{-}\fract{1}{2\lambda}$
& \\[4pt]
&
& \mc{${\scriptstyle B_2}~~~\fract{1}{\lambda}$}
& 
& \mc{${\scriptstyle B_5}~~~\fract{3}{2\lambda}{-}\fract{1}{2}$} 
&
& \\[4pt]
\cline{1-3}
&\multicolumn{1}{c}{}&&&\mc{}&&\\[-12pt]
\hline
&\multicolumn{1}{c}{}&&&\mc{}&&\\[-10pt]
$K_2$ 
& \multicolumn{1}{c}{}
& 
& ${\scriptstyle K_2}~~~\fract{2}{3}$
& \mc{${\scriptstyle K_1}~~~\fract{2}{3}{+}\fract{1}{2\lambda}$}
&
& ${\scriptstyle K_2}~~~\fract{1}{2}{+}\fract{1}{\lambda}$\\[4pt]
& \multicolumn{1}{c}{}
&
& ${\scriptstyle B_5}~~~1{-}\fract{2}{\lambda}$
& \mc{}
& 
& \\[4pt]
\hline
&\multicolumn{2}{c}{}&&\mc{}&&\\[-10pt]
$B_3$ 
& \multicolumn{2}{c}{}
& 
& \mc{${\scriptstyle B_3}~~~\fract{2}{3}$}
& ${\scriptstyle B_1}~~~\fract{5}{6}$
& ${\scriptstyle B_1}~~~\fract{7}{6}{-}\fract{1}{2\lambda}$\\[8pt]
\cline{1-5}
&\multicolumn{3}{c}{}&&&\\[-12pt]
\hline
&\multicolumn{3}{c}{}&&&\\[-10pt]
$B_2$ 
& \multicolumn{3}{c}{}
& 
& ${\scriptstyle B_2}~~~\fract{2}{3}$
& ${\scriptstyle B_2}~~~\fract{2}{3}{+}\fract{1}{2\lambda}$\\[4pt]
& \multicolumn{3}{c}{}
& 
& ${\scriptstyle B_5}~~~\fract{2}{3}{-}\fract{1}{\lambda}$
& \\[4pt]
\hline
&\multicolumn{4}{c}{}&&\\[-10pt]
$B_5$ 
& \multicolumn{4}{c}{}
& 
& ${\scriptstyle B_5}~~~\fract{2}{3}$\\[6pt]
\hline
\hline
\end{tabular}
\\[20pt]
\end{center}
As mentioned in \S\ref{otherwork}, exceptional couplings 
appear at the points
$\lambda=2,\frac{9}{4},\frac{5}{2}$. Rather than tabulate them here, we
refer the reader to \cite{BCDSa} for complete listings.

\newpage 

%
%
\resection{Extra fusings and fusing angles for $\lambda>9/4$}
\label{efusings}
Our results are incomplete for $\lambda>9/4$, and we cannot be sure
that the bootstrap closes at all. However, there is good evidence for
the existence of at least three further particles, 
namely $B_6$, $B_7$ and $B_8$. The first two we
expect to be present for all $\lambda>9/4$, while
$B_8$ may only enter the spectrum for $\lambda>12/5$. 
The following table summarises the additional
fusings and fusing angles involving these new
particles. Further particles cannot be ruled out (indeed, they are
most probably required if the bootstrap is to close) and so we do not
claim that this list is exhaustive.

\begin{center}
\begin{tabular}{|l||l|l|l|l|l|l||}
\hline
\hline
 & & &\mc{} & & & \\[-7pt]
 &$B_3$ &$B_2$ &\mc{$B_5$}&$B_6$ &$B_7$ &$B_8$ \\[5pt]
\hline
 & & &\mc{} & & & \\[-12pt]
\hline
 & & &\mc{} & & & \\[-10pt]
$B_1$ 
& 
& 
& \mc{${\scriptstyle B_7}~~~\fract{2}{\lambda}{-}\fract{2}{3}$}
& 
& ${\scriptstyle B_5}~~~\fract{3}{2}{-}\fract{3}{2\lambda}$
& \\[4pt]
\hline
&&&\mc{}&&&\\[-10pt]
$B_3$ 
& ${\scriptstyle B_6}~~~\fract{1}{\lambda}$
& 
& \mc{}
& ${\scriptstyle B_3}~~~1{-}\fract{1}{2\lambda}$
& ${\scriptstyle B_3}~~~\fract{4}{3}{-}\fract{1}{\lambda}$
& \\[4pt]
& ${\scriptstyle B_7}~~~\fract{2}{\lambda}{-}\fract{2}{3}$
&
& \mc{}
& 
& 
& \\[4pt]
\hline
&&&\mc{}&&&\\[-10pt]
$B_2$ 
&
& ${\scriptstyle B_6}~~~\fract{1}{\lambda}-\fract{1}{3}$
& \mc{${\scriptstyle B_6}~~~\fract{1}{3}{+}\fract{1}{2\lambda}$}
& ${\scriptstyle B_2}~~~\fract{7}{6}{-}\fract{1}{2\lambda}$
& 
& ${\scriptstyle B_6}~~~\fract{4}{3}{-}\fract{1}{\lambda}$\\[4pt]
& 
&
& \mc{}
& ${\scriptstyle B_5}~~~\fract{5}{6}{-}\fract{1}{2\lambda}$
& 
& \\[4pt]
& 
&
& \mc{}
& ${\scriptstyle B_8}~~~\fract{3}{2\lambda}{-}\fract{1}{2}$
& 
& \\[4pt]
\hline
&\multicolumn{1}{c}{}&&\mc{}&&&\\[-10pt]
$B_5$ 
& \multicolumn{1}{c}{}
& 
& \mc{${\scriptstyle B_8}~~~\fract{1}{\lambda}$}
& ${\scriptstyle B_2}~~~\fract{5}{6}$
& ${\scriptstyle B_1}~~~\fract{7}{6}{-}\fract{1}{2\lambda}$
& ${\scriptstyle B_5}~~~1{-}\fract{1}{2\lambda}$\\[8pt]
\cline{1-4}
&\multicolumn{2}{c}{}&&&&\\[-12pt]
\hline
&\multicolumn{2}{c}{}&&&&\\[-10pt]
$B_6$ 
& \multicolumn{2}{c}{}
& 
& ${\scriptstyle B_6}~~~\fract{2}{3}$
&
& ${\scriptstyle B_2}~~~\fract{7}{6}{-}\fract{1}{2\lambda}$\\[4pt]
\hline
&\multicolumn{3}{c}{}&&&\\[-10pt]
$B_7$ 
& \multicolumn{3}{c}{}
& 
& ${\scriptstyle B_7}~~~\fract{2}{3}$
& \\[4pt]
\hline
&\multicolumn{4}{c}{}&&\\[-10pt]
$B_8$ 
& \multicolumn{4}{c}{}
&
& ${\scriptstyle B_8}~~~\fract{2}{3}$
 \\[6pt]
\hline
\hline
\end{tabular}\\[40pt]
\end{center}
\resection{Bound state poles of the S-matrix for
$\lambda<9/4$}
\label{polediags}
In this appendix we summarise the bound states of the scattering
amplitudes which appear for 
$\lambda\leq\frac{9}{4}$. For the kink-kink amplitudes, the 
physical strip
pole contents of each separate amplitude are given in tables 
\ref{t:psp}, \ref{t:pspk1k2} and \ref{t:pspk2k2} of the main text;
here we show their locations (though not their orders)
with a symbol $\pbl{x}$, indicating
that the combined set of amplitudes has poles at $t=x$ and at
$t=1{-}x$.
The poles of the other S-matrix elements can be read off quite easily
from their explicit formulae -- in the following, we add a subscript
$a$ to a block $[x]$ to signify that $a$ appears as forward-channel
bound state of that scattering process, at $\theta=i \pi x$.
A question mark indicates that, at least for some range of
$\lambda\in (\frac{9}{4},3]$,
the corresponding pole is currently unexplained.
In spite of the problems
in completing the bootstrap in the interval
 $\frac{9}{4}<\lambda<\frac{5}{2}$, at 
$\lambda=\frac{5}{2}$ (ie. $q=1$) the S-matrix is 
well behaved and reduces to that of
the $E_8$ related model. 

In the plots, $t=\frac{\theta}{i\pi}$ as in the main text, 
and $l=\lambda$. Double lines denote poles of even order. Grey shading
denotes presence of a higher order scattering process for that pole. In
general, direct channel poles have solid lines and cross
channel poles have dashed lines - for poles with no associated bound
states the choice is arbitrary. Forward-channel bound state poles
are identified on the graphs by the relevant particle, while dotted
grey shading indicates poles for which we have yet to find a
satisfactory explanation.
The crossings-over of poles
in the $K_1\,K_1$ and $K_1\,K_2$ 
scattering amplitudes cause no problems for the assignment of forward
and crossed channels, as the amplitudes
affected -- $\mathcal{S}_1$ 
and $\mathcal{S}_2$ at $\lambda=\frac{3}{2}$, and
$\mathcal{S}_3$ at $\lambda=2$ -- 
vanish at these points. This extra zero in the
non-scalar factors changes the 
analytic continuation, preserving the signs of the 
residues.

In each graph, the value $\lambda=\fra{9}{4}$ is indicated by a
vertical line. Beyond this point the problems mentioned in
\S\ref{probsec}
set in, and our results must be considered to be incomplete.

\subsection{Kink-kink S-matrix elements}

\bigskip

\begin{center}

\vspace{2mm}

\fa{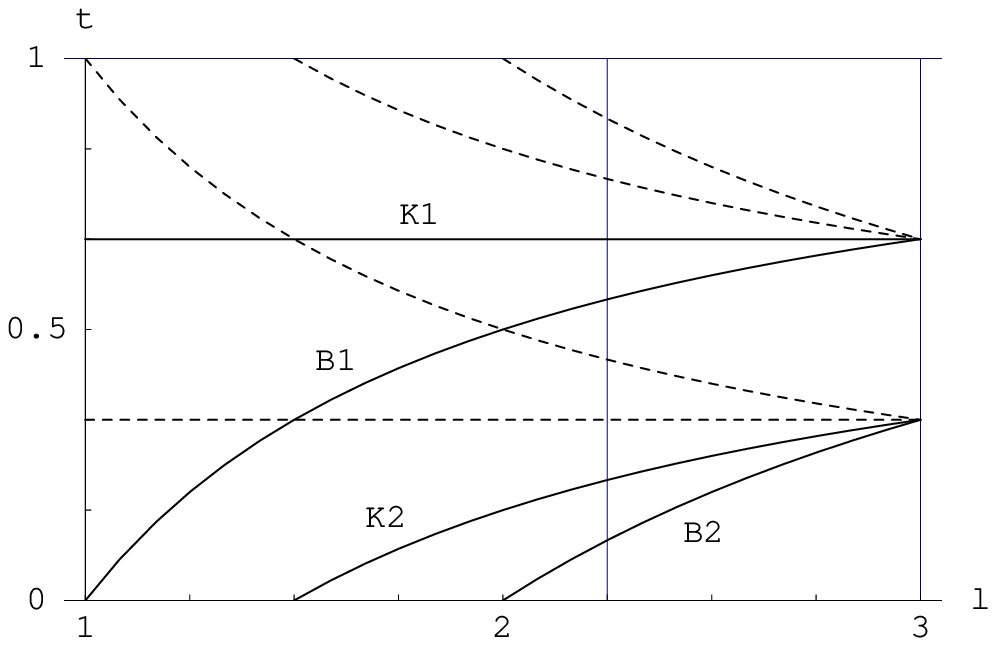}
\vspace{.2in}
{~~Bound state poles of $\CS^0_{K_1K_1}$, $\CS^1_{K_1K_1}$,
$\CS^2_{K_1K_1}$ and $\CS^3_{K_1K_1}$}\\[4pt]

Pole locations: 
$\pbl{\fra{2}{3}}\pl{K_1}
 \pbl{\fra{2}{3}{-}\fra{1}{\lambda}}\pl{K_2}
 \pbl{1{-}\fra{1}{\lambda}}\pl{B_1}
 \pbl{1{-}\fra{2}{\lambda}}\pl{B_2}
$

\newpage

\vspace{2mm}

\fb{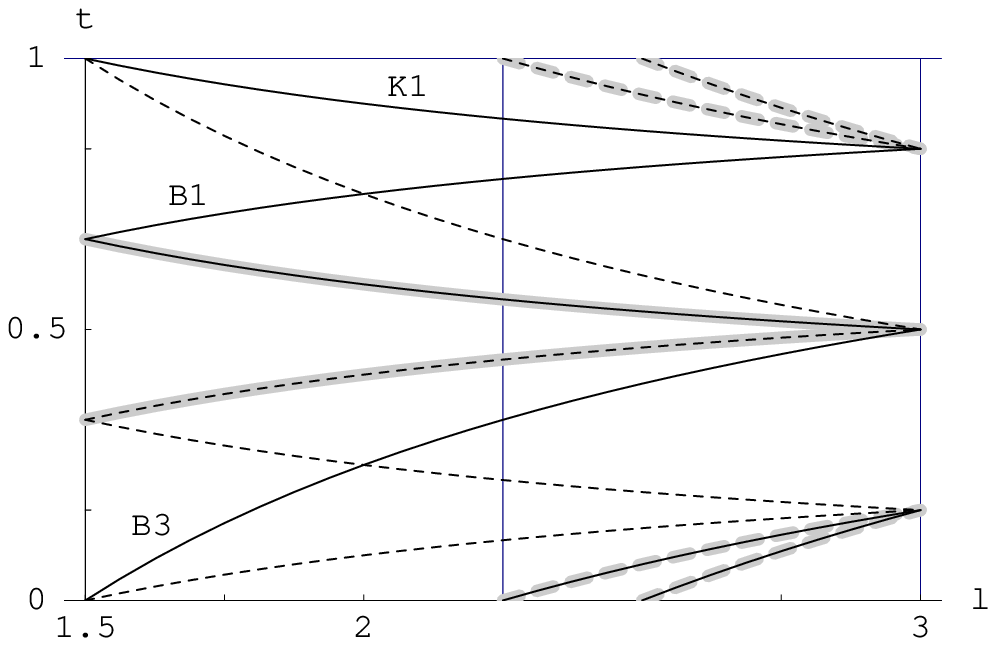}

\vspace{.2in}
{~~Bound state poles of $\CS^0_{K_2K_1}$, $\CS^1_{K_2K_1}$,
$\CS^2_{K_2K_1}$ and $\CS^3_{K_2K_1}$}\\[4pt]

Pole locations: 
$\pbl{\fra{2}{3}{+}\fra{1}{2\lambda}}\pl{K_1}
 \pbl{1{-}\fra{1}{2\lambda}}\pl{B_1}
 \pbl{1{-}\fra{3}{2\lambda}}\pl{B_3}
 \pbl{\fra{1}{3}{+}\fra{1}{2\lambda}}
 \pbl{\fra{2}{3}{-}\fra{3}{2\lambda}}\pl{?}
 \pbl{1{-}\fra{5}{2\lambda}}\pl{?}
$

\bigskip
\vspace{4mm}

\fb{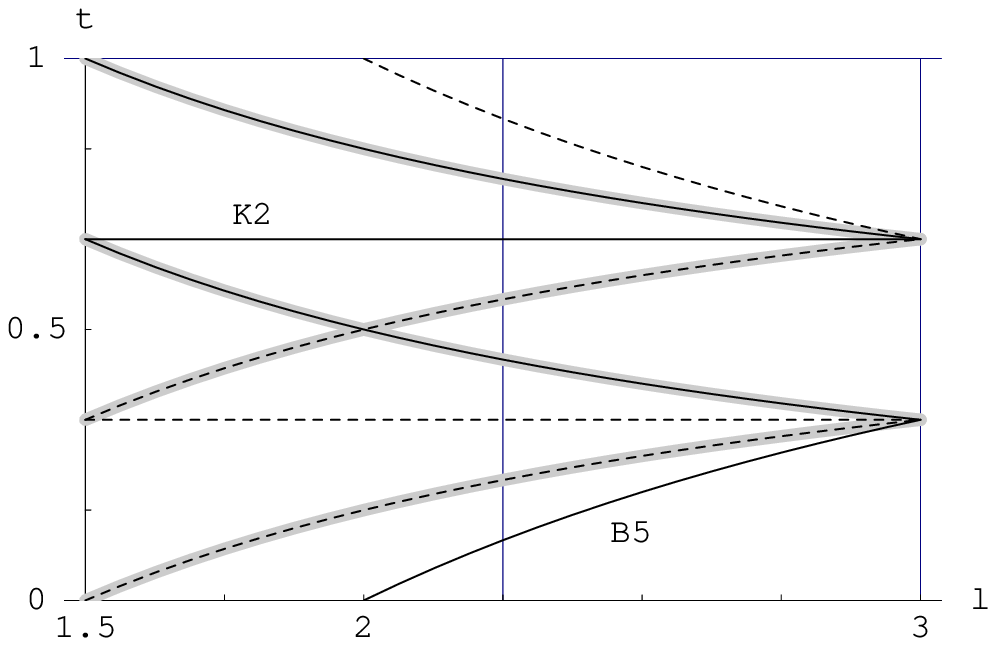}
\vspace{.2in}
{~~Bound state poles of $\CS^0_{K_2K_2}$, $\CS^1_{K_2K_2}$,
$\CS^2_{K_2K_2}$ and $\CS^3_{K_2K_2}$}\\[4pt]

Pole locations: 
$\pbl{\fra{2}{3}}\pl{K_2}
 \pbl{1{-}\fra{2}{\lambda}}\pl{B_5}
 \pbl{\fra{2}{3}{-}\fra{1}{\lambda}}
 \pbl{1{-}\fra{1}{\lambda}}
$

\newpage

\subsection{Kink-breather and breather-breather S-matrix elements,
$1<\lambda\le\frac{3}{2}$}

\fa{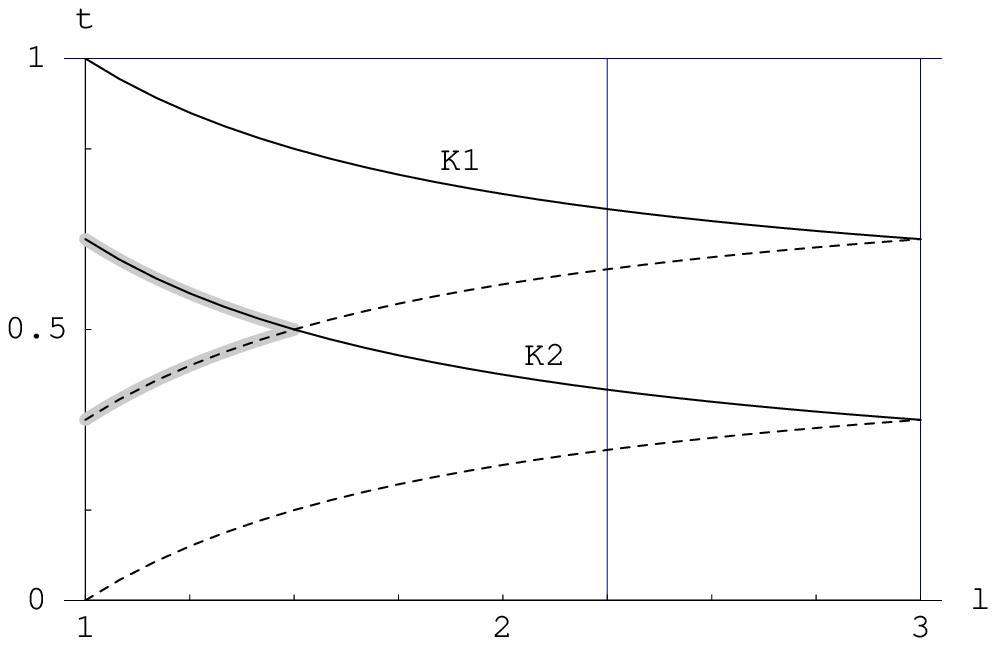}
\[
{\cal S}_{B_1K_1} = [\fra{1}{2}+\fra{1}{2\lambda}]\pl{K_1}
[\fra{1}{6}+\fra{1}{2\lambda}]\pl{K_2}
\]

\vspace{.2in}

\vspace{.1in}

\blnkt{
\begin{tabular}{|l|c|c|} \hline
& & \\
Pole: & $[\frac{1}{2}+\frac{1}{2\lambda}]$ &
$[\frac{1}{6}+\frac{1}{2\lambda}]$ \\
& & \\ \hline
& & \\
Bound state: & $K_1$ & $K_2\ \lambda>\frac{3}{2}$ \\
& & \\ \hline
\multicolumn{3}{c}{ } \\
\multicolumn{3}{c}{ } \\ \hline
 & & \\
l $=1$ & $1$ & $\frac{2}{3}$\\
& & \\
l $=3$ & $\frac{2}{3}$ & $\frac{1}{3}$\\
& & \\ \hline
\end{tabular}
}

\fa{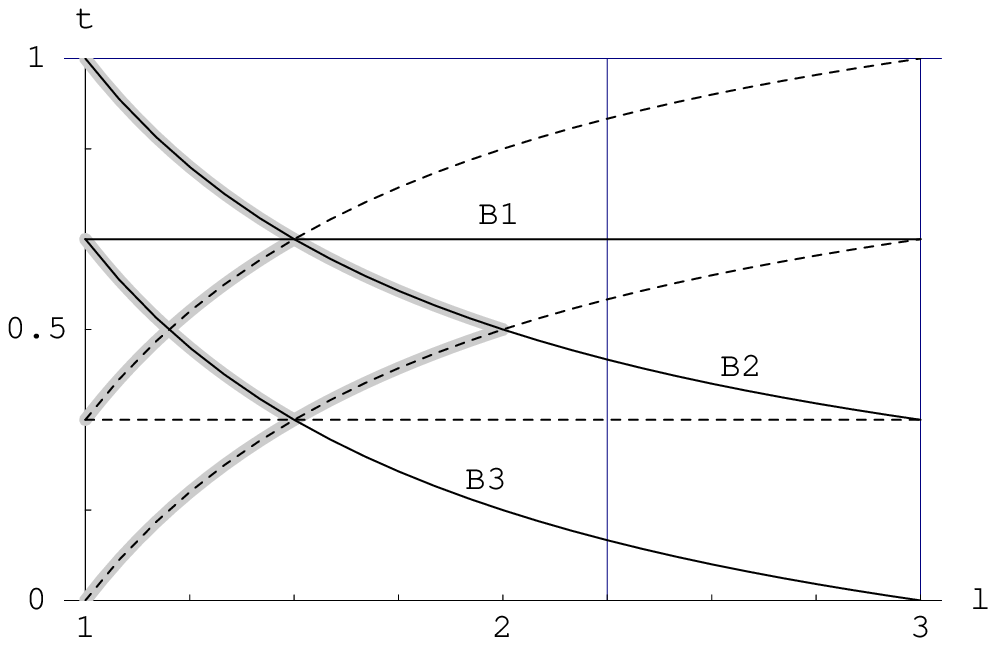}
\[
{\cal S}_{B_1B_1} = [\fra{2}{3}]\pl{B_1}[\fra{1}{\lambda}]\pl{B_2}
[\fra{1}{\lambda}-\fra{1}{3}]\pl{B_3}
\]

\vspace{.1in}

\blnkt{
\begin{tabular}{|l|c|c|c|} \hline
& & & \\  
Pole: & $[\fra{2}{3}]$ &
$[\fra{1}{\lambda}]$ & $[\fra{1}{\lambda}-\fra{1}{3}]$ \\
& & & \\ \hline
& & & \\ 
Bound state: & $B_1$ & $B_2\ \lambda>2$ & $B_3\ \lambda>\frac{3}{2}$\\
& & & \\ \hline
\multicolumn{4}{c}{ } \\
\multicolumn{4}{c}{ } \\ \hline
 & & & \\
l $=1$ & $\frac{2}{3}$ & $1$ & $\frac{2}{3}$\\
& & & \\
l $=3$ & $\frac{2}{3}$ & $\frac{1}{3}$ & $0$\\
& & & \\ \hline
\end{tabular}
}

\newpage

\subsection{Additional kink-breather and breather-breather S-matrix elements
for $\frac{3}{2}<\lambda\le{2}$}

\fb{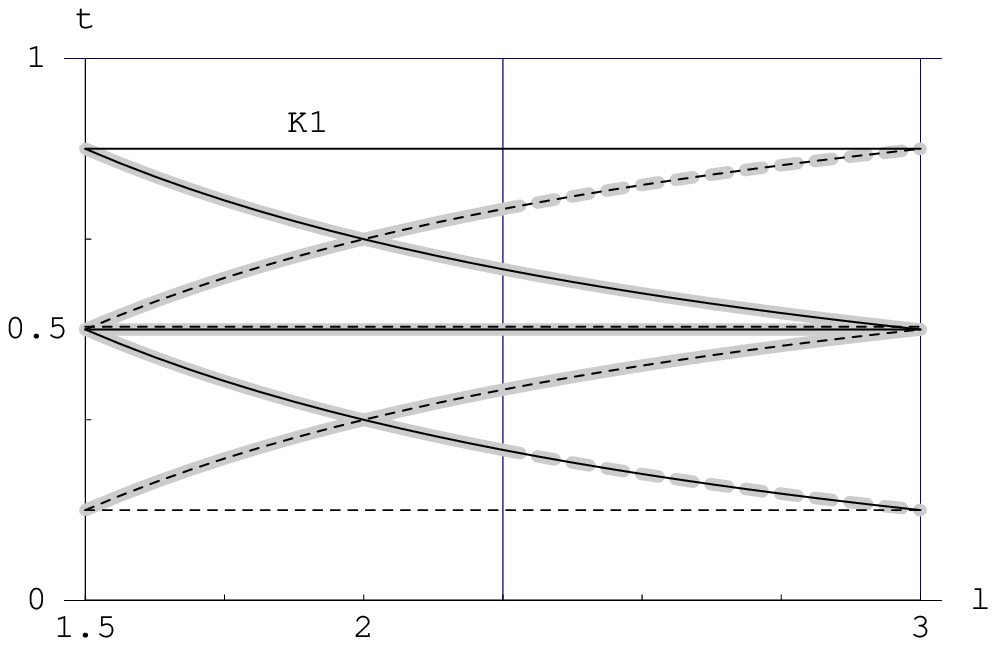}
\[
{\cal S}_{B_1K_2} =
[\fra{1}{2}][\fra{5}{6}]\pl{K_1}
[\fra{1}{6}+\fra{1}{\lambda}]
[\fra{1}{\lambda}-\fra{1}{6}]\pl{?}
\]
\vspace{.2in}

\vspace{.1in}

\blnkt{
\begin{tabular}{|l|c|c|c|c|} \hline
& & & & ? \\
Pole: & $\left[\frac{1}{2}\right]$ & $\left[\frac{5}{6}\right]$ &
$\left[\frac{1}{6}+\frac{1}{\lambda}\right]$ &
$\left[\frac{1}{\lambda}-\frac{1}{6}\right]$ \\
& & & & \\ \hline
& & & & \\
Bound state: & & $K_1$ & $K_2\ \lambda=\frac{9}{4}$ &
$K_3\ \lambda=\frac{9}{4}$ \\
& & & & \\ \hline
\multicolumn{5}{c}{ } \\
\multicolumn{5}{c}{ } \\ \hline
 & & & & \\
l $=\frac{3}{2}$ & $\frac{1}{2}$ & $\frac{5}{6}$ & $\frac{5}{6}$ &
$\frac{1}{2}$\\
& & & & \\
l $=3$ & $\frac{1}{2}$ & $\frac{5}{6}$ & $\frac{1}{2}$ & $\frac{1}{6}$\\
& & & & \\ \hline
\end{tabular}
}

\fb{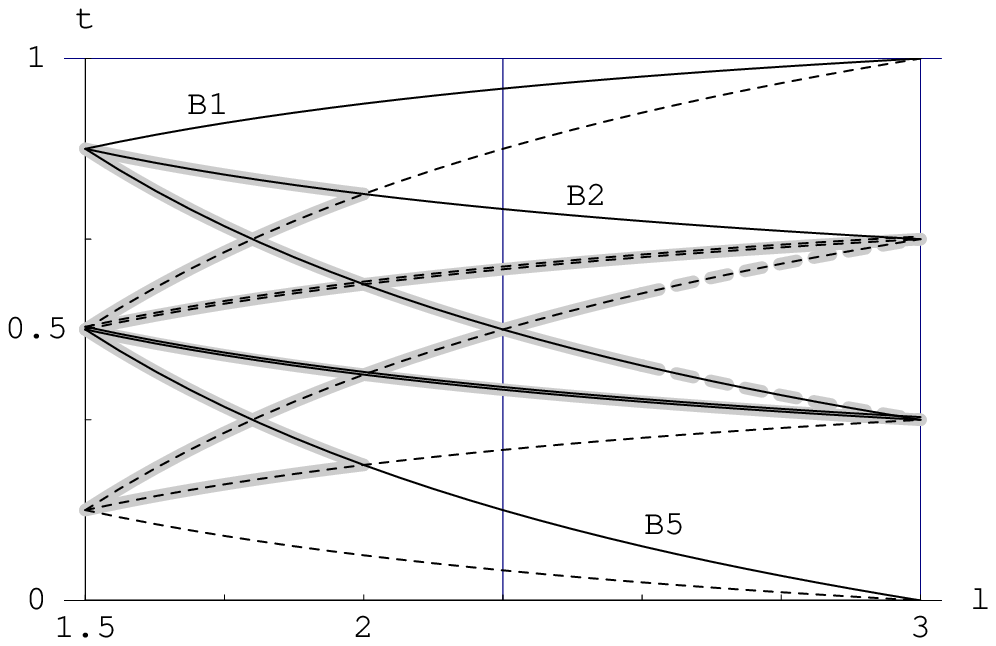}
\[
{\cal S}_{B_1B_3} = [\fra{1}{6}+\fra{1}{2\lambda}]^2
[\fra{1}{2}+\fra{1}{2\lambda}]\pl{B_2}
[\fra{7}{6}-\fra{1}{2\lambda}]\pl{B_1}
[\fra{3}{2\lambda}-\fra{1}{2}]\pl{B_5}
[\fra{3}{2\lambda}-\fra{1}{6}]\pl{?}
\]

\vspace{.1in}

\blnkt{
\begin{tabular}{|l|c|c|c|c|c|} \hline
& & & & ? & \\
Pole: & $\left[\frac{7}{6}-\frac{1}{2\lambda}\right]$ &
$\left[\frac{1}{2}+\frac{1}{2\lambda}\right]$ &
$\left[\frac{3}{2\lambda}-\frac{1}{2}\right]$ &
$\left[\frac{3}{2\lambda}-\frac{1}{6}\right]$ &
$\left[\frac{1}{6}+\frac{1}{2\lambda}\right]^2$ \\
& & & & & \\ \hline
& & & &\multicolumn{2}{c|}{} \\
& & & &\multicolumn{2}{c|}{$B_3\ \lambda=2$} \\
Bound & $B_1$ & & &\multicolumn{2}{c|}{} \\ \cline{3-6}
state: & & & & & \\
&  & $B_2\ \lambda>2$ &
$B_5\ \lambda>2$ & $B_4\ \lambda=\frac{5}{2}$ & \\
& & & & & \\ \hline
\multicolumn{6}{c}{ } \\
\multicolumn{6}{c}{ } \\ \hline
 & & & & & \\
l $=\frac{3}{2}$ & $\frac{5}{6}$ & $\frac{5}{6}$ & $\frac{1}{2}$ &
$\frac{5}{6}$ & $\frac{1}{2}$\\
 & & & & & \\
l $=3$ & $1$ & $\frac{2}{3}$ & $0$ & $\frac{1}{3}$ & $\frac{1}{3}$\\
 & & & & & \\ \hline
\end{tabular}
}

\newpage

\fb{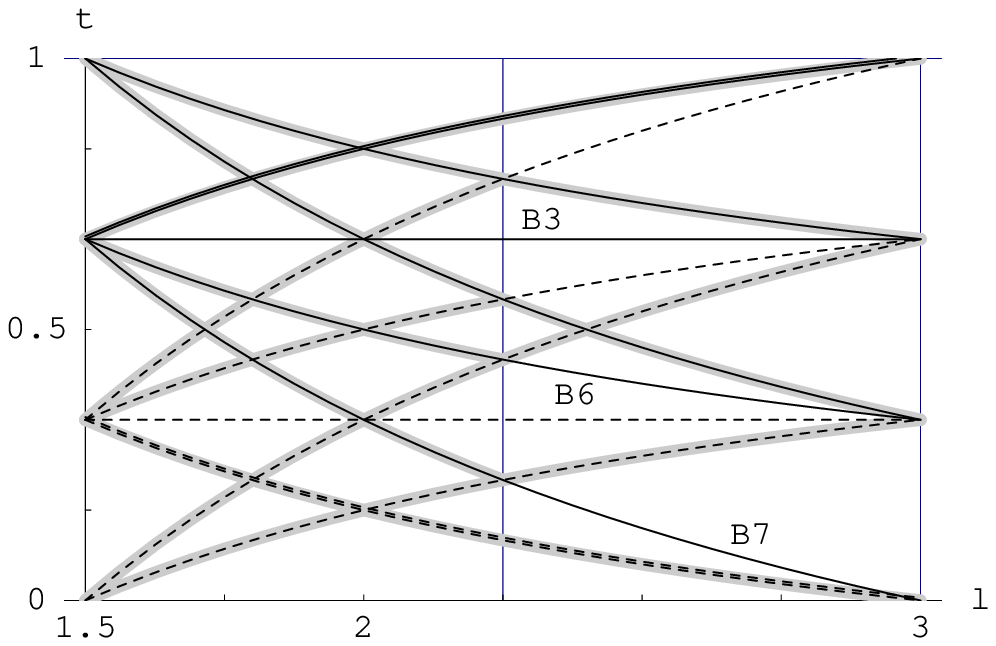}
\[
{\cal S}_{B_3B_3} = 
[\fra{2}{3}]^3\pll{B_3}
[\fra{1}{\lambda}]^3\pll{B_6}
[\fra{4}{3}-\fra{1}{\lambda}]^2
[\fra{1}{3}+\fra{1}{\lambda}]
[\fra{2}{\lambda}-\fra{2}{3}]\pl{B_7}
[\fra{2}{\lambda}-\fra{1}{3}]
\]

\vspace{.1in}

\blnkt{
\begin{tabular}{|l|c|c|c|c|c|c|} \hline
& & & & & & \\
Pole: & $\left[\frac{4}{3}-\frac{1}{\lambda}\right]^2$ &
$\left[\frac{1}{3}+\frac{1}{\lambda}\right]$ &
$\left[\frac{2}{3}\right]^3$ &
$\left[\frac{2}{\lambda}-\frac{1}{3}\right]$ &
$\left[\frac{2}{\lambda}-\frac{2}{3}\right]$ &
$\left[\frac{1}{\lambda}\right]^3$ \\
& & & & & & \\ \hline
& & & & & & \\
& & & $B_3\ \lambda<2$ & & & \\
& & & & & & \\ \cline{2-6}
Bound &\multicolumn{2}{c|}{} &\multicolumn{3}{c|}{} & \\
state: &\multicolumn{2}{c|}{ $B_1\ \lambda=2$} &
\multicolumn{3}{c|}{$B_3\ \lambda=2$} & \\
&\multicolumn{2}{c|}{} &\multicolumn{3}{c|}{} & \\ \cline{2-7}
& & & & & & \\
& &$B_2\ \lambda=\fra{5}{2}$ & $B_3\ \lambda>2$ &
$B_5\ \lambda=\fra{5}{2}$ & $B_7\ \lambda>\frac{9}{4}$ &
$B_6\ \lambda>\frac{9}{4}$ \\
& & & & & & \\ \hline
\multicolumn{7}{c}{ } \\
\multicolumn{7}{c}{ } \\ \hline
 & & & & & & \\
l $=\frac{3}{2}$ & $\frac{2}{3}$ & $1$ & $\frac{2}{3}$ & $1$ &
$\frac{2}{3}$ & $\frac{2}{3}$\\
 & & & & & & \\
l $=3$ & $1$ & $\frac{2}{3}$ & $\frac{2}{3}$ & $\frac{1}{3}$ & $0$ &
$\frac{1}{3}$\\
 & & & & & & \\ \hline
\end{tabular}
}

\fb{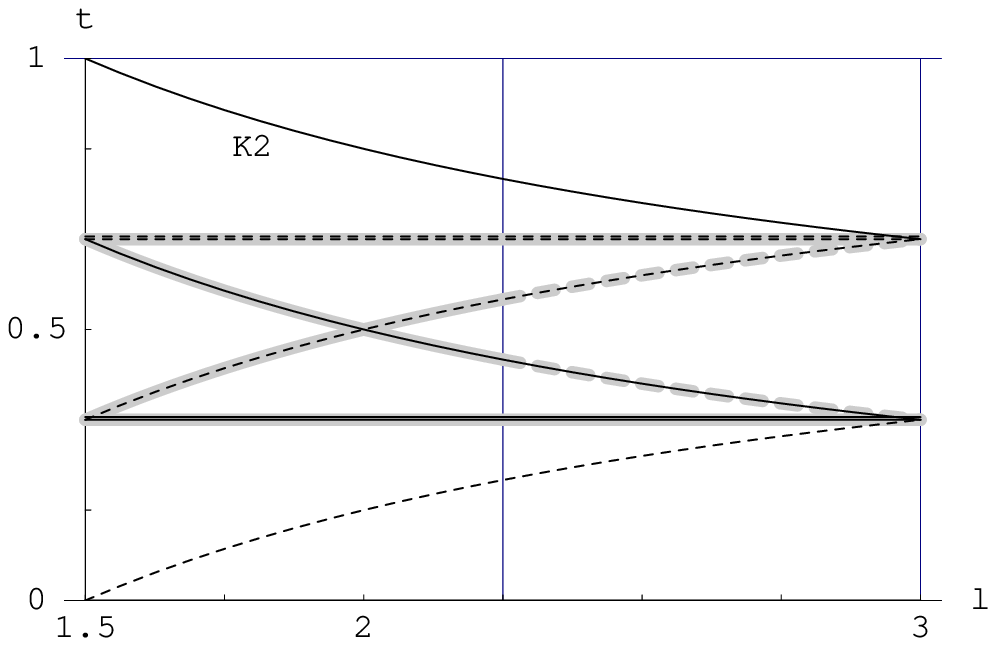}
\[
{\cal S}_{B_3K_1} = 
[\fra{1}{3}]^2
[\fra{1}{\lambda}]\pl{?}%
[\fra{1}{\lambda}+\fra{1}{3}]\pl{K_2}
\]
\blnkt{
\begin{tabular}{|l|c|c|c|} \hline
& & & ? \\
Pole: & $\left[\frac{1}{3}\right]^2$ &
$\left[\frac{1}{3}+\frac{1}{\lambda}\right]$ &
$\left[\frac{1}{\lambda}\right]$\\
& & & \\ \hline
& & & \\
Bound state: & & $K_2$ & $K_3\ \lambda=\frac{9}{4}$ \\
& & & \\ \hline
\multicolumn{4}{c}{ } \\
\multicolumn{4}{c}{ } \\ \hline
 & & & \\
l $=\frac{3}{2}$ & $\frac{1}{3}$ & $1$ & $\frac{2}{3}$\\
 & & & \\
l $=3$ & $\frac{1}{3}$ & $\frac{2}{3}$ & $\frac{1}{3}$\\
 & & & \\ \hline
\end{tabular}
}

\newpage

\fb{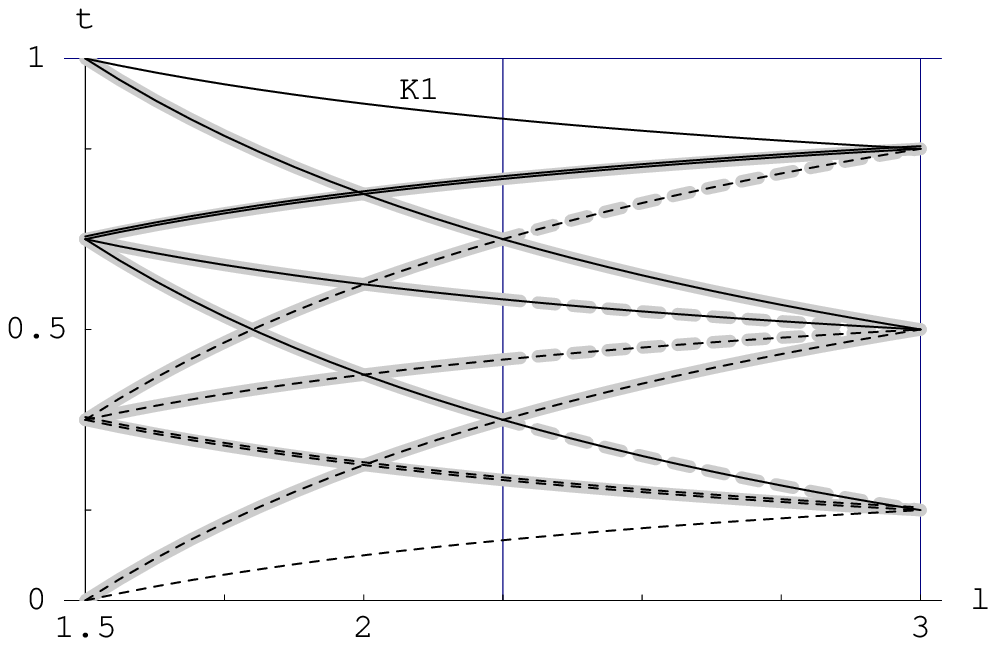}
\[
{\cal S}_{B_3K_2} = 
[1-\fra{1}{2\lambda}]^2
[\fra{1}{3}+\fra{1}{2\lambda}]^3\pll{?}
[\fra{2}{3}+\fra{1}{2\lambda}]\pl{K_1}
[\fra{3}{2\lambda}-\fra{1}{3}]\pl{?}
[\fra{3}{2\lambda}]
\]

\vspace{.1in}

\blnkt{
\begin{tabular}{|l|c|c|c|c|c|} \hline
& & & & ? & ? \\
Pole: & $\left[\frac{2}{3}+\frac{1}{2\lambda}\right]$ &
$\left[1-\frac{1}{2\lambda}\right]^2$ &
$\left[\frac{3}{2\lambda}\right]$ &
$\left[\frac{1}{3}+\frac{1}{2\lambda}\right]^3$ &
$\left[\frac{3}{2\lambda}-\frac{1}{3}\right]$ \\
& & & & & \\ \hline
& &\multicolumn{2}{c|}{ }& & \\
Bound state: & $K_1$ & \multicolumn{2}{c|}{$K_2\ \lambda=2$} &
$K_3\ \lambda=\frac{9}{4}$ & \\
& &\multicolumn{2}{c|}{ }& & \\ \hline
\multicolumn{6}{c}{ } \\
\multicolumn{6}{c}{ } \\ \hline
 & & & & & \\
l $=\frac{3}{2}$ & $1$ & $\frac{2}{3}$ & $1$ & $\frac{2}{3}$ &
$\frac{2}{3}$\\
 & & & & & \\
l $=3$ & $\frac{5}{6}$ & $\frac{5}{6}$ & $\frac{1}{2}$ & $\frac{1}{2}$ &
$\frac{1}{6}$\\
 & & & & & \\ \hline
\end{tabular}
}

\subsection{Additional kink-breather and breather-breather S-matrix elements
for ${2}<\lambda\le\frac{9}{4}$}

\fc{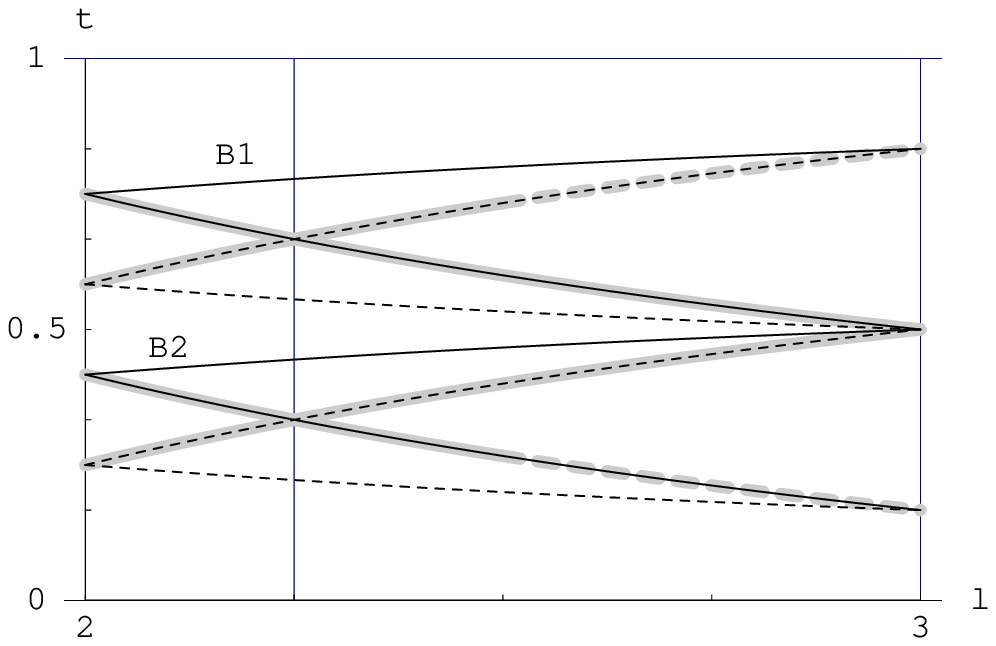}
\[
{\cal S}_{B_1B_2} =  
[1-\fra{1}{2\lambda}]\pl{B_1}
[\fra{2}{3}-\fra{1}{2\lambda}]\pl{B_3}
[\fra{3}{2\lambda}]
[\fra{3}{2\lambda}-\fra{1}{3}]\pl{?}
\]
\blnkt{
\begin{tabular}{|l|c|c|c|c|} \hline
& & & & ? \\
Pole: & $[1-\frac{1}{2\lambda}]$ & $[\frac{2}{3}-\frac{1}{2\lambda}]$ &
$[\frac{3}{2\lambda}]$ & $[\frac{3}{2\lambda}-\frac{1}{3}]$ \\
& & & & \\ \hline
& & & & \\
Bound state: & $B_1$ & $B_3$ & $B_2\ \lambda=\frac{5}{2}$ &
$B_4\ \lambda=\frac{5}{2}$ \\
& & & & \\ \hline
\multicolumn{5}{c}{ } \\
\multicolumn{5}{c}{ } \\ \hline
 & & & & \\
l $=2$ & $\frac{3}{4}$ & $\frac{5}{12}$ & $\frac{3}{4}$ & $\frac{5}{12}$\\
& & & & \\
l $=3$ & $\frac{5}{6}$ & $\frac{1}{2}$ & $\frac{1}{2}$ & $\frac{1}{6}$\\
& & & & \\ \hline
\end{tabular}
\vspace{8mm}
}

\newpage

\fc{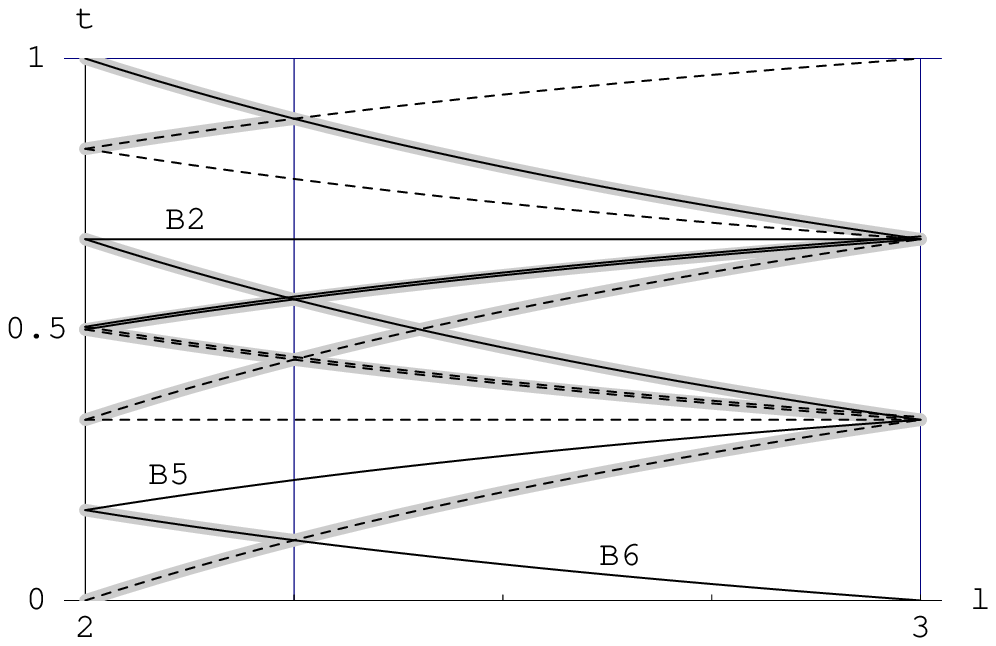}
\[
{\cal S}_{B_2B_2} = 
[\fra{2}{3}]\pl{B_2}
[\fra{2}{3}-\fra{1}{\lambda}]\pl{B_5}
[\fra{1}{\lambda}-\fra{1}{3}]\pl{B_6}
[\fra{2}{\lambda}]
[\fra{2}{\lambda}-\fra{1}{3}]
[1-\fra{1}{\lambda}]^2
\]
\blnkt{
\begin{tabular}{|l|c|c|c|c|c|c|} \hline
& & & & & & \\
Pole: & $[\frac{2}{3}]$ & $[\frac{2}{3}-\frac{1}{\lambda}]$ &
$[\frac{1}{\lambda}-\frac{1}{3}]$ & $[\frac{2}{\lambda}]$ &
$[\frac{2}{\lambda}-\frac{1}{3}]$ & $[1-\frac{1}{\lambda}]^2$ \\
& & & & & & \\ \hline
& & & & &\multicolumn{2}{c|}{} \\
& & & & &\multicolumn{2}{c|}{$B_3\ \lambda=\frac{9}{4}$} \\
Bound & $B_2$ & $B_5$ & & &\multicolumn{2}{c|}{} \\ \cline{4-7}
state: & & & & & & \\
& & & $B_6\ \lambda>\frac{9}{4}$ & $B_1\ \lambda=\frac{5}{2}$
& $B_4\ \lambda=\frac{5}{2}$ & \\
& & & & & & \\ \hline
\multicolumn{7}{c}{ } \\
\multicolumn{7}{c}{ } \\ \hline
 & & & & & & \\
l $=2$ & $\frac{2}{3}$ & $\frac{1}{6}$ & $\frac{1}{6}$ & $1$ &
$\frac{2}{3}$ & $\frac{1}{2}$\\
 & & & & & & \\
l $=3$ & $\frac{2}{3}$ & $\frac{1}{3}$ & $0$ & $\frac{2}{3}$ &
$\frac{1}{3}$ & $\frac{2}{3}$\\
 & & & & & & \\ \hline
\end{tabular}
}

\vspace{.2in}

\fc{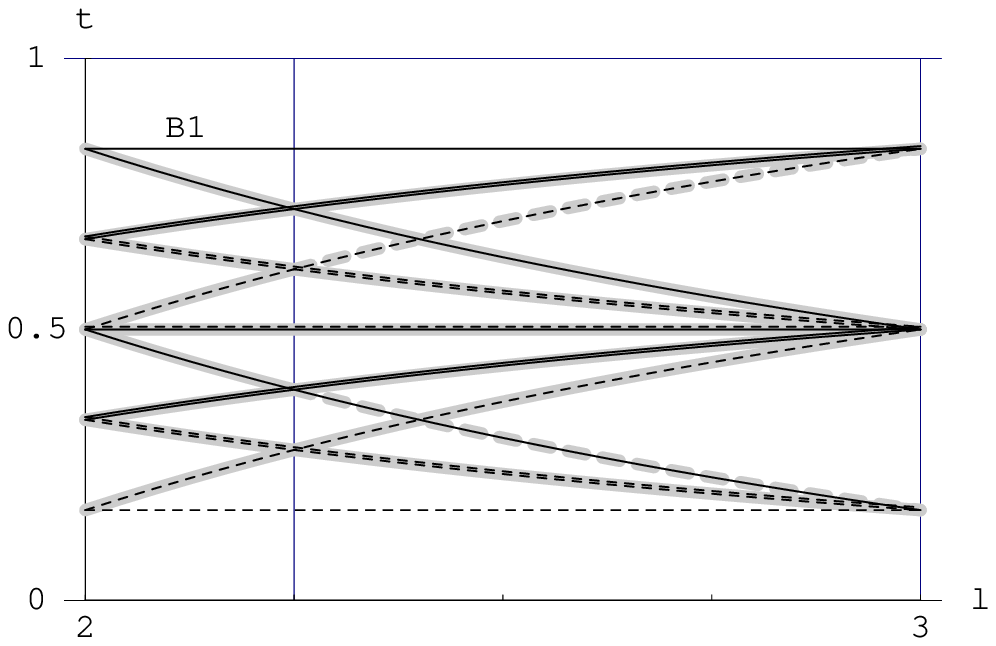}
\[
{\cal S}_{B_2B_3} = 
[\fra{1}{2}]
[\fra{5}{6}]\pl{B_1}
[\fra{2}{\lambda}-\fra{1}{6}]
[\fra{7}{6}-\fra{1}{\lambda}]^2
[\fra{2}{\lambda}-\fra{1}{2}]\pl{?}
[\fra{5}{6}-\fra{1}{\lambda}]^2
\]
\blnkt{
\begin{tabular}{|l|c|c|c|c|c|c|} \hline
& & & & & ? & \\
Pole: & $[\frac{1}{2}]$ & $[\frac{5}{6}]$ &
$[\frac{2}{\lambda}-\frac{1}{6}]$ & $[\frac{7}{6}-\frac{1}{\lambda}]^2$
& $[\frac{2}{\lambda}-\frac{1}{2}]$ & $[\fra{5}{6}-\fra{1}{\lambda}]^2$ \\
& & & & & & \\ \hline
& & &\multicolumn{2}{c|}{} &\multicolumn{2}{c|}{} \\
& & &\multicolumn{2}{c|}{$B_2\ \lambda=\frac{9}{4}$} &
\multicolumn{2}{c|}{$B_5\ \lambda=\frac{9}{4}$} \\
Bound state: & & $B_1$ &\multicolumn{2}{c|}{} &
\multicolumn{2}{c|}{} \\ \cline{4-7}
& & & & & & \\
& & & $B_3\ \lambda=\frac{5}{2}$ & &
$B_6\ \lambda=\frac{5}{2}$ & \\
& & & & & & \\ \hline
\multicolumn{7}{c}{ } \\
\multicolumn{7}{c}{ } \\ \hline
 & & & & & & \\
l $=2$ & $\frac{1}{2}$ & $\frac{5}{6}$ & $\frac{5}{6}$ & $\frac{2}{3}$ &
$\frac{1}{2}$ & $\frac{1}{3}$\\
 & & & & & & \\
l $=3$ & $\frac{1}{2}$ & $\frac{5}{6}$ & $\frac{1}{2}$ & $\frac{5}{6}$ &
$\frac{1}{6}$ & $\frac{1}{2}$\\
 & & & & & & \\ \hline
\end{tabular}
}

\newpage

\fc{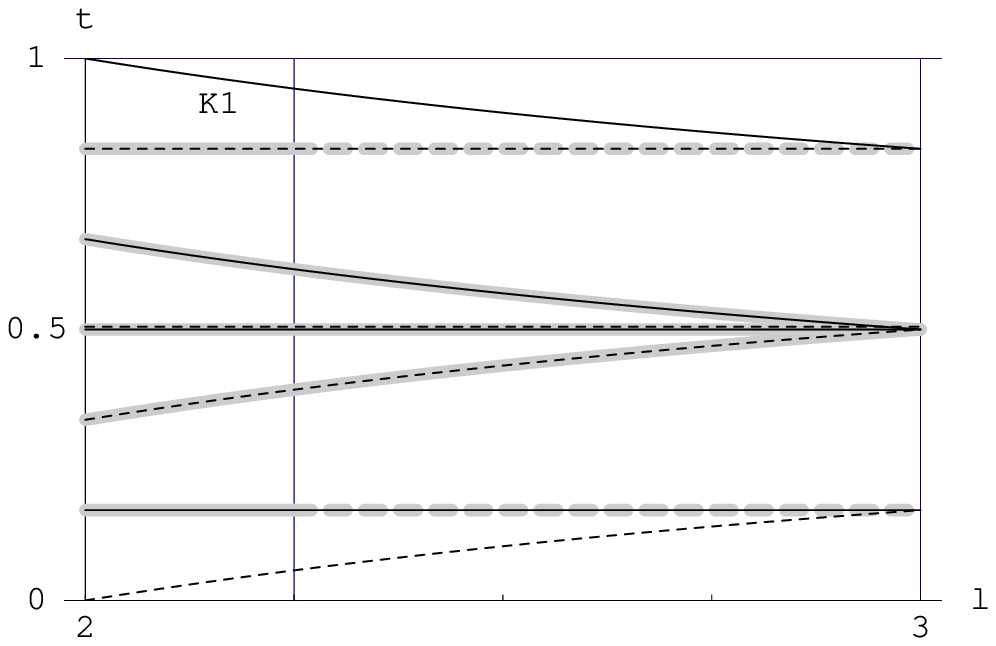}
\[
{\cal S}_{B_2K_1} = 
[\fra{1}{2}]
[\fra{1}{6}]\pl{?}
[\fra{1}{2}+\fra{1}{\lambda}]\pl{K_1}
[\fra{1}{6}+\fra{1}{\lambda}]
\]
\blnkt{
\begin{tabular}{|l|c|c|c|c|} \hline
& & ? & & \\
Pole: & $[\frac{1}{2}]$ & $[\frac{1}{6}]$ &
$[\frac{1}{2}+\frac{1}{\lambda}]$ & $[\frac{1}{6}+\frac{1}{\lambda}]$ \\
& & & & \\ \hline
& & & & \\
Bound state: & & $K_3\ \lambda=\frac{9}{4}$ & $K_1$ &
$K_2\ \lambda=\frac{9}{4}$ \\
& & & & \\ \hline
\multicolumn{5}{c}{ } \\
\multicolumn{5}{c}{ } \\ \hline
 & & & & \\
l $=2$ & $\frac{1}{2}$ & $\frac{1}{6}$ & $1$ & $\frac{2}{3}$\\
& & & & \\
l $=3$ & $\frac{1}{2}$ & $\frac{1}{6}$ & $\frac{5}{6}$ & $\frac{1}{2}$\\
& & & & \\ \hline
\end{tabular}
}

\vspace{.2in}

\fc{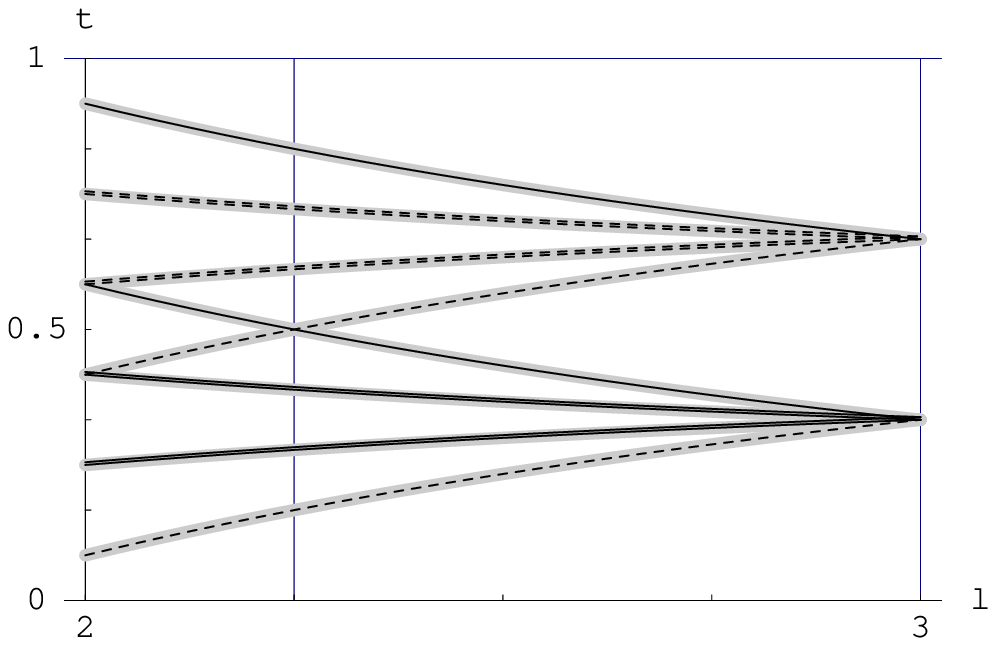}
\[
{\cal S}_{B_2K_2} = 
[\fra{1}{2}-\fra{1}{2\lambda}]^2
[\fra{1}{6}+\fra{1}{2\lambda}]^2
[\fra{1}{6}+\fra{3}{2\lambda}]
[\fra{3}{2\lambda}-\fra{1}{6}]
\]
\blnkt{
\begin{tabular}{|l|c|c|c|c|} \hline
& & & & \\
Pole: & $[\frac{1}{2}-\frac{1}{2\lambda}]^2$ &
$[\frac{1}{6}+\frac{1}{2\lambda}]^2$ &
$[\frac{1}{6}+\frac{3}{2\lambda}]$ & $[\frac{3}{2\lambda}-\frac{1}{6}]$ \\
& & & & \\ \hline
& & & & \\
Bound state: & & & $K_1\ \lambda=\frac{9}{4}$ & \\
& & & & \\ \hline
\multicolumn{5}{c}{ } \\
\multicolumn{5}{c}{ } \\ \hline
 & & & & \\
l $=2$ & $\frac{1}{4}$ & $\frac{5}{12}$ & $\frac{11}{12}$ &
$\frac{7}{12}$\\
& & & & \\
l $=3$ & $\frac{1}{3}$ & $\frac{1}{3}$ & $\frac{2}{3}$ & $\frac{1}{3}$\\
& & & & \\ \hline
\end{tabular}
}

\newpage

\fc{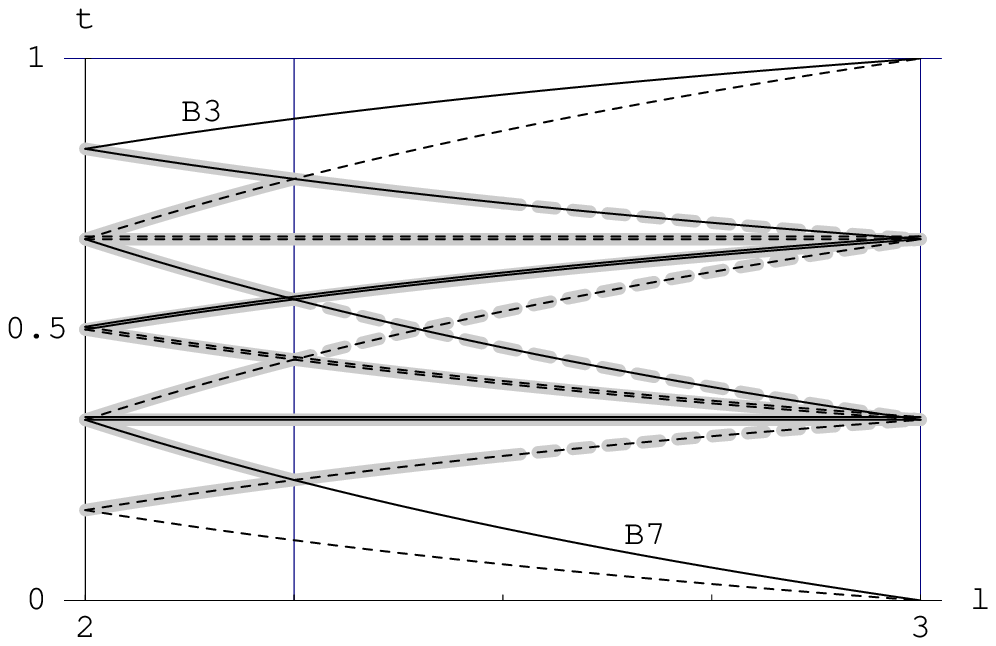}
\[
{\cal S}_{B_1B_5} = 
[\fra{1}{3}]^2
[\fra{4}{3}-\fra{1}{\lambda}]\pl{B_3}
[\fra{1}{3}+\fra{1}{\lambda}]\pl{?}
[\fra{2}{\lambda}-\fra{2}{3}]\pl{B_7}
[\fra{2}{\lambda}-\fra{1}{3}]\pl{?}
[1-\fra{1}{\lambda}]^2
\]
\blnkt{
\begin{tabular}{|l|c|c|c|c|c|c|} \hline
& & & ? & & ? & \\
Pole: & $[\frac{1}{3}]^2$ & $[\frac{4}{3}-\frac{1}{\lambda}]$ &
$[\frac{1}{3}+\frac{1}{\lambda}]$ & $[\frac{2}{\lambda}-\frac{2}{3}]$ &
$[\frac{2}{\lambda}-\frac{1}{3}]$ & $[1-\fra{1}{\lambda}]^2$ \\
& & & & & & \\ \hline
& & & & & \multicolumn{2}{c|}{} \\
& & & & &\multicolumn{2}{c|}{$B_5\ \lambda=\frac{9}{4}$} \\
Bound & & $B_3$ & & & \multicolumn{2}{c|}{} \\ \cline{4-7}
state: & & & & & & \\
& & & $B_4\ \lambda=\frac{5}{2}$ & $B_7\ \lambda>\frac{9}{4}$ &
$B_6\ \lambda=\frac{5}{2}$ & \\
& & & & & & \\ \hline
\multicolumn{7}{c}{ } \\
\multicolumn{7}{c}{ } \\ \hline
 & & & & & & \\
l $=2$ & $\frac{1}{3}$ & $\frac{5}{6}$ & $\frac{5}{6}$ & $\frac{1}{3}$ &
$\frac{2}{3}$ & $\frac{1}{2}$\\
& & & & & & \\
l $=3$ & $\frac{1}{3}$ & $1$ & $\frac{2}{3}$ & $0$ & $\frac{1}{3}$ &
$\frac{2}{3}$\\
& & & & & & \\ \hline
\end{tabular}
}

\vspace{.2in}
\fc{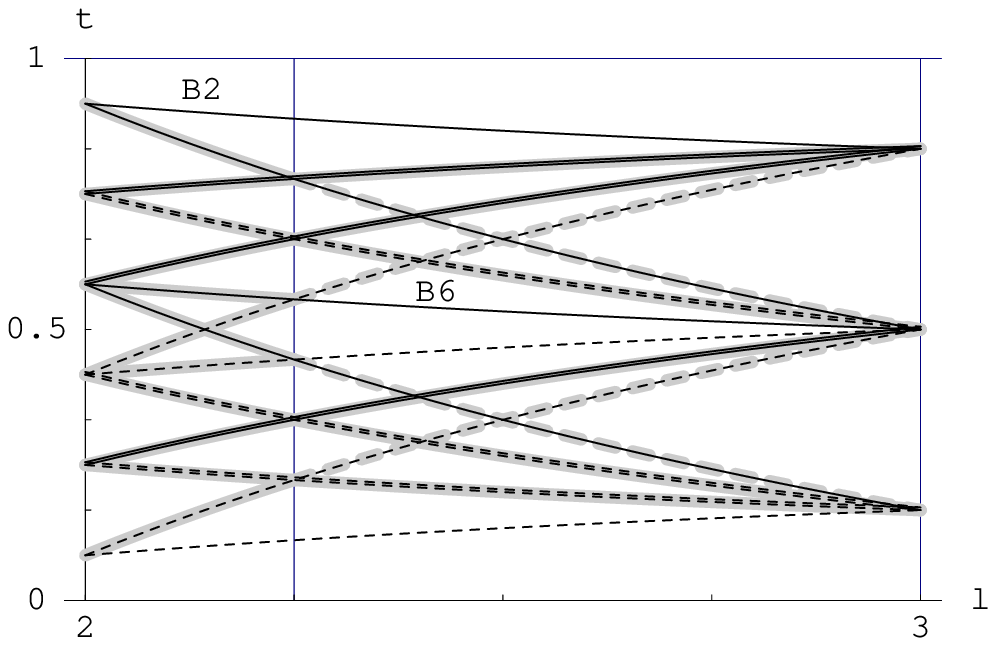}
\[
{\cal S}_{B_2B_5} = 
[\fra{4}{3}-\fra{3}{2\lambda}]^2
[1-\fra{3}{2\lambda}]^2
[\fra{2}{3}+\fra{1}{2\lambda}]\pl{B_2}
[\fra{1}{3}+\fra{1}{2\lambda}]^3\pll{B_6}
[1-\fra{1}{2\lambda}]^2
[\fra{5}{2\lambda}-\fra{1}{3}]\pl{?}
[\fra{5}{2\lambda}-\fra{2}{3}]\pl{?}
\]
\blnkt{
\begin{tabular}{|l|c|c|c|c|c|c|c|} \hline
& & & & & & ? & ? \\
Pole: & $[\fra{4}{3}-\fra{3}{2\lambda}]^2$ &
$[1-\fra{3}{2\lambda}]^2$ & $[\fra{2}{3}+\fra{1}{2\lambda}]$ &
$[\fra{1}{3}+\fra{1}{2\lambda}]^3$  & $[1-\fra{1}{2\lambda}]^2$ &
$[\fra{5}{2\lambda}-\fra{1}{3}]$ & $[\fra{5}{2\lambda}-\fra{2}{3}]$ \\
 & & & & & & & \\ \hline
 & & & & &\multicolumn{2}{c|}{} & \\
Bound state: & & & $B_2$ & $B_6\ \lambda>\fra{9}{4}$ &
\multicolumn{2}{c|}{$B_3\ \lambda=\fra{9}{4}$} & \\
 & & & & &\multicolumn{2}{c|}{} & \\ \hline
\multicolumn{8}{c}{ } \\
\multicolumn{8}{c}{ } \\ \hline
 & & & & & & & \\
l $=2$ & $\frac{7}{12}$ & $\frac{1}{4}$ & $\frac{11}{12}$ & $\frac{7}{12}$
& $\frac{3}{4}$ & $\frac{11}{12}$ & $\frac{7}{12}$\\
& & & & & & & \\
l $=3$ & $\frac{5}{6}$ & $\frac{1}{2}$ & $\frac{5}{6}$ & $\frac{1}{2}$ &
$\frac{5}{6}$ & $\frac{1}{2}$ & $\frac{1}{6}$\\
& & & & & & & \\ \hline
\end{tabular}
}

\newpage

\fc{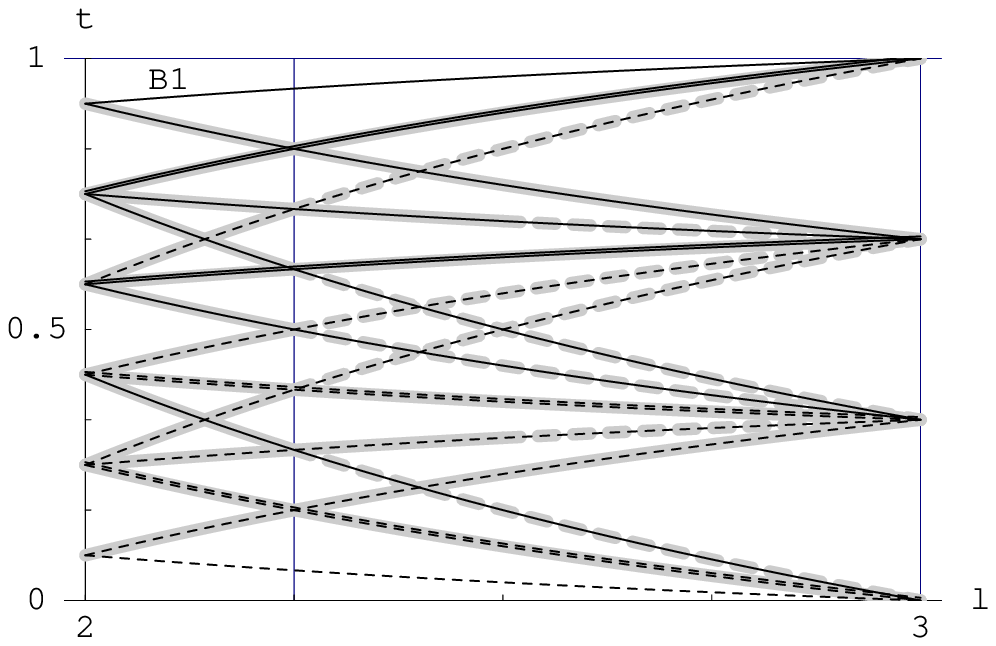}
\[
{\cal S}_{B_3B_5} = 
[\fra{7}{6}-\fra{1}{2\lambda}]\pl{B_1}
[\fra{1}{6}+\fra{3}{2\lambda}]
[\fra{3}{2}-\fra{3}{2\lambda}]^2
[\fra{5}{2\lambda}-\fra{5}{6}]\pl{?}
[\fra{1}{2}+\fra{1}{2\lambda}]^3\pll{?}
[\fra{3}{2\lambda}-\fra{1}{6}]^3\pll{?}
[\fra{5}{2\lambda}-\fra{1}{2}]\pl{?}
[\fra{5}{6}-\fra{1}{2\lambda}]^4
\]
\blnkt{
\begin{tabular}{|l|c|c|c|c|c|c|c|c|} \hline
& & & & ? & ? & ? & ? & \\
Pole: & $[\frac{7}{6}-\frac{1}{2\lambda}]$ &
$[\frac{1}{6}+\frac{3}{2\lambda}]$ & $[\frac{3}{2}-
\frac{3}{2\lambda}]^2$ & $[\frac{5}{2\lambda}-\frac{5}{6}]$ &
$[\frac{1}{2}+\frac{1}{2\lambda}]^3$ &
$[\frac{3}{2\lambda}-\frac{1}{6}]^3$ & $[\frac{5}{2\lambda}-
\frac{1}{2}]$ & $[\frac{5}{6}-\frac{1}{2\lambda}]^4$ \\
& & & &  & & & & \\ \hline
& &\multicolumn{2}{c|}{} & & & &\multicolumn{2}{c|}{} \\
& &\multicolumn{2}{c|}{$B_2\ \lambda=\frac{9}{4}$} & & &
&\multicolumn{2}{c|}{$B_5\ \lambda=\frac{9}{4}$} \\
Bound & $B_1$ &\multicolumn{2}{c|}{} & & & & \multicolumn{2}{c|}{} \\
\cline{3-9}
state: & & & & & & & & \\
& & $B_3\ \lambda=\frac{5}{2}$ & & $B_8\ \lambda=\frac{5}{2}$ &
$B_4\ \lambda=\frac{5}{2}$ & $B_7\ \lambda=\frac{5}{2}$ & & \\
& & & & & & & & \\ \hline
\multicolumn{9}{c}{ } \\
\multicolumn{9}{c}{ } \\ \hline
 & & & & & & & & \\
l $=2$ & $\frac{11}{12}$ & $\frac{11}{12}$ & $\frac{3}{4}$ &
$\frac{5}{12}$ & $\frac{3}{4}$ & $\frac{7}{12}$ & $\frac{3}{4}$ &
$\frac{7}{12}$\\
& & & & & & & & \\
l $=3$ & $1$ & $\frac{2}{3}$ & $1$ & $0$ & $\frac{2}{3}$ & $\frac{1}{3}$ &
$\frac{1}{3}$ & $\frac{2}{3}$\\
& & & & & & & & \\ \hline
\end{tabular}
}

\vspace{.2in}

\fc{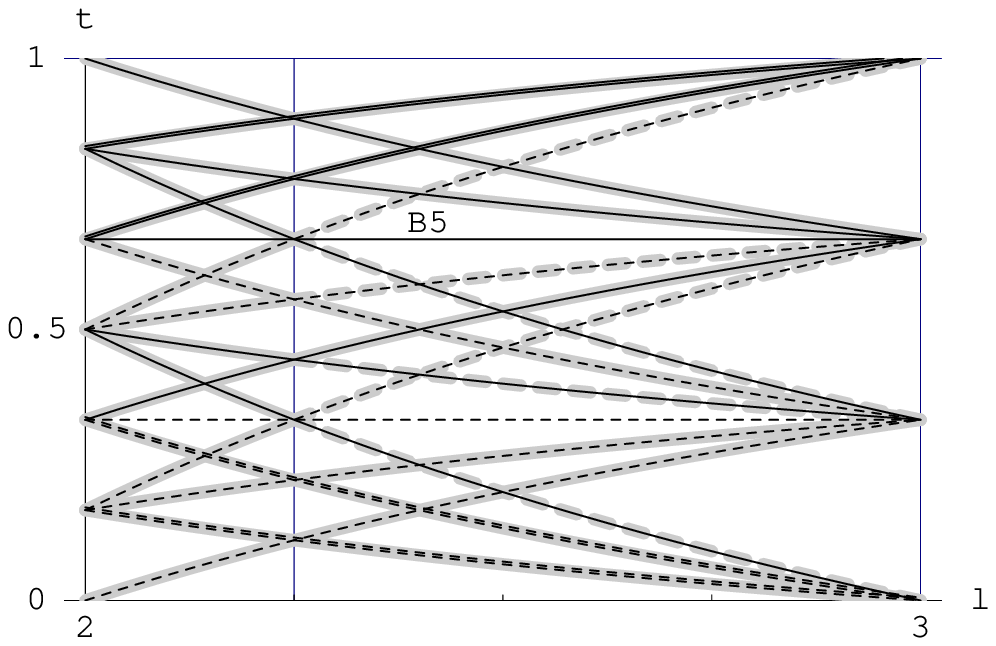}
\[
{\cal S}_{B_5B_5} = 
[\fra{5}{3}-\fra{2}{\lambda}]^2
[\fra{2}{3}]^5\pll{B_5}
[\fra{3}{\lambda}-1]\pl{?}
[\fra{3}{\lambda}-\fra{2}{3}]\pl{?}
[\fra{1}{\lambda}]^5\pll{?}
[\fra{1}{3}+\fra{1}{\lambda}]^3
[\fra{4}{3}-\fra{2}{\lambda}]^3
[\fra{2}{\lambda}]
[\fra{4}{3}-\fra{1}{\lambda}]^2
\]
\blnkt{
\begin{tabular}{|l|c|c|c|c|c|c|c|c|c|} \hline
& & & ? & ? & ? & & & & \\
Pole: & $[\frac{5}{3}-\frac{2}{\lambda}]^2$ & $[\frac{2}{3}]^5$ &
$[\frac{3}{\lambda}-1]$ & $[\frac{3}{\lambda}-\frac{2}{3}]$ &
$[\frac{1}{\lambda}]^5$ & $[\frac{1}{3}+\frac{1}{\lambda}]^3$ &
$[\frac{4}{3}-\frac{2}{\lambda}]^3$ & $[\frac{2}{\lambda}]$ &
$[\frac{4}{3}-\frac{1}{\lambda}]^2$ \\
& & & & & & & & & \\ \hline
& & & & & & & & & \\
& &$B_5\ \lambda<\frac{9}{4}$ & & & & & & & \\
& & & & & & & & & \\ \cline{3-10}
Bound & &\multicolumn{3}{c|}{ }& & & &\multicolumn{2}{c|}{} \\
state: & &\multicolumn{3}{c|}{$B_5\ \lambda=\frac{9}{4}$} &
 & $B_3\ \lambda=\frac{9}{4}$ & &
\multicolumn{2}{c|}{$B_1\ \lambda=\frac{9}{4}$} \\
& &\multicolumn{3}{c|}{ }& & & &\multicolumn{2}{c|}{} \\ \cline{3-10}
& & & & & & & & & \\
& &$B_5\ \lambda>\frac{9}{4}$ & & & $B_8\ \lambda=\frac{5}{2}$ &
$B_4\ \lambda=\frac{5}{2}$ & & & \\
& & & & & & & & & \\ \hline
\multicolumn{10}{c}{ } \\
\multicolumn{10}{c}{ } \\ \hline
 & & & & & & & & & \\
l $=2$ & $\frac{2}{3}$ & $\frac{2}{3}$ & $\frac{1}{2}$ & $\frac{5}{6}$ &
$\frac{1}{2}$ & $\frac{5}{6}$ & $\frac{1}{3}$ & $1$ & $\frac{5}{6}$\\
& & & & & & & & & \\
l $=3$ & $1$ & $\frac{2}{3}$ & $0$ & $\frac{1}{3}$ & $\frac{1}{3}$ &
$\frac{2}{3}$ & $\frac{2}{3}$ & $\frac{2}{3}$ & $1$ \\
& & & & & & & & & \\ \hline
\end{tabular}
}

\newpage

{~}\vspace{-25pt}

\fc{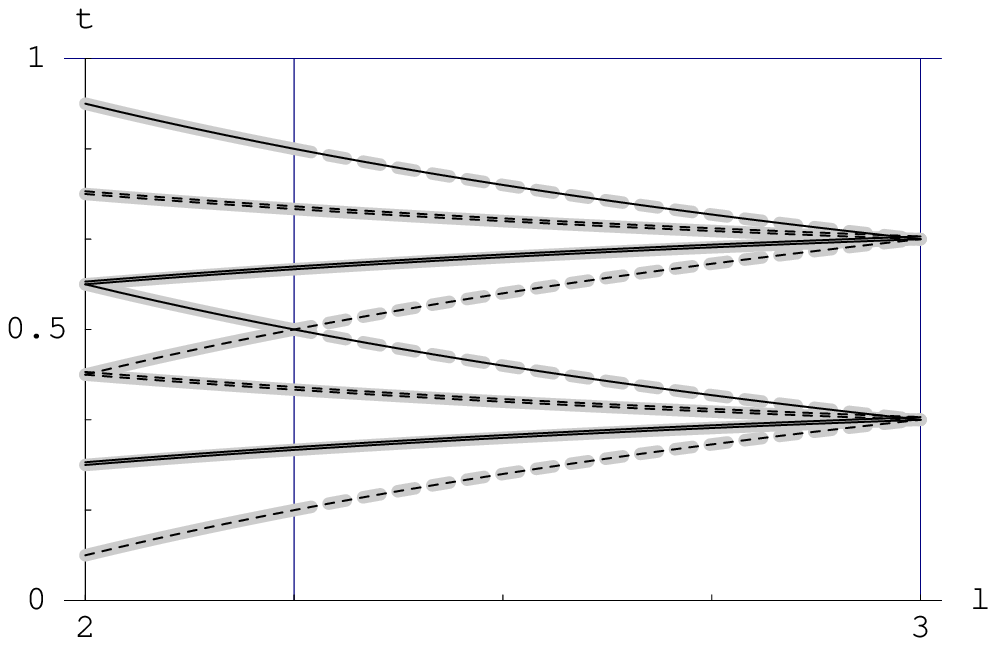}
\[
{\cal S}_{B_5K_1} = 
[\fra{1}{2}-\fra{1}{2\lambda}]^2
[\fra{5}{6}-\fra{1}{2\lambda}]^2
[\fra{1}{6}+\fra{3}{2\lambda}]\pl{?}
[\fra{3}{2\lambda}-\fra{1}{6}]\pl{?}
\]
\blnkt{
\begin{tabular}{|l|c|c|c|c|} \hline
& & & ? & ? \\
Pole: & $[\frac{1}{2}-\frac{1}{2\lambda}]^2$ &
$[\frac{5}{6}-\frac{1}{2\lambda}]^2$ &
$[\frac{1}{6}+\frac{3}{2\lambda}]$ & $[\frac{3}{2\lambda}-\frac{1}{6}]$ \\
& & & & \\ \hline
& & & & \\
Bound state: & & & $K_3\ \lambda=\frac{9}{4}$ & \\
& & & & \\ \hline
\multicolumn{5}{c}{ } \\
\multicolumn{5}{c}{ } \\ \hline
 & & & & \\
l $=2$ & $\frac{1}{4}$ & $\frac{7}{12}$ & $\frac{11}{12}$ &
$\frac{7}{12}$\\
& & & & \\
l $=3$ & $\frac{1}{3}$ & $\frac{2}{3}$ & $\frac{2}{3}$ & $\frac{1}{3}$\\
& & & & \\ \hline
\end{tabular}
}

\vspace{.1in}

\fc{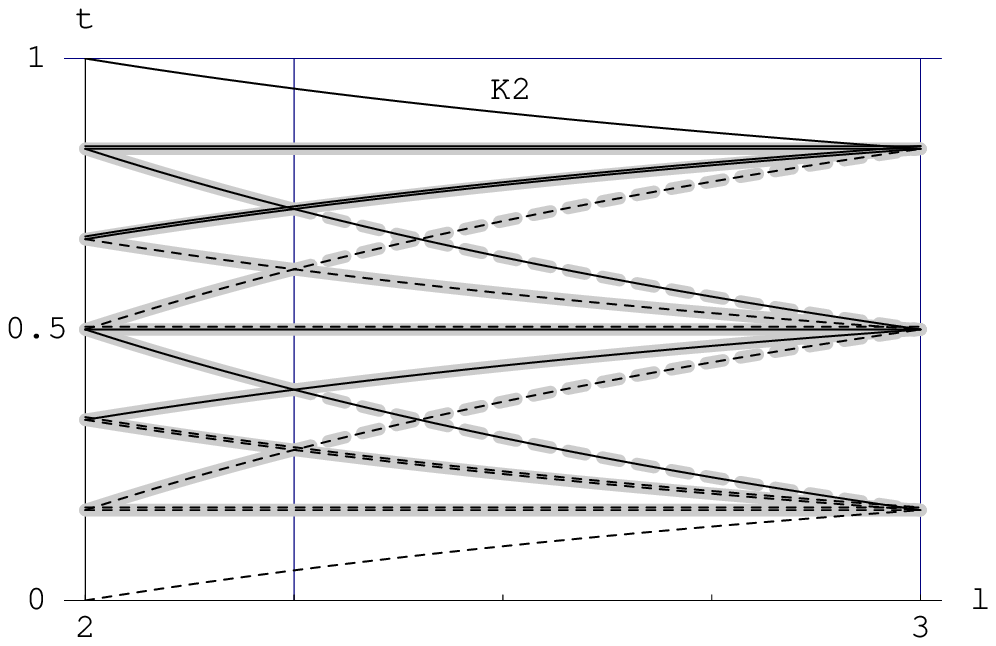}
\[
{\cal S}_{B_5K_2} = 
[\fra{5}{6}]^2
[\fra{1}{2}]^2  
[\fra{7}{6}-\fra{1}{\lambda}]^2
[\fra{5}{6}-\fra{1}{\lambda}]^3
[\fra{1}{2}+\fra{1}{\lambda}]\pl{K_2}
[\fra{2}{\lambda}-\fra{1}{6}]\pl{?}
[\fra{2}{\lambda}-\fra{1}{2}]\pl{?}
\]
\blnkt{
\begin{tabular}{|l|c|c|c|c|c|c|c|} \hline
& & & & & & ? & ? \\
Pole: & $[\frac{5}{6}]^2$ & $[\frac{1}{2}]^2$ &
$[\frac{7}{6}-\frac{1}{\lambda}]^2$ & $[\frac{5}{6}-
\frac{1}{\lambda}]^3$ & $[\frac{1}{2}+\frac{1}{\lambda}]$ &
$[\frac{2}{\lambda}-\frac{1}{6}]$ & $[\frac{2}{\lambda}-\frac{1}{2}]$ \\
& & & & & & & \\ \hline
& & & & & & & \\
Bound state: & & & & & $K_2$ & $K_3\ \lambda=\frac{9}{4}$ & \\
& & & & & & & \\ \hline
\multicolumn{8}{c}{ } \\
\multicolumn{8}{c}{ } \\ \hline
 & & & & & & & \\
l $=2$ & $\frac{5}{6}$ & $\frac{1}{2}$ & $\frac{2}{3}$ & $\frac{1}{3}$ &
$1$ & $\frac{5}{6}$ & $\frac{1}{2}$\\
& & & & & & & \\
l $=3$ & $\frac{5}{6}$ & $\frac{1}{2}$ & $\frac{5}{6}$ & $\frac{1}{2}$ &
$\frac{5}{6}$ & $\frac{1}{2}$ & $\frac{1}{6}$ \\
& & & & & & & \\ \hline
\end{tabular}
}

\end{center}
%
%

%


\begin{thebibliography}{99}
%
\raggedright
\parskip 1pt
%
\bibitem{CZ}
L.\ Chim and A.B.\ Zamolodchikov,
{`Integrable field theory of q-state Potts model with $0<q<4$',}
Int. J. Mod. Phys. A7 (1992) 5317\toline{5335}
%
\bibitem{Sm}
F.A.\ Smirnov, 
{`Exact S-matrices for $\phi_{12}$-Perturbated minimal models of conformal
field theory'},
Int. J. Mod. Phys. A6 (1991) 1407\toline{1428}
%
\bibitem{KF}
E.M.\ Fortuin and P.\ Kasteleyn,
{`On the random-cluster model.1. Introduction  and  relation  to  other  
models.',}
Physica 57, (1972) 536\toline{564}
%
%
\bibitem{FR}
P.~Fendley and N.~Read,
`Exact S-matrices for supersymmetric sigma models and the Potts
model',
{\tt hep-th/0207176}
%
\bibitem{pottsII}
P.\ Dorey,  A.\ Pocklington and R.\ Tateo, {`Integrable aspects of
the scaling q-state Potts models II: finite-size effects'},
preprint DCPT-02/53
%
\bibitem{Dot}
V.S.\ Dotsenko,
{`Critical behaviour and associated conformal algebra of the $\Zth$
Potts model',}
Nucl. Phys. B235 (1984) 54\toline{74}
%
\bibitem{Nienh}
B.\ Nienhuis, A.N.\ Berker, E.K.\ Riedel and M.\ Schick, `First-
and second-order phase transitions in Potts models:
renormalization-group solution', Phys.\ Rev.\ Lett.\  43 (1979) 737
%
\bibitem{Bu}
T.W.\ Burkhardt,
'Critical and tricritical exponents of the Potts lattice gas',
Z. Phys. B (1980) 159
%
\bibitem{Na}
B.\ Nienhuis,
{`Analytical calculation of two leading exponents of the dilute
Potts model'},
J. Phys. A15 (1982) 199\toline{213}
%
\bibitem{dFSZ}
P.\ di Francesco, H.\ Saleur and J.-B.\ Zuber,
{`Relations between the Coulomb Gas Picture and Conformal Invariance of 
Two-Dimensional Critical Models'}
J. Stat. Phys. 49 (1987), 57\toline{79} 
%
\bibitem{Zam1}
A.B.\ Zamolodchikov,
{`Integrable Field Theory from Conformal Field Theory',}
Proceedings of Taniguchi Symposium, Kyoto (1988)
%
\bibitem{CD}
G.\ Delfino and J.L.\ Cardy,
{`Universal amplitude ratios in the two-dimensional q-state
Potts model and percolation from quantum field theory',}
Nucl. Phys. B519 (1998) 551, {\tt hep-th/9712111}
%
\bibitem{DF}
V.S.\ Dotsenko and V.A.\ Fateev,
{`Conformal algebra and multipoint correlation functions in
 two-dimensional statistical models',}
Nucl. Phys. B240 (1984) 312
%
\bibitem{parke}
S.~Parke,
`Absence Of Particle Production And Factorization Of The S Matrix In
(1+1)-Dimensional Models',
Nucl.\ Phys.\ B174 (1980) 166.
%
\bibitem{CT}
S.\ Coleman and H.J.\ Thun,
{`On the prosaic origin of the double poles in the sine-Gordon
S-matrix',}
Commun. Math. Phys. 61 (1978) 31
%
\bibitem{pedrev}
P.~Dorey,
{`Exact S-matrices'},
Proceedings of the 1996
E\"otv\"os Graduate School (Springer 1997)
85--125, {\tt hep-th/9810026}
%
\bibitem{CDS}
E.\ Corrigan, P.E.\ Dorey and R.\ Sasaki,
{`On a generalised bootstrap principle',}
Nucl. Phys. B408 (1993) 579, {\tt hep-th/9304065} 
%
\bibitem{DTW}
P.\ Dorey, R.\ Tateo and G.M.T.\ Watts,
`Generalisations of the Coleman-Thun mechanism and boundary
reflection factors', Phys. Lett. B448 (1999) 249, 
{\tt hep-th/9810098}
%
\bibitem{D4}
H.W.\ Braden, E.\ Corrigan, P.E.\ Dorey and R.\ Sasaki,
{`Extended Toda field theory and exact S-matrices',}
Phys. Lett. B227 (1989) 411
%
\bibitem{AP}
A.\ Pocklington,
`Bulk and boundary scattering in the q-state Potts model',
Ph.D thesis, Durham, 1998
%
%
\bibitem{BCDSb}
H.W.\ Braden, E.\ Corrigan, P.E.\ Dorey and R.\ Sasaki,
{`Multiple poles and other features of affine Toda field theory'}
Nucl. Phys. B356 (1991) 469\toline{498}
%
\bibitem{BCDSa}
H.W.\ Braden, E.\ Corrigan, P.E.\ Dorey and R.\ Sasaki,
{`Affine Toda field theory and exact S-matrices'},
Nucl. Phys. B338 (1990) 689\toline{746}
%
\bibitem{Martins:1991jj}
M.~J.~Martins,
`The off critical behavior of the multicritical Ising models',
Int.\ J.\ Mod.\ Phys.\ A7 (1992) 7753
%
\bibitem{Ko}
A.\ Koubek, 
{`S matrices of $\phi_{1,2}$-perturbed minimal models: IRF formulation and 
bootstrap program',} Int. J. Mod. Phys. A9 (1994) 1909
%
\bibitem{Koubek:1991qq}
A.~Koubek, M.~J.~Martins and G.~Mussardo,
`Scattering matrices for $\phi_{1,2}$ perturbed conformal minimal models
in absence of kink states',
Nucl.\ Phys.\ B368 (1992) 591
%
\bibitem{PED}
P.\ Dorey,
{`Root systems and purely elastic S matrices',}
Nucl. Phys. B358 (1991) 654\toline{676}
%
\bibitem{PEDb}
P.\ Dorey,
{`Root systems and purely elastic S matrices II',}
Nucl. Phys. B374 (1992) 741\toline{761},
{\tt hep-th/9110058}
%
\bibitem{Oota}
T.\ Oota, `q deformed Coxeter element in nonsimply
    laced affine Toda field theories',
    Nucl.\ Phys.\ B504 (1997) 738\toline{752}, {\tt hep-th/9706054}
%
\bibitem{Fring:1991gh}
A.\ Fring and D.I.\ Olive,
`The Fusing rule and the scattering matrix of affine Toda theory',
Nucl.\ Phys.\ B379 (1992) 429\toline{447}
%
\bibitem{Dorey:1992gr}
P.~Dorey,
`Hidden geometrical structures in integrable models',
in the proceedings of the NATO ARW {\em Integrable quantum field
theories}, Como (Plenum 1993),
{\tt hep-th/9212143}
%
\bibitem{Fring:1999jk}
A.\ Fring, C.\ Korff and B.J.\ Schulz,
`On the universal representation of the scattering matrix of affine
Toda  field theory',
Nucl.\ Phys.\ B567 (2000) 409\toline{453},
{\tt hep-th/9907125}
%
\bibitem{SalKau}
H.\ Saleur and B.\ Wehefritz-Kaufmann,
{`Thermodynamics of the complex SU(3) Toda  theory',}
Phys. Lett. B481 (2000),  419\toline{426},
{\tt hep-th/0003217}
%
\bibitem{mussgab}
G.\ Mussardo and G.\ Takacs,
unpublished
%
\bibitem{bigyellowbook}
P.~Di Francesco, P.~Mathieu and D.~Senechal,
{\it Conformal Field Theory},
New York, USA: Springer (1997)
%
%
\bibitem{Vogel}
P.\ Vogel,
`Algebraic structures on modules of diagrams',
{\it preprint} 
%
\bibitem{Del}
P.\ Deligne, `La s\'erie exceptionnelle des groupes de Lie',
C. R. Acad. Sci. Paris 322, S\'erie 1 (1996) 321\toline{326}
%
\bibitem{Cohman}
A.M.\ Cohen and R.\ de Man, `Computational evidence for Deligne's
conjecture regarding exceptional groups',
C. R. Acad. Sci. Paris 322, S\'erie 1 (1996) 427\toline{432}
%
\bibitem{DelII}
P.\ Deligne et R.\ de Man, `La s\'erie exceptionelle des groupes de
Lie II',
C. R. Acad. Sci. Paris 323, S\'erie 1 (1996) 577\toline{582}
%
\bibitem{Cvit}
P.\ Cvitanovic, {\em Classics illustrated: Group Theory}, Nordita
notes (1984); and {\em Group Theory} webbook at {\tt
http://www.nbi.dk/GroupTheory/}
%
\bibitem{landsberg}
J.M.\ Landsberg and L.\ Manivel,
`Representation theory and projective geometry',
{\tt math.AG/0203260}
%
\bibitem{Macf}
A.J.\ Macfarlane and H.\ Pfeiffer,
`Representations of the exceptional and other Lie algebras with
integral eigenvalues of the Casimir operator',
{\tt math-ph/0208014}
%
\bibitem{niall}
N.J.\ MacKay,
`Rational K-matrices and representations of twisted Yangians',
J. Phys. A35 (2002) 7865\toline{7876},
{\tt math.qa/0205155}
%
\bibitem{Ennes:1997vt}
I.P.~Ennes, A.V.~Ramallo and J.M.~Sanchez de Santos,
`$osp(1|2)$ conformal field theory',
in the proceedings of the CERN-Santiago de Compostela-La Plata Meeting
{\em Trends in Theoretical Physics}, La Plata 1997,
{\tt hep-th/9708094}



\end{thebibliography}
\end{document}